\documentclass[12pt]{article}
\usepackage[utf8]{inputenc}
\usepackage[english,activeacute]{babel}
\usepackage[margin=0.6in]{geometry}
\usepackage[longnamesfirst,numbers,square]{natbib}
\usepackage[pdfpagemode=UseNone,pdfstartview={XYZ null null null}]{hyperref}
\usepackage{xcolor}
\usepackage{amsthm}
\usepackage{graphicx}
\usepackage{bbm}
\usepackage{longtable,multirow}
\usepackage{amssymb}
\usepackage{stmaryrd}
\usepackage[intlimits]{amsmath}
\usepackage{amsfonts}
\usepackage{subcaption}
\usepackage{float}
\usepackage{color}
\usepackage{comment}
\usepackage{makecell}
\usepackage{enumitem}
\usepackage[affil-it]{authblk}
\numberwithin{equation}{section}
\newcommand{\R}{\mathbb{R}}
\newtheorem{remark}{Remark}
\newcommand{\mb}[1]{ \mbox{\boldmath$#1$} }

\newcommand{\beq}{\begin{eqnarray}}
\newcommand{\eeq}{\end{eqnarray}}
\newcommand{\beqq}{\begin{eqnarray*}}
\newcommand{\eeqq}{\end{eqnarray*}}
\newcommand{\p}{\partial}

\newcommand{\eps}{\varepsilon}
\newcommand{\x}{\mbox{\boldmath$x$}}
\newcommand{\bu}{\mbox{\boldmath$u$}}
\newcommand{\bp}{\mbox{\boldmath$p$}}
\newcommand{\bgamma}{\mbox{\boldmath$\gamma$}}
\newcommand{\bbm}{\mbox{\boldmath$m$}}

\newcommand{\y}{\mbox{\boldmath$y$}}

\newcommand{\n}{\mbox{\boldmath$n$}}

\newcommand{\w}{\mbox{\boldmath$w$}}

\newcommand{\pt}{\tilde{p}}
\newcommand{\jt}{\tilde{j}}

\newtheorem{Defi}{Definition}

\newtheorem{lem}{Lemma}

\newtheorem{prop}{Proposition}
\newtheorem{coro}{Corollary}

\font\bb=msbm10 at 12pt

\newcommand{\hp}{\hat{p}}

\def\eE{\hbox{\bb E}}
\begin{document}
\title{From Distribution of Extremes With Degradation to the Stress Redundancy Principle}
\author[1]{F.~Paquin-Lefebvre}
\author[1]{É.~Delbreil}
\author[1]{D.~Holcman}
\affil[1]{Group of Data Modeling and Computational Biology, IBENS, Ecole Normale Supérieure -- Université PSL, 75005 Paris, France.}
\maketitle
\begin{abstract}
How a degradation source affects the statistics of the first among many identical independent stochastic particles to successfully reach a narrow target remains unclear. Here, we investigate the impact of degradation on the distributions of first-passage times and exit locations by exploiting the time-dependent fluxes associated with the corresponding Fokker–Planck equation. We also compute the escape probability by solving a mixed boundary value problem. In the large initial population $N_0$ limit, we derive an asymptotic formula for the mean first passage time associated with the fastest particle, revealing how the killing field modifies the classical $1/\log(N_0)$ scaling. However, when considering the population size at the time of the first exit, we obtain a specific dependency with respect to degradation, a phenomenon that we call the stress redundancy principle. For a degradation field restricted to a narrow subregion, we use an optimal control approach to derive the optimal trajectory where the fastest escape particles will be concentrated: interestingly, we report here that the optimal trajectories avoid the degradation area depending on killing intensity. Finally, we obtain analytical expressions for the distribution of first exit points along a narrow absorbing arc in two-dimensional regular domains. In addition, for a killing subregion tangent to the target boundary, our computations reveal how geometrical parameters influence the exit distribution. In summary, a killing source can significantly affect the statistics of the fastest diffusing particles, relevant for defining cellular and molecular time scales.
\end{abstract}
\section{Introduction}
Extreme narrow escape theory describes statistical moments and distributions of the first among many independent identical redundant stochastic particles, drawn from the same initial density, to reach a small hidden absorbing target \cite{schuss2019redundancy, basnayake2020extremecusp,basnayake2018PLA,basnayake2019asymptotic,lawley2020distribution,lawley2020proba}. Such fastest first passage problems \cite{basnayake2019fastplos,Grebenkov2024chapter} have been widely investigated in the limit of a large number of searchers, for instance on graphs and networks \cite{dora2020,lawley2020PRevE}, with initial particles injected at a continuous rate \cite{akame2026accelerated,grebenkov2025} and under the competitive effect of slow and fast diffusive regimes \cite{lawley2020JPhysA}. \\
In cell biology, these fast events define activation time scales for many cellular and physiological processes \cite{schuss2019redundancy}, including calcium signal amplification in dendritic spines by the first few ions \cite{basnayake2019fastplos}, transmembrane channel activation \cite{katz2011} and gene transcription within the nucleus \cite{lucas2014}. Events for the fastest have been used to define the time where a cell lineage terminates (senescence), when the shortest telomere reach a given threshold following each cell division \cite{daoducPRL,Kimsy2018}. This modeling paradigm, along with the redundancy principle assessing that large numbers are not a waste, but fulfill an optimal search strategy \cite{schuss2019redundancy}, is also relevant for characterizing the large number of spermatozoa in fertility: this large number of sperm cells is necessary to guarantee a compatible small egg search with the physiological constraints and size of the uterus \cite{reynaud2015so,yang2016search}. But spermatozoa could be affected by external degradation sources \cite{sakkas2015}, thereby affecting their probability of reaching the egg within a reasonable time. A similar situation appears during the early cellular viral infection phase, where in some cases viral particles need to reach small nuclear pores \cite{lagache2009,Holcman2015,lagache2009physical} before being degraded by cellular defense mechanisms \cite{alberts2013essential}. In the endoplasmic reticulum organelle, degradation is an important response to diffusing unfolded proteins before they can reach exit sites, thus correcting a stress that could eventually lead to cellular death and has been linked to several human diseases \cite{tsai2010unfolded,zhang2020endoplasmic}. Calcium ions diffusing within the endoplasmic reticulum are also affected by extrusion and long binding processes: these are caused by buffers such as calmodulin or SERCA pumps, which could prevent fast signal activation \cite{basnayake2019fastplos}. \\
Modeling target search with evanescent or mortal diffusing agents has used one-dimensional geometries \cite{marchewka2005,meerson2015mortality,Abad2024} or focused on uniform degradation field \cite{ma2020strong}. The goal of this manuscript is to study how two-dimensional heterogeneous killing sources can affect the exit time of the fastest, the optimal paths where the fastest particles can be found and the associated exit point distribution. We will consider the fast arrival of the first particle when the initial pool of size $N_0$ is large, surviving to a killing field $k(\x)$ giving at each position $\x$ the rate per unit time at which stochastic trajectories are terminated. The killing term thus reduces the single particle escape probability and the number of surviving particles. Interestingly, when the killing is restricted to a subregion, we shall see how the spatial distribution of the fastest escape trajectories is affected although the fastest random particle can enter multiple times the killing zone without disappearing, contrasting with an absorbing target where trajectories are ended with probability one. We focus on two-dimensional diffusion models, but with appropriate Green's functions many of our results can be extended to arbitrary dimensions. \\
The manuscript is organized as follows: we start with the stochastic modeling framework and calculate the distribution of surviving particles when the first particle exits, and the probability of observing an escape versus a killing as first event in sections \S \ref{sec:def_stoch_model}. We also show in section \S \ref{sec:exitpoint_flux} how the distribution of first exit points can be computed by normalizing the outward exit flux through a narrow absorbing arc (target). In section \S \ref{sec:FPE}, we study the Fokker-Planck equation with uniform killing, and derive a formula for the mean first passage time of the fastest in the large initial population limit. In section \S \ref{sec:simulations}, we perform stochastic simulations that we compare with our asymptotic formula for the single-particle escape probabilities, obtained by detailed Green's function computations (appendix \S \ref{sec:escapeProba}). For a killing field restricted to a localized subregion between the target and the source (section \S \ref{sec:optimal_path}), we use a classical large deviation principle \cite{freidlin1998random} to derive the variational problem, leading to optimal paths associated with the fastest escape trajectories that agrees with our stochastic simulation results (\S \ref{sec:num_shooting}). In section \S \ref{sec:explicit_exitpoint}, we develop an asymptotic theory for the exit points distribution of the fastest, along an arc that uses a half-plane approximation for the exit flux when the killing is uniform (\S \ref{sec:rho_uniform_killing}). For non-uniform killing sources, such as localized killing zones hiding the absorbing target, a WKB asymptotic analysis reveals how the distribution is modified by the shortest geodesic, obtained by solving am optimal control problem (\S \ref{sec:rate-function-killing}). Finally, we conclude in section \S \ref{sec:discuss} by discussing a few open questions and some applications.
\subsection*{Summary of main results}
We summarize here key results that can be found in our manuscript.
\begin{itemize}[topsep=0pt,itemsep=0pt,partopsep=0pt,parsep=0pt]
\item We derive formulas for the fastest among $N_0$ particles to exit from a 2-D regular domain for both uniform (Eq.~\eqref{eq:tau_narrow}) and nonuniform (Eq.~\eqref{eq:tNk_explicit_quadratic_clean}) killing field $k(\x)$, revealing how killing influences extreme statistics.
\item We obtain explicit expressions (Eq.~\eqref{eq:escapeProbaNeu} and \eqref{eq:split_prob}) for the escape probability in terms of geometrical parameters and killing rates (detailed Green's function computations presented in appendix \S \ref{sec:escapeProba}).
\item We compute the distribution of survival particles at the time of first exit (eq.~\eqref{eq:distribution_first_exit}) and the expectation (Eq.~\eqref{eq:expected_nb_survival}).
\item We derive a formula for the probability of observing a killing as a first event (Eq.~\eqref{eq:proba_escape_vs_killing}). In particular for a uniform killing this probability converges to zero exponentially fast (Eq.~\eqref{eq:proba_escape_before_killing_equiv}) in the large population limit $N_0 \to \infty$.
\item We compute optimal trajectories where the fastest particles get concentrated using an optimal control approach. The trajectories of the fastest are located in a close neighborhood of the optimal trajectories  (Eq.~\eqref{eq:canonical}), where the associated Hamiltonian is constant (along each optimal trajectory), as revealed by Eq.~\eqref{eq:cte_hamiltonian}.
\item We analyze the distribution of exit points for the fastest among $N_0$ particles and show that it depends on the location of the minimizers of the optimal-control problem on the narrow absorbing arc. This result follows from a WKB asymptotic analysis of the Fokker-Planck equation with a non-uniform killing (section \S \ref{sec:rate-function-killing}).
\item For a localized killing subregion between the source and the target, we compute the distribution of exit points and show that it is controlled by the position of the killing region. In fact, what matters is the position of the disk with respect to the narrow absorbing arc (section \S \ref{sec:tangent_two_optimal} and \ref{sec:simulated_exit_point}).
\item Our asymptotic formulas are validated against stochastic simulations performed with the classical Euler's scheme as described in appendix \S \ref{sec:algo_simu}.
\end{itemize}
\section{Stochastic properties of Brownian escape with killing}\label{sec:def_stoch_model}
\subsection{Background on Fokker-Planck equation}
We recall here some known results \cite{karlin1981,Schuss2009Book,Holcman2015} and present also some new properties of stochastic events of a process $\x(t)$, representing the motion of random particles evolving in a domain $\Omega$, that satisfies
\beq\label{eqstochastic}
d\x = \mb{b}(\x)\,dt + \sqrt{2}\mb{B}(\x)\,d{\mb{w}}(t) \quad \mbox{for} \quad \x\in\Omega,
\eeq
where $\mb{b}(\x)$ is a smooth drift vector, $\mb{B}(\x)$ is a diffusion tensor, and $\mb{w}(t)$ is a vector of independent standard Brownian motions. The domain boundary is decomposed as $\p\Omega=\p\Omega_a\cup\p\Omega_r$, where $\p\Omega_r$ is reflecting while $\p\Omega_a$ consists of a small absorbing part. \\
A killing measure $k(\x)$ is added to the process \eqref{eqstochastic} to model the removal of particles inside the domain \cite{karlin1981}. The transition probability density function (pdf) of the process $\x(t)$ with killing and absorption is the pdf of trajectories that have neither been killed nor absorbed in $\p\Omega_a$ at time $t$, this is \cite{marchewka2005}
\beq \label{pxTtau}
p(\x,t\mid\x_0)\,d\x=\Pr\{\x(t)\in\x+d\x,\,\tau^k>t,\, \tau^e>t \mid \x(0) = \x_0\},
\eeq
and $\tau^k$ and $\tau^e$ are random times at which the particle is either killed or absorbed. This pdf satisfies the Fokker-Planck equation (FPE) \cite{Schuss2009Book}
\beq\label{FPEp}
\frac{\p p(\x,t\mid\x_0)}{\p t}={\cal L}_{\x} p(\x,t\mid\x_0)-k(\x)p(\x,t\mid\x_0)\hspace{0.5em}\ \mbox{for}\ \x,\x_0 \in \Omega,
\eeq
where ${\cal L}_{\x}$ is the forward operator
\beq\label{FO}
{\cal L}_{\x}p(\x,t\mid\x_0)=\sum_{i,j=1}^d\frac{\p^2\sigma^{i,j}(\x)p(\x,t\mid\x_0)}{\p x^i\p x^j} -\sum_{i=1}^d\frac{\p b^i(\x)p(\x,t\mid\x_0)}{\p x^i},
\eeq
and $\mb{\sigma}(\x)=\frac12\mb{B}(\x)\mb{B}^T(\x)$. With the divergence the forward operator can be written as ${\cal L}_{\x}p(\x,t\mid\x_0)=-\nabla\cdot\mb{J}(\x,t\mid\x_0)$, where the components of the flux density vector $\mb{J}(\x,t\mid\x_0)$ are
\beq
J^i(\x,t\mid\x_0)=-\sum_{j=1}^d\frac{\p\sigma^{i,j}(\x)p(\x,t\mid\x_0)}{\p x^j}+b^i(\x)p(\x,t\mid\x_0),\hspace{0.5em}(i=1,2,\ldots,d)\,,
\eeq
Initial and boundary conditions for the FPE \eqref{FPEp} are
\beq
p(\x,0\mid\x_0)&=&\,\delta(\x-\x_0) \quad \mbox{for} \quad \x,\x_0 \in \Omega \label{ICdD} \\
p(\x,t\mid\x_0)&=&\,0 \quad \mbox{for} \quad  t>0, \quad \x\in\p \Omega_a, \quad \x_0 \in \Omega \label{BCdD}\\
j(\x,t\mid\x_0)&=&\,0 \quad \mbox{for} \quad  t>0, \quad \x\in\p\Omega_r, \quad \x_0 \in \Omega \,, \label{noflux}
\eeq
where $j(\x,t\mid\x_0)$ is the scalar outward flux density defined as
\begin{equation}\label{eq:scalar_flux_density}
j(\x,t\mid\x_0)=\mb{J}(\x,t\mid\x_0) \cdot \n(\x), \quad \x \in \p\Omega_a\,,
\end{equation}
with $n(\x)$ the outward normal unit vector to the boundary. The total flux through $\p\Omega_a$ is thus given by
\beq\label{eq:flux1}
J(t\mid\x_0)=\oint_{\p\Omega_a} j(\x,t\mid\x_0) \,dS_{\x}\,,
\eeq
while the survival function $S(t\mid\x_0)$ of trajectories still diffusing and alive at time $t$ is
\beq
S(t\mid\x_0) = \Pr\{\tau^e > t \, , \, \tau^k > t \mid \x(0) = \x_0\} = \int_\Omega p(\x,t\mid\x_0) d\x \,.
\eeq
By then integrating over past history we get the fundamental identity \cite{marchewka2005},
\beq\label{eq:identity}
F(t\mid \x_0) + S(t \mid \x_0) + K(t \mid \x_0) = 1\,,
\eeq
where $F(t \mid \x_0)$ and $K(t \mid \x_0)$ are the probabilities to either escape or be killed before time $t$, defined as
\beq
F(t\mid \x_0) = \int\limits_0^t J(s\mid\x_0) ds, \quad K(t\mid \x_0) = \int_0^t \int_{\Omega} k(\x) p(\x,s\mid\x_0)d\x ds\,.
\eeq
Hence $F(t\mid \x_0)$ corresponds to the single particle escape success cumulative distribution function: the limit $t\to \infty$ gives the splitting probability $P_e(\x_0)$ to reach the target before being killed
\[
P_e(\x_0) = \Pr\{\tau^e < \tau^k \mid \x(0) = \x_0\} = \lim_{t\to \infty} F(t\mid\x_0) = 1 - \lim_{t\to \infty} K(t\mid \x_0)\,,
\]
which can be expressed as
\beq\label{eq:Pe}
P_e(\x_0) = 1 - \int_\Omega k(\x) \pt(\x\mid\x_0)d\x, \qquad \text{with} \qquad \pt(\x\mid\x_0) = \int\limits_0^\infty p(\x,t\mid\x_0) dt\,.
\eeq
Here $\pt(\x\mid\x_0)$ is solution of a steady mixed boundary value problem \cite{holcman2007stat,lagache2009},
\begin{subequations}\label{eq:BVP}
\begin{align}
\mathcal{L}_{\x}\pt(\x\mid\x_0) - k(\x)\pt(\x\mid\x_0) = -\delta(\x-\x_0)\,, \quad \x,\,\x_0 \in \Omega\,,
\end{align}
subject to the boundary conditions
\begin{align}
\pt(\x\mid\x_0) \,\, = \,\, 0, \quad \x \in \p\Omega_a, \,\, \x_0 \in \Omega, \qquad \jt(\x \mid \x_0) = \int\limits_0^\infty j(\x,t \mid \x_0) dt \,\, = \,\, 0\,, \quad \x \in \p\Omega_r, \,\, \x_0 \in \Omega\,.
\end{align}
\end{subequations}
\subsection{Probability distribution of the fastest escape time} \label{sec:EMFPT}
We shall now present a generic expression for the mean escape time for the fastest to reach the target, starting with $N_0$ independent identically distributed particles driven by \eqref{eqstochastic} and the killing measure $k(\x)$. For $i=1,\ldots,N_0$, using the time $\tau_i^e$ at which the $i$-th particle escapes the domain $\Omega$ for the first time and the time $\tau_i^k$ at which it is killed, the effective exit time $\tau_i$ is
\beq \label{eq:effective}
\tau_i =
\left\{
\begin{array}{ll}
\tau_i^e & \mbox{if } \tau_i^k > \tau_i^e \\
+ \infty & \mbox{otherwise}
\end{array}\,,
\right. \quad i = 1, \ldots, N_0 \,,
\eeq
hence $\tau_i$ is infinite if the particle is killed before exiting otherwise it is equal to the exit time. The first exit time is thus the minimum of all the $\tau_i$,
\beq\label{eq:tau_exit_N_0}
\tau^{\rm exit} = \min \left\{\tau_1, \ldots, \tau_{N_0} \right\}\,,
\eeq
with corresponding survival probability
\begin{equation}\label{eq:tau_exit_survival}
S_{\tau^{\rm exit}}(t) = \Pr(\tau^{\rm exit} > t) = \left( 1 - F(t\mid\x_0) \right)^{N_0} \,,
\end{equation}
and probability density function
\begin{equation}\label{eq:tau_exit_density}
f_{\tau^{\rm exit}}(t) = -S_{\tau^{\rm exit}}'(t) = N_0 J(t\mid\x_0) \left( 1 - F(t\mid\x_0) \right)^{N_0 - 1}\,.
\end{equation}
The probability of no successful escape is
\beq
\Pr\left(\tau^{\rm exit} = \infty\right) = \left(1 - F(\infty\mid\x_0)\right)^{N_0} = \left(1 - P_e(\x_0) \right)^{N_0}\,,
\eeq
and thus the probability of observing at least one success is $\Pr\left(\tau^{\rm exit} < \infty\right) = 1 - \left(1 - P_e(\x_0) \right)^{N_0}$, with which we define the conditional expectation of $\tau^{\rm exit}$ assuming at least one successful escape: for $N_0$ large this corresponds to the extreme mean first passage time (EMFPT) with killing,
\beq
\bar\tau_{\rm EMFPT}(N_0) = \mathbb{E}(\tau^{\rm exit}\mid\tau^{\rm exit} < + \infty) = \int\limits_0^{+\infty} \frac{\Pr(\tau^{\rm exit} > t) - \Pr(\tau^{\rm exit} = \infty)}{\Pr(\tau^{\rm exit} < \infty)} dt \,.
\eeq
In terms of boundary fluxes this expectation formula becomes \cite{lagache2009physical,Holcman2015},
\beq\label{eq:tau1}
\bar\tau_{\rm EMFPT}(N_0) = \int\limits_0^{+\infty}\left(\frac{\left(1 - F(t\mid\x_0)\right)^{N_0} -\left(1-P_e(\x_0)\right)^{N_0}}{1-\left(1-P_e(\x_0)\right)^{N_0}}\right)\, dt\,.
\eeq
\subsection{Distribution of surviving particles at first exit}\label{sec:distribution_first_exit}
We derive an expression for the full distribution and expectation of surviving particles at first successful escape, i.e.~when $t = \tau^{\rm exit}$. We consider the number $n(t)$ of particles at time $t$ whose trajectories have not been terminated by the killing field: these are either still undergoing diffusion or have escaped already. If the first escape occurs within the interval $[t,t+dt]$, then $q-1$ must still be diffusing while $N_0-q$ were killed: this gives the density
\beq\label{eq:distribution_first_exit}
\Pr\left(n\left(\tau^{\rm exit}\right) = q, \,\, \tau^{\rm exit} \in dt \right) = N_0\binom{N_0-1}{q-1}J(t\mid\x_0) S(t\mid\x_0)^{q-1}K(t\mid\x_0)^{N_0-q} dt\,,
\eeq
and integrating with respect to time gives the probability that $q$ particles avoided killing at time of first successful exit,
\beq
\Pr\left(n\left(\tau^{\rm exit}\right) = q\right) = \int\limits_0^{+\infty} N_0 \binom{N_0-1}{q-1} J(t \mid \x_0)S(t\mid\x_0)^{q-1}K(t\mid\x_0)^{N_0-q} dt\,.
\eeq
The expectation $\mathbb{E}\left(n(\tau^{\rm exit})\right)$ of the survival distribution at first exit then corresponds to the sum
\begin{align*}
\mathbb{E}\left(n\left(\tau^{\rm exit}\right)\right) &= \sum\limits_{q=1}^{N_0}q \Pr\left(n\left(\tau^{\rm exit}\right)=q\right) \\
&= \int\limits_0^{+\infty} N_0 J(t \mid \x_0) \left( \sum_{q=1}^{N_0}q\binom{N_0-1}{q-1}S(t\mid\x_0)^{q-1}K(t\mid\x_0)^{N_0-q} \right) dt \,,
\end{align*}
which explicitly evaluates as
\beq
\mathbb{E}\left(n\left(\tau^{\rm exit}\right)\right) = \int_0^\infty N_0 J(t\mid \x_0) \left( S(t \mid \x_0) + K(t \mid \x_0) \right)^{N_0-2}\left(K(t\mid \x_0) + N_0 S(t\mid \x_0)\right) dt \,,
\eeq
and then using identity \eqref{eq:identity} we get
\beq
\mathbb{E}\left(n\left(\tau^{\rm exit}\right)\right) = \int_0^\infty N_0 J(t\mid \x_0) \left( 1 - F(t \mid \x_0) \right)^{N_0-2}\left(1 - F(t\mid \x_0) + (N_0-1) S(t\mid \x_0)\right) dt \,.
\eeq
and finally integrating by parts yields the formula
\beq\label{eq:expected_nb_survival}
\mathbb{E}\left(n\left(\tau^{\rm exit}\right)\right) = N_0 \left(1  - \int\limits_0^\infty \left( 1 - F(t\mid\x_0) \right)^{N_0-1} \left(\int_\Omega k(\x) p(\x,t\mid\x_0) d\x\right) dt  \right) \,.
\eeq
Interestingly when $N_0=1$ we recover the single particle escape probability \cite{Holcman2015},
\beq
\mathbb{E}\left(n\left(\tau^{\rm exit}\right)\right) = 1 - \int\limits_0^\infty \int_\Omega k(\x) p(\x,t\mid\x_0) d\x dt = \int \limits_0^\infty J(t \mid \x_0) dt \,.
\eeq
\subsection{Probability of escape versus killing as a first event}\label{sec:prob_escape_first}
To compute the probability of observing an escape rather than a killing as a first event, we introduce the first escape $\tau_{(N_0)}^e$ and killing $\tau_{(N_0)}^k$ times
\beq
\tau_{(N_0)}^e = \min \left(\tau_1^e, \ldots, \tau_{N_0}^e \right) \quad \text{and} \quad \tau_{(N_0)}^k = \min \left(\tau_1^k, \ldots, \tau_{N_0}^k\right)\,.
\eeq
The probability that none of the $N_0$ initial particles undergoes escape or killing before $t$ is
\beq
\Pr\left( \tau_{(N_0)}^e > t, \tau_{(N_0)}^k > t \right) = S(t\mid\x_0)^{N_0}\,,
\eeq
while the probability of observing a single escape and no killing within the interval $[t,t+dt]$ is
\beq
\Pr\left( t < \tau_{(N_0)}^e < t + dt, t + dt < \tau_{(N_0)}^k \right) = N_0J(t\mid \x_0)S(t \mid \x_0)^{N_0-1}dt
\eeq
and integrating with respect to time gives the probability of the first event being an escape
\beq
\Pr\left(\tau_{(N_0)}^e < \tau_{(N_0)}^k\right) = N_0 \int_0^\infty J(t\mid \x_0) S(t\mid \x_0)^{N_0 - 1}dt\,.
\eeq
Alternatively this is
\beq
\Pr\left(\tau_{(N_0)}^e < \tau_{(N_0)}^k\right) = - N_0 \int_0^\infty \left( \frac{d}{dt}S(t \mid \x_0) + \int_\Omega k(\x) p(t \mid \x_0) d\x \right) S(t\mid \x_0)^{N_0 - 1}dt\,,
\eeq
and upon integrating we get
\beq\label{eq:proba_escape_vs_killing}
\Pr\left(\tau_{(N_0)}^e < \tau_{(N_0)}^k\right)= 1 - N_0 \int\limits_0^\infty S(t\mid \x_0)^{N_0-1} \left( \int_\Omega k(\x) p(t \mid \x_0) d\x \right) dt \,,
\eeq
which corresponds to the probability complementary to that of observing a killing event first,
\beq
\Pr\left(\tau_{(N_0)}^k < \tau_{(N_0)}^e\right) = N_0 \int\limits_0^\infty S(t\mid \x_0)^{N_0-1} \left( \int_\Omega k(\x) p(t \mid \x_0) d\x \right) dt\,,
\eeq
since the first event is necessarily either an escape or a killing,
\beq\label{eq:twoEvents}
\Pr\left(\tau_{(N_0)}^e < \tau_{(N_0)}^k\right) + \Pr\left(\tau_{(N_0)}^k < \tau_{(N_0)}^e\right) = 1\,.
\eeq
Finally for a single initial particle, Eq.~\eqref{eq:twoEvents} reduces to identity \eqref{eq:identity} evaluated in the limit $t \to \infty$:
\beq
\Pr\left(\tau_{(1)}^e < \tau_{(1)}^k\right) + \Pr\left(\tau_{(1)}^k < \tau_{(1)}^e\right) = \int_0^\infty J(t\mid \x_0) dt + \int\limits_0^\infty \int_\Omega k(\x) p(t \mid \x_0) d\x dt = 1\,.
\eeq
\subsection{Distribution of first exit points along a narrow absorbing boundary}\label{sec:exitpoint_flux}
We derive the exit points distribution on the narrow absorbing boundary $\p\Omega_a$ of the first successful escape among $N_0$ i.i.d.~particles evolving according to \eqref{eqstochastic} in the presence of killing $k(\x)$. We start from the escape probability through a small infinitesimal boundary element $dS_{\x}$ within the time interval $[t,t+dt]$, for a single particle $i$ this is
\begin{equation}\label{eq:single_joint_exit}
\Pr\{\tau_i^e\in[t,t+dt],\, \x(\tau_i^e)\in dS_{\x},\, \tau_i^e<\tau_i^k \mid \x(0) = \x_0,\,\, i = 1,\ldots N_0\} = j(\x,t\mid\x_0)\,dS_{\x}\,dt\,,
\end{equation}
and the probability that particle $i$ exits through $\x \in dS_{\x}$ during $[t,t+dt]$, with the remaining $N_0 - 1$ having no successful escape before time $t$, is
\begin{align}
\Pr\{\tau_i^e\in[t,t+dt],\,\, & \x(\tau_i^e)\in dS_{\x},\,\, \tau_i^e<\tau_i^k,\,\, \tau_i^e < \tau_j^e \,\, \text{for} \,\, j \neq i \mid \x(0) = \x_0,\,\, i = 1,\ldots N_0\} \nonumber \\
&= N_0 j(\x,t\mid\x_0) \left(1-F(t\mid\x_0)\right)^{N_0-1}dS_{\x}dt \,.
\end{align}
With $\y \in \p\Omega_a$ parametrizing the position along the narrow target, the joint time and space density $f_{N_0}(\y,t\mid\x_0)$ of the first successful escape is given by
\beq\label{eq:joint_fastest_xt}
f_{N_0}(\y,t\mid\x_0) = N_0 j(\y,t\mid\x_0) \left(1-F(t\mid\x_0)\right)^{N_0-1}, \quad \y \in \p\Omega_a, \quad t > 0\,,
\eeq
and integrating with respect to time yields the first exit-point density on $\p\Omega_a$,
\begin{equation}\label{eq:exitpoint_density_fastest}
p_{N_0}(\y\mid\x_0) = \int_0^\infty f_{N_0}(\y,t\mid\x_0)\,dt\,.
\end{equation}
We then define the spatial distribution of first exit points $p_{N_0}(\y \mid \tau^{\rm exit}=t,\x_0)$ conditioned on $\tau_{\rm exit} = t$,
\begin{equation}\label{eq:conditional_exitpoint_time}
p_{N_0}(\y \mid \tau^{\rm exit}=t,\x_0) = \frac{f_{N_0}(\y,t\mid\x_0)}{f_{\tau^{\rm exit}}(t)} = \frac{j(\y,t\mid\x_0)}{J(t\mid\x_0)}, \quad \y \in \p\Omega_a\,,
\end{equation}
which is independent of $N_0$, normalized on the narrow exit arc and denoted with $\rho(\y \mid t,\x_0)$ as follows
\beq\label{eq:rho_formula}
\rho(\y \mid t,\x_0) = \frac{j(\y,t\mid\x_0)}{J(t\mid\x_0)}, \quad \text{with} \quad \int_{\p \Omega_a} \rho(\y \mid t,\x_0) dS_{\y} = 1\,.
\eeq
We now seek an asymptotic description for the first exit points distribution $p_{N_0}(\y \mid \x_0)$ when $N_0$ is large, by rewriting Eq.~\eqref{eq:exitpoint_density_fastest} as
\beq\label{eq:pN-start-direct}
p_{N_0}(\y\mid\x_0) = \int_0^\infty N_0\,j(\y,t\mid\x_0)\Big(1-F(t \mid \x_0)\Big)^{N_0-1}dt\,,
\eeq
where for each $t>0$ we have $\left(1-F(t\mid \x_0)\right)^{N_0-1} \to 0$ when $N_0 \to \infty$, causing the integrand to vanish pointwise. This implies that dominant contributions come from very short times $t$, for which we write
\begin{equation}\label{eq:exp-approx}
\Big(1-F(t\mid \x_0)\Big)^{N_0-1} = \exp\!\Big((N_0-1)\log(1-F(t\mid \x_0))\Big) \approx \exp\!\Big(-N_0\,F(t\mid \x_0)\Big)\,,
\end{equation}
which is a uniform approximation given that $N_0\, F(t\mid\x_0)^2 \ll 1$: in  particular this is the case when $F(t\mid\x_0)\sim 1/N_0$. After substituting Eq.~\eqref{eq:exp-approx} within \eqref{eq:pN-start-direct} we get the Laplace-type integral
\begin{equation}\label{eq:pN-laplace-form}
p_{N_0}(\y \mid \x_0) \approx \int_0^\infty N_0 \,j(\y,t\mid \x_0)\,\exp\!\Big(-N_0\,F(t\mid \x_0)\Big)\,dt\,,
\end{equation}
from which we extract the leading order for large $N_0$ by changing variable with $u = N_0F(t\mid\x_0)$,
\begin{equation}\label{eq:pN-mixture-direct}
p_{N_0}(\y \mid \x_0) \approx \int_{0}^{N_0 P_e(\x_0)} \rho\big(\y\mid t(u),\x_0\big)\,e^{-u}\,du, \quad \text{with} \quad \rho\big(\y\mid t(u),\x_0\big) = \frac{j(\y,t(u)\mid \x_0)}{J(t(u)\mid \x_0)}\,,
\end{equation}
where $P_e(\x_0)$ is the single particle escape probability and $\rho(\y \mid t,\x_0)$ is the time-slice exit location density, normalized on $\partial\Omega_a$ for each fixed $t$ with $J(t \mid \x_0)>0$. Since $e^{-u}$ concentrates on $u=O(1)$, the relevant times satisfy $F(t\mid \x_0) = u/N_0 \to 0$, hence $t(u)\downarrow 0$. In particular, if $\rho(\y \mid t,\x_0)$ varies slowly on the time scale where $u=O(1)$, we can freeze it at the characteristic time $t_{N_0}$ satisfying $N_0 \, F(t_{N_0} \mid \x_0) \asymp 1$, and we get a formula for the first exit-point distribution,
\begin{equation}\label{eq:pN-leading-direct}
p_{N_0}(\y \mid \x_0) \approx \rho(\y \mid t_{N_0},\x_0), \quad \text{with} \quad \rho(\y \mid t_{N_0},\x_0) = \frac{j(\y,t_{N_0}\mid \x_0)}{J(t_{N_0} \mid \x_0)},
\qquad \y \in \p\Omega_a, \quad N_0 \gg 1\,.
\end{equation}
Finally, conditioning on at least one escape among $N_0$ particles yields the density $p_{N_0}^{\rm succ}(\y\mid\x_0)$,
\begin{equation}\label{eq:exitpoint_density_conditional_success}
p_{N_0}^{\rm succ}(\y\mid\x_0) = \frac{p_{N_0}(\y\mid\x_0)}{1 -\left(1-P_e(\x_0)\right)^{N_0}} \approx \rho(\y \mid t_{N_0},\x_0), \quad N_0 \gg 1\,,
\end{equation}
but for $N_0$ large both conditioned and unconditioned distributions are identical.
\section{Asymptotic expression for the extreme mean first passage time with killing}\label{sec:FPE}
Here we derive an asymptotic formula for the extreme mean first passage time of the fastest trajectory from an initial pool of $N_0$ stochastic particles subject to a uniform killing measure $k(x) = V$ and narrow exit $\p\Omega_a$ of radius $\eps$ and centered in $\x_a \in \p\Omega$. This computation relies on evaluating Eq.~\eqref{eq:tau1} in the large $N_0$ limit for a fixed geometry. The strategy is to solve the Fokker-Planck equation for the pdf $p(\x,t\mid\x_0)$ in the short-time asymptotic regime,
\begin{subequations}\label{eq:FPE}
\begin{align}
\frac{\p p(\x,t\mid\x_0)}{\p t} = D \Delta p(\x,t\mid\x_0) - V p(\x,t\mid\x_0), \quad \x \in \Omega, \quad t > 0 \,,
\end{align}
with initial $p(\x,0\mid\x_0) = \delta(\x-\x_0)$ and boundary conditions
\begin{align}
p(\x,t\mid\x_0) = 0, \quad \x \in \p\Omega_a, \quad \n \cdot \nabla p(\x,t\mid\x_0) = 0, \quad \x \in \p\Omega_r.
\end{align}
\end{subequations}
We use the frequency domain with the Laplace transform $\hp(\x,q\mid\x_0)$ given by
\beq
\hp(\x,q\mid\x_0) = \mathcal{L}\left\{ p(\x,t\mid\x_0) \right\} = \int_0^\infty e^{-q t} p(\x,t\mid\x_0) dt\,.
\eeq
This allows us to reformulate Eq.~\eqref{eq:FPE} as a Helmholtz equation
\begin{subequations}\label{eq:BVPq}
\begin{align}
\Delta \hp(\x,q\mid\x_0) - \omega(q)^2 \hp(\x,q|\x_0) = - \frac{\delta(\x-\x_0)}{D}\,, \quad \x \in \Omega, \quad \text{with} \quad \omega(q) = \sqrt{\frac{q + V}{D}}\,,
\end{align}
with mixed boundary conditions
\begin{align}
\hp(\x,q\mid\x_0) = 0, \quad \x \in \p\Omega_a, \quad \n \cdot \nabla \hp(\x,q\mid\x_0) = 0, \quad \x \in \p\Omega_r\,.
\end{align}
\end{subequations}
To solve system \eqref{eq:BVPq}, we use the reduced-wave Green's function $\hat{G}(\x,q;\y)$, which satisfies \cite{chen2011stability,basnayake2019asymptotic}
\beq\label{eq:rwgreen_pde}
\Delta\hat{G}(\x,q;\y) - \omega(q)^2 \hat{G}(\x,q;\y)  = -\frac{\delta(\x-\y)}{D}, \quad \x, \y \in \Omega \,, \quad \n \cdot \nabla \hat{G}(\x,q;\y) = 0, \quad \x \in \p\Omega, \y \in \Omega\,.
\eeq
and gets solved as
\beq\label{eq:rwgreen}
\hat{G}(\x,q;\y)  = \frac{1}{\pi D}K_0\left(\omega(q) \|\x-\y\| \right) + U(\x,q;\y), \quad \x \in \Omega, \y \in \p\Omega\,,
\eeq
with $K_0(z)$ the order zero modified Bessel function of the second kind and $U(\x,q;\y)$ the regular part of the reduced-wave Green's function. Using Green's identity, we solve the boundary value problem \eqref{eq:BVPq} as
\beq\label{eq:greenID}
\hat{p}(\y,q\mid\x_0) = \hat{G}(\x_0,q;\y) - \int_{\partial \Omega_a} \hat{G}(\x,q;\y) \hat{j}(\x,q|\x_0) d\x \,,
\eeq
where $\hat{j}(\x,q|\x_0)$ is the flux defined as
\beq
\hat{j}(\x,q\mid\x_0) = - D \left( \n \cdot\nabla \hp(\x,q\mid\x_0)\right)\,, \quad \x \in \Omega, \quad q > 0 \,.
\eeq
We use a constant flux approximation,
\beq
\hat{j}(\x,q\mid\x_0) = \hat{j}(\x_a,q\mid\x_0), \quad \x \in \p\Omega_a\,,
\eeq
and then by setting $\y = \x_a$ in Eq.~\eqref{eq:greenID} and using the absorbing boundary condition we get
\beq
\hat{j}(\x_a,q\mid\x_0) = \frac{\hat{G}(\x_0,q;\x_a)}{\int_{\partial \Omega_a} \hat{G}(\x,q;\x_a) d\x} \,,
\eeq
which can be further evaluated using the solution \eqref{eq:rwgreen}
\beq
\hat{j}(\x_a,q\mid\x_0) = \frac{K_0(\omega(q)\|\x_0 - \x_a\|) + \pi D U(\x_0,q;\x_a)}{\int_{\p\Omega_a} \left( K_0(\omega(q)\|\x - \x_a\|) + \pi D U(\x,q;\x_a) \right) d\x} \,.
\eeq
On the target $\p\Omega_a$ we have $\omega(q) \|\x - \x_a\| < \omega(q) \eps \ll 1$ and the modified Bessel function behaves as \cite{chen2011stability,basnayake2019asymptotic},
\beq
K_0(\omega(q)\|\x - \x_a\|) \approx -\log(\omega(q)\|\x - \x_a\|) + O(1), \quad \x \in \p\Omega_a\,,
\eeq
which allows us to evaluate the boundary integral as
\beq
\int_{\partial \Omega_a} \hat{G}(\x,q;\x_a) d\x \approx 2\int_0^\eps \left( -\log\left(\omega(q) \rho \right) + O(1) \right)d\rho = 2\eps \left( -\log\left( \omega(q) \eps \right) + O(1) \right) \,,
\eeq
which yields
\beq\label{eq:fluxq}
\hat{j}(\x_a,q|\x_0) \approx \frac{K_0(\omega(q)\|\x_0 - \x_a\|)}{2\eps\left( \log\left( \frac{1}{\omega(q) \eps} \right) + O(1) \right)}\,.
\eeq
We note that $\hp(\x,0\mid\x_0) = \pt(\x\mid\x_0)$ is solution of Eq.~\eqref{eq:BVP}, thus we get a formula for the escape probability $P_e(\x_0)$,
\beq\label{eq:escapeProba}
P_e(\x_0) = \left. \left( \int_{\p\Omega_a} \hat{j}(\x,q\mid\x_0) d\x \right)\right|_{q = 0} \approx \frac{K_0(\omega(0) \left\|\x_0 - \x_a\right\|)}{\log\left( \frac{1}{\omega(0) \eps} \right)} \,,
\eeq
which holds if $\omega(0)\eps \ll 1$. The total flux $J(t\mid\x_0)$ through the narrow target is obtained by taking the inverse Laplace transform of Eq.~\eqref{eq:fluxq},
\beq
J(t\mid\x_0) = |\p\Omega_a| \mathcal{L}^{-1} \left\{ \hat{j}(\x_a,q\mid\x_0) \right\} \approx \mathcal{L}^{-1} \left\{ \frac{1}{\log\left( \frac{1}{\omega(q)\eps}\right) + O(1)} K_0\left(\omega(q)\|\x_0 - \x_a\|\right)\right\} \,,
\eeq
with the expansion,
\beq
\frac{1}{\log\left( \frac{1}{\omega(q)\eps}\right) + O(1)} \approx \frac{1}{\log\left( \frac{R}{\eps} \right)} \left( 1 - O\left( \left( \log\left( \frac{R}{\eps} \right)\right)^{-1} \right) \right)\,,
\eeq
which yields, with Mathematica \cite{Mathematica},
\beq\label{eq:flux_J}
J(t\mid\x_0) \approx \frac{1}{\log\left( \frac{R}{\eps} \right)} \mathcal{L}^{-1} \left\{ K_0\left(\omega(q)\|\x_0 - \x_a\|\right) \right\} = \frac{1}{2\log\left( \frac{R}{\eps}\right)t} \exp\left( -Vt - \frac{\|\x_0 - \x_a\|^2}{4Dt} \right) \,.
\eeq
By then integrating \eqref{eq:flux_J} on the range $[0,t]$ we get the probability $F(t\mid\x_0)$ to escape before $t$,
\beq
F(t\mid\x_0) = \int\limits_0^t J(s\mid\x_0) ds \approx \frac{1}{2\log\left( \frac{R}{\eps}\right)} \int_0^t \frac{1}{s}\exp\left(-Vs - \frac{\|\x_0 - \x_a\|^2}{4Ds}\right) ds \,,
\eeq
and by neglecting contributions from the killing term $e^{-Vs}$ assuming small time $t$ and with the change of variables $u = \|\x_0 - \x_a\|^2/(4Ds)$ we get
\beq
F(t\mid\x_0) \approx \frac{1}{2\log\left( \frac{R}{\eps}\right)} \int_{\|\x_0 - \x_a\|^2/(4Dt)}^\infty \frac{\exp\left( -u \right)}{u} du = \frac{\Gamma\left( 0, \frac{\|\x_0-\x_a\|^2}{4Dt} \right)}{2\log\left( \frac{R}{\eps} \right)}\,,
\eeq
where $\Gamma(0,z)$ is the incomplete Gamma function of order zero. With the expansion $\Gamma(0,z) \sim e^{-z}/z$ for $z \to \infty$ \cite{stegun1964}, for small time $t$ we find that $F(t\mid \x_0)$ behaves as
\beq\label{eq:intJsmalltime}
F(t\mid\x_0) &\approx& C t\exp\left(- \frac{a}{t} \right), \quad \text{with} \quad C = \frac{2D}{\|\x_0 - \x_a\|^2\log\left( \frac{R}{\eps}\right)}, \quad a = \frac{\|\x_0-\x_a\|^2}{4D} \,.
\eeq
We then recall formula \eqref{eq:tau1} for the extreme mean first passage time,
\beq\label{eq:tau_temp}
\bar\tau_{\rm EMFPT}(N_0) &=& \int\limits_0^\infty \frac{\left(1 - F(t\mid\x_0) \right)^{N_0} - \left(1 - P_e(\x_0) \right)^{N_0}}{1 - \left(1 - P_e(\x_0)\right)^{N_0}} dt\,,
\eeq
where $N_0$ and $P_e(\x_0) = F(\infty \mid \x_0)$ are the initial population size and single particle escape probability. Dominant contributions for this integral come from short times $t$, for which we expand the integrand as,
\beq
\frac{\left(1 - F(t\mid\x_0) \right)^{N_0} - \left(1 - P_e(\x_0) \right)^{N_0}}{1 - \left(1 - P_e(\x_0)\right)^{N_0}} \approx 1 - \frac{N_0F(t\mid\x_0)}{1 - \left(1 - P_e(\x_0)\right)^{N_0}} \approx \exp\left(-\frac{N_0F(t\mid\x_0)}{1 - \left(1 - P_e(\x_0)\right)^{N_0}} \right)\,.
\eeq
With $\delta$ as a small upper integration bound and $N_0 \gg 1$ large, we split the integral in Eq.~\eqref{eq:tau_temp} as,
\beq\label{eq:tau_delta}
\bar\tau_{\rm EMFPT}(N_0) &=& \int\limits_0^\delta \exp\left(-\frac{N_0F(t\mid\x_0)}{1 - \left(1 - P_e(\x_0)\right)^{N_0}} \right) dt + O(\exp(-N_0)), \quad N_0 \gg 1\,,
\eeq
with an exponentially small remainder term that can be neglected as in \cite{basnayake2019asymptotic}. Here the characteristic time $t_{N_0}$ introduced in section \S \ref{sec:exitpoint_flux} satisfies
\beq
\frac{N_0F(t_{N_0}\mid\x_0)}{1 - \left(1 - P_e(\x_0)\right)^{N_0}} = O(1) \quad \Longrightarrow \quad Ct_{N_0}\exp\left( -\frac{a}{t_{N_0}}\right) = \frac{1 - \left(1 - P_e(\x_0)\right)^{N_0}}{N_0}\,,
\eeq
which gets solved with the Lambert function $W(z)$ \cite{corless1996},
\beq
t_{N_0} = \frac{a}{W\left(\frac{N_0Ca}{1 - \left(1 - P_e(\x_0)\right)^{N_0}}\right)}\,.
\eeq
The integral \eqref{eq:tau_delta} is of the same order as the characteristic time $t_{N_0}$, as seen upon evaluating
\beq
\bar\tau_{\rm EMFPT}(N_0) \approx \int\limits_0^\delta \exp\left(-N_0 \left(\frac{ Ct \exp\left(- \frac{a}{t} \right)}{1 - \left(1 - P_e(\x_0)\right)^{N_0}} \right) \right) dt\,, \nonumber
\eeq
using the change of variable $x = a/t$,
\beq
\bar\tau_{\rm EMFPT}(N_0) \approx a\int_{a/\delta}^\infty \frac{1}{x^2}\exp\left( - \left(\frac{N_0Ca}{1 - \left(1 - P_e(\x_0)\right)^{N_0}} \right) \frac{e^{-x}}{x} \right) dx\,.
\eeq
By introducing $w = W\left(N_0Ca/\left(1 - \left(1 - P_e(\x_0)\right)^{N_0}\right)\right)$ we obtain
\beq
\bar\tau_{\rm EMFPT}(N_0) \approx a\int_{a/\delta}^\infty \frac{1}{x^2} \exp\left( - \frac{we^{w-x}}{x} \right) dx \approx a \int_{w}^\infty \frac{1}{x^2} dx = \frac{a}{w}, \quad w \gg 1 \,,
\eeq
since for $w$ large $we^{w-x}/x$ acts as a step function around $w$ \cite{basnayake2019asymptotic}. With the Lambert function $W(z)$ we get
\beq
\bar\tau_{\rm EMFPT}(N_0) \approx t_{N_0} =  \frac{a}{W\left(\frac{N_0Ca}{1 - \left(1 - P_e(\x_0)\right)^{N_0}}\right)}, \quad N_0 \gg 1\,,
\eeq
and with the expansion $W(z) \approx \log(z) - \log(\log(z)) + O(1)$ for $z\to \infty$ \cite{corless1996} we obtain
\beq
\bar\tau_{\rm EMFPT}(N_0) \approx \frac{a}{\log\left(\frac{N_0Ca}{\left(1 - \left(1 - P_e(\x_0)\right)^{N_0}\right)}\right) + \log\left(\log\left(\frac{N_0Ca}{\left(1 - \left(1 - P_e(\x_0)\right)^{N_0}\right)}\right)\right)}, \quad N_0 \gg 1\,.
\eeq
Retaining the leading logarithm and with original model parameters we get the following formula for the extreme mean first passage time,
\beq\label{eq:tau_narrow}
\bar\tau_{\rm EMFPT}(N_0) \approx \frac{\|\x_0 - \x_a\|^2}{4D \left(\log\left( \frac{N_0}{2\log\left( \frac{R}{\eps}\right)\left(1 - \left(1 - P_e(\x_0) \right)^{N_0} \right)} \right) + \alpha \right)}, \quad N_0 \gg 1\,,
\eeq
where $\alpha$ is a constant that accounts for neglected high order terms including the subleading double logarithm. However in the limit $N_0 \to \infty$ with fixed model parameters, a uniform killing field does not influence exit statistics of the fastest particle since the probability $\left(1 - \left(1 - P_e(\x_0) \right)^{N_0}\right)$ converges to one exponentially fast. But for finite $N_0$ killing effects are captured by the fitting constant $\alpha$ as revealed in the next section.
\section{Asymptotic vs stochastic simulations for the extreme mean first time with a uniform and non-uniform killing}\label{sec:simulations}
In this section we explore how different killing affects the escape dynamics on a disk of radius $R$. We shall compare stochastic simulation results with asymptotic predictions for escape probabilities (appendix \S \ref{sec:escapeProba}) and extreme mean first passage times (section \S \ref{sec:FPE}). The simulation algorithm is described in appendix \S \ref{sec:algo_simu}.
\subsection{Extreme narrow escape for a uniform killing field}\label{sec:uniform}
\begin{figure}[H]
\begin{center}
\includegraphics[width=0.75\textwidth]{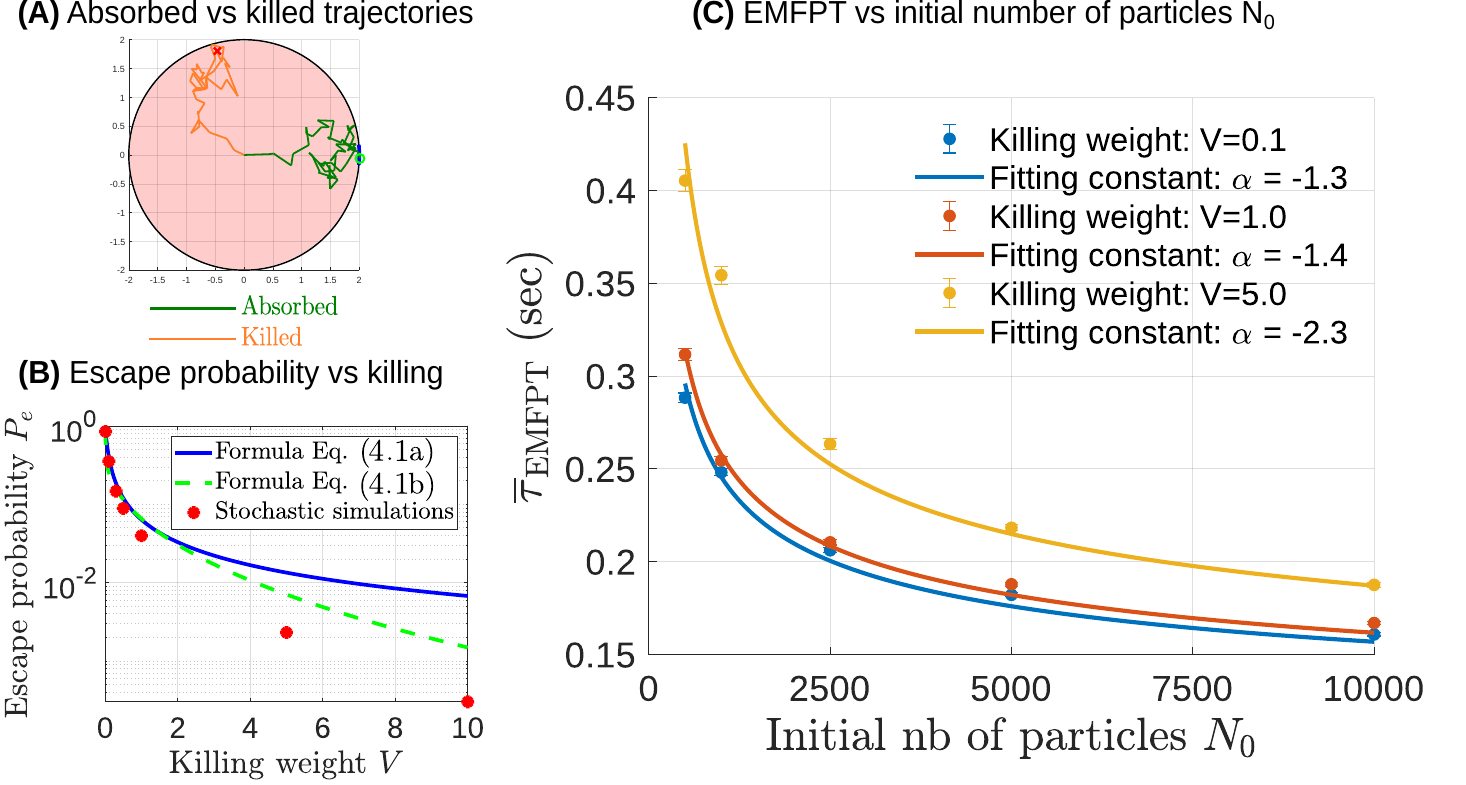}
\caption{\textbf{Narrow exit statistics and uniform killing field.} \textbf{(A)} Schematic showing two trajectories: one killed (orange) while the other reaches the narrow boundary exit $\p\Omega_a$ (green). \textbf{(B)} Escape probability vs killing weight $V$, stochastic simulation (red dots) vs asymptotic formulas (blue and green dashed lines) given in Eq.~\eqref{eq:escapeProbaTwo}. \textbf{(C)} EMFPT vs initial number of particles $N_0$, fitting Eq.~\eqref{eq:tau_narrow_fit} (full lines) against simulation results (colored disks) with killing rates $V = 0.1,\,1,\,5$. Error bars indicate the standard error of the mean calculated over 1000 repeated runs. Parameter values are $D=1$, $R=2$ and $\eps = \pi/18$.}
\label{graph2}
\end{center}
\end{figure}

\begin{figure}[H]
\begin{center}
\includegraphics[width=0.75\textwidth]{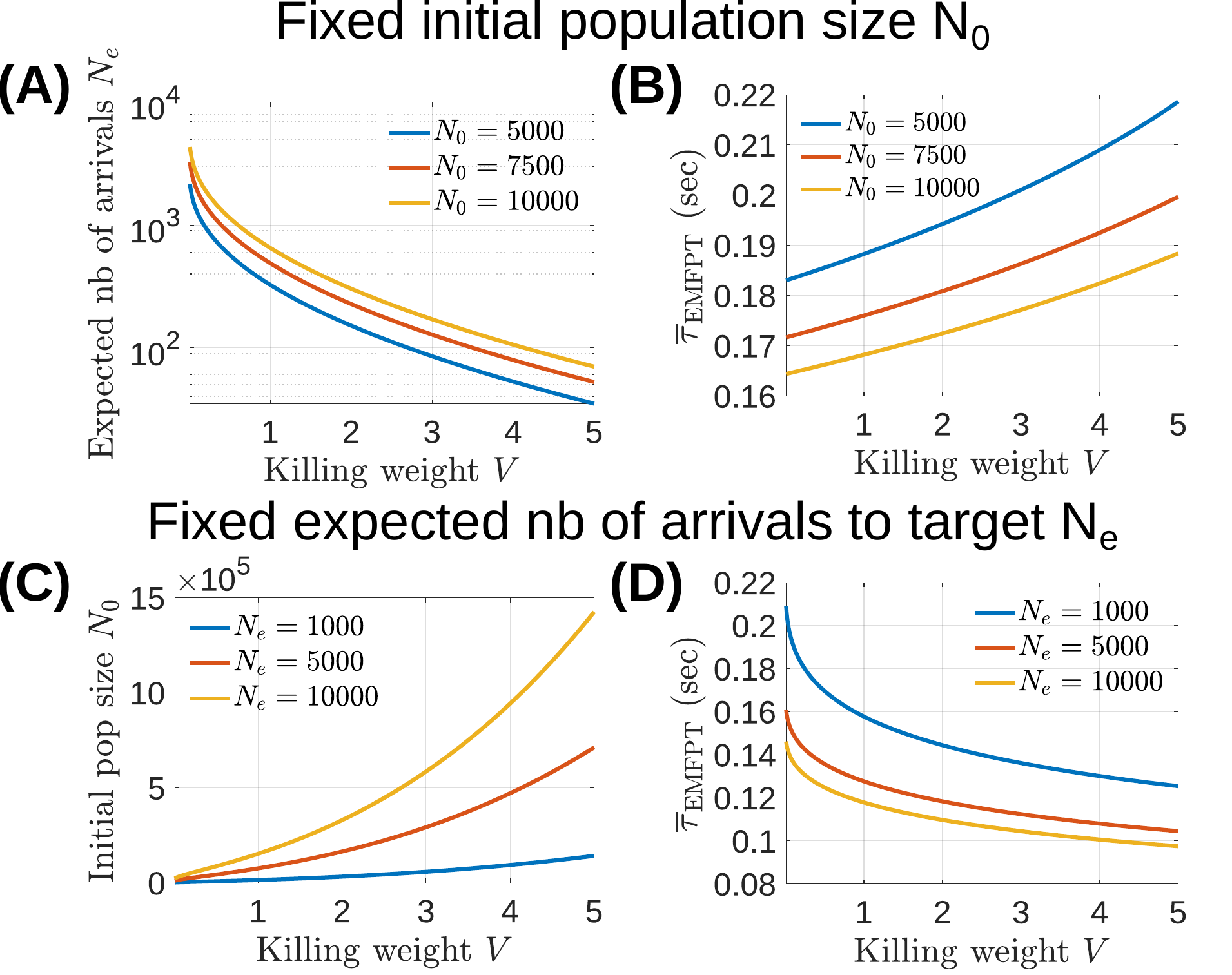}
\caption{\textbf{Killing affects extreme mean first passage times.} \textbf{(A)}-\textbf{(B)}. Fixed initial population size $N_0$. \textbf{(C)}-\textbf{(D)} Fixed expected number of successful arrivals $N_e = P_e(V)N_0$. Results obtained upon numerically integrating formula \eqref{eq:tau_quad} with parameters $R = 2$, $D=1$, $\eps = \pi/18$.}
\label{graph3}
\end{center}
\end{figure}

We first set $k(\x) = V$ and perform stochastic simulations of the uniform killing case (Fig.~\ref{graph2}\textbf{(A)}) with particles released at the origin $\x_0=(0,0)$, i.e.~with killing region corresponding to the full domain. In appendix \S \ref{sec:escapeProbaUnif} we derive an asymptotic formula for the single-particle escape probability using the Neumann Green's function, that we compare with the expression obtained in section \S \ref{sec:FPE},
\begin{subequations}\label{eq:escapeProbaTwo}
\begin{align}
\text{Eq.~\eqref{eq:escapeProbaNeu}}: \qquad P_e(0,0) &= \frac{1}{1 + (\omega R)^2\left(\log\left(\frac{R}{\eps} \right) + \frac{5}{4}\right)}, \quad \text{with} \quad \omega = \sqrt{\frac{V}{D}}\,, \label{eq:escapeProbaNeu_x0_0} \\
\text{Eq.~\eqref{eq:escapeProba}}: \qquad P_e(0,0) &= \frac{K_0\left(\omega R\right)}{\log\left( \frac{1}{\omega \eps} \right)}\,, \label{eq:escapeProba_x0_0}
\end{align}
\end{subequations}
against stochastic simulations on the killing range $V \in [0, 10]$ (Fig.~\ref{graph2}\textbf{(B)}). Both asymptotic approximations agree well with simulation results for small killing weights, however for larger values, formula \eqref{eq:escapeProba_x0_0} which uses the modified Bessel function $K_0(x)$ becomes more appropriate to reveal the fast decay to zero of the escape probability (green dashed curve in Fig.~\ref{graph2}\textbf{(B)}). By then substituting Eq.~\eqref{eq:escapeProba_x0_0} within formula \eqref{eq:tau_narrow} for the extreme mean first passage time, we get
\beq\label{eq:tau_narrow_fit}
\bar\tau_{\rm EMFPT}(N_0) &=& \frac{R^2}{4D \left( \log\left( N_0\left( 2\log\left( \frac{R}{\eps}\right)\left(1 - \left(1 - \frac{K_0\left(\omega R\right)}{\log\left( \frac{1}{\omega \eps} \right)} \right)^{N_0} \right)\right)^{-1} \right) + \alpha \right) },
\eeq
where $\alpha$ is determined by fitting Eq.~\eqref{eq:tau_narrow_fit} against stochastic simulations as $N_0$ is increased (Fig.~\ref{graph2}\textbf{(C)}). Interestingly by repeating the fit for a few different killing weight values, we find that increasing $V$ leads to longer exit times of the first particle. This phenomenon means that for $N_0$ large and fixed, high killing weight $V$ corresponds to less particles reaching the absorbing target: some of the ``killed trajectories'' may have been short and fast, thereby increasing average first exit times. Finally, the extreme statistics diffusion modeling paradigm holds by first fixing modeling and geometrical parameters, and then exploring the large-$N_0$ limit: in particular initial population and killing limits cannot be interchanged. \\
\noindent To circumvent this issue we numerically integrate (Fig.~\ref{graph3}) formula \eqref{eq:tau1} prior to taking the limit $N_0 \gg 1$,
\beq\label{eq:tau_quad}
\bar\tau_{\rm EMFPT}(V,N_0) = \int\limits_0^\infty \frac{\left(1 -  F(t,V)\right)^{N_0} - \left(1 - P_e(V) \right)^{N_0}}{1 - \left(1 - P_e(V)\right)^{N_0}} dt, \quad \text{with} \quad F(t,V) = \int\limits_0^t \frac{e^{-\left(Vs + \frac{R^2}{4Ds} \right)}}{2\log\left(\frac{R}{\eps}\right)s}ds\,,
\eeq
with $P_e(V)$ as defined in Eq.~\eqref{eq:escapeProba_x0_0}. We vary the killing rate $V$ by either keeping constant the initial population size $N_0$ (Fig.~\ref{graph3}\textbf{(A)}-\textbf{(B)}) or the expected number of successful arrivals (Fig.~\ref{graph3}\textbf{(C)}-\textbf{(D)}) here given by $N_e = P_e(V)N_0$. Interestingly these two cases lead to opposite trends, with a slowing down of the EMFPT due to killing in the first scenario, consequence of the fewer expected arrivals $N_e$ (Fig.~\ref{graph3}\textbf{(A)}). Alternatively when $N_e$ is kept fixed (instead of $N_0$), higher killing rates decrease the EMFPT due to the initial population size that massively increases to compensate the small escape probability, as observed by computing the ratio $N_0 = N_e/P_e(V)$ (Fig.~\ref{graph3}\textbf{(C)}).
\subsection{Distribution of surviving particles at first exit for a uniform killing field}\label{sec:simu_poissonian}
\begin{figure}[H]
\centering
\includegraphics[width=\textwidth]{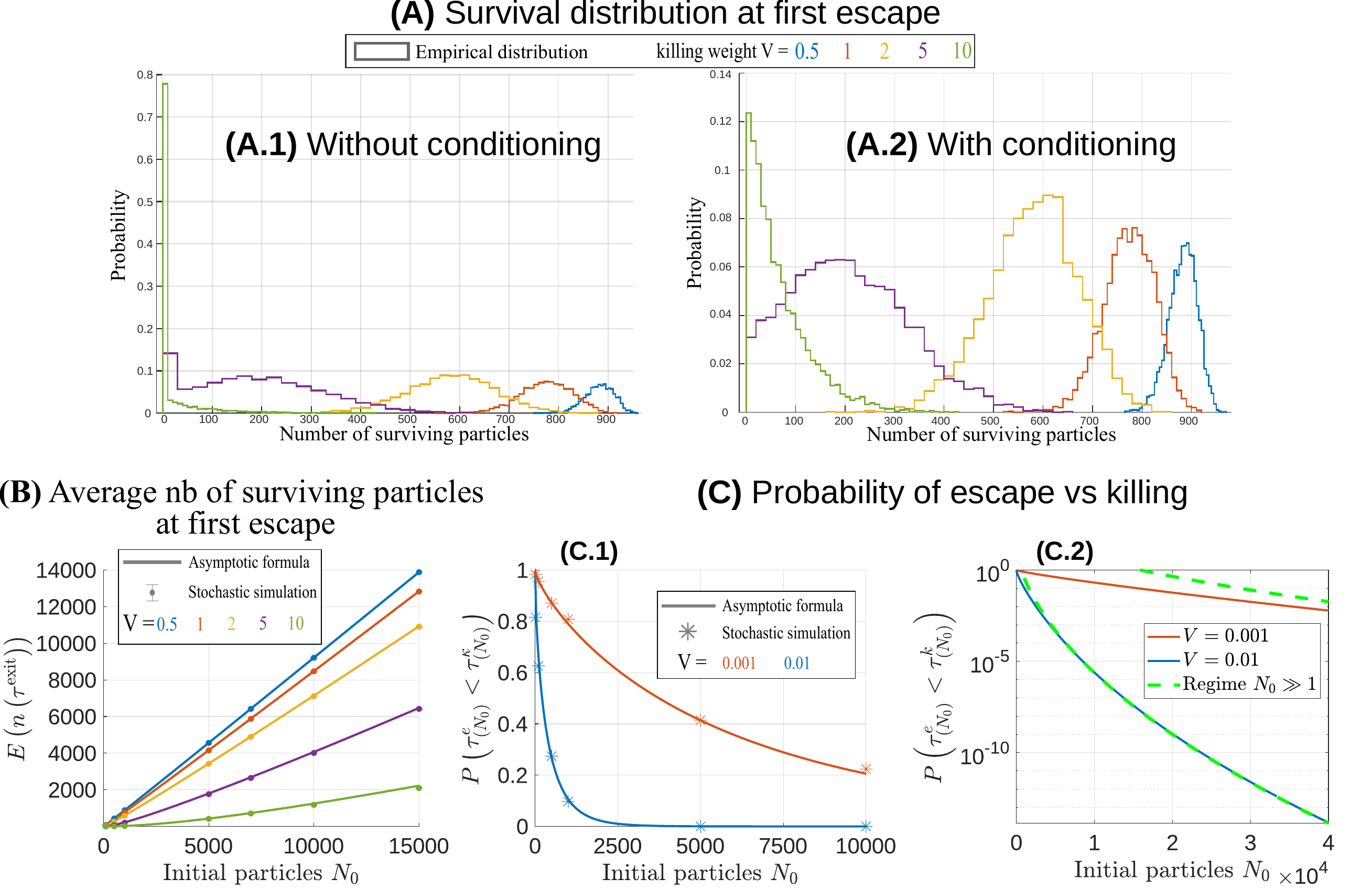}
\caption{\textbf{Escape vs killing in a disk with uniform killing field and narrow absorbing target.} \textbf{(A)} Empirical distributions of surviving particles at first exit, obtained from $5000$ repeated simulations with $N_0 = 1000$ initial particles released. \textbf{(A.1)} Without conditioning: number of survivors set to zero if all particles are destroyed prior to the first escape. \textbf{(A.2)} With conditioning: simulations without successful exits are discarded. \textbf{(B)} Expected number of surviving particles at first exit vs initial population size, comparing formula \eqref{eq:esp_numeric} (full curves) with stochastic simulation results (dots) obtained from $1000$ repeated runs. \textbf{(C)} Probabilities of escape vs killing as first event vs initial population $N_0$ ($1000$ repeated runs). \textbf{(C.1)} Formula Eq.~\eqref{eq:proba_escape_before_killing_integral} (full curves) vs stochastic simulations (star dots) \textbf{(C.2)} Same plot in log scale, comparing Eq.~\eqref{eq:proba_escape_before_killing_integral} (full curves) with explicit asymptotic formula \eqref{eq:proba_escape_before_killing_equiv} for $N_0 \gg 1$ (dashed green curves). Parameter values are $R=2$, $D=1$, $\eps = \pi/18$, $\x_0 = (0,0)$, $\x_a = (R,0)$, with time-step $\Delta t = 0.01$.}
\label{graph4}
\end{figure}

We investigate the number of surviving particles at first exit for a uniform killing field (Fig.~\ref{graph4}), focusing on the expectation of the distribution $\mathbb{E}\left(n\left(\tau^{\rm exit}\right)\right)$ and on the probability $\Pr\left(\tau_{(N_0)}^e < \tau_{(N_0)}^k\right)$ of observing an escape as a first event. These quantities are defined in sections \S \ref{sec:distribution_first_exit} and \S \ref{sec:prob_escape_first}: in particular setting $k(\x) = V$ in Eq.~\eqref{eq:expected_nb_survival} for the expectation $\mathbb{E}\left(n\left(\tau^{\rm exit}\right)\right)$ yields,
\beq\label{eq:expected_1st}
\mathbb{E}\left(n\left(\tau^{\rm exit}\right)\right) = N_0\left(1 - V\int\limits_0^{+\infty}\left( 1 - \int\limits_0^t J(s\mid\x_0)ds \right)^{N_0-1} S(t\mid\x_0)dt \right)\,,
\eeq
with the survival probability $S(t \mid \x_0)$ solution of
\beq\label{eq:dS}
\frac{d}{dt}S(t\mid\x_0) + VS(t\mid\x_0) = - J(t\mid\x_0), \quad \text{with} \quad S(0\mid\x_0) = 1\,,
\eeq
which gives
\beq\label{eq:S}
S(t\mid \x_0) = e^{-Vt}\left(1 - F_e(t\mid \x_0)\right), \quad \text{with} \quad F_e(t\mid\x_0) = \int\limits_0^t e^{Vs}J(s\mid\x_0)ds\,.
\eeq
Here $F_e(t\mid\x_0)$ corresponds to the cumulative distribution function of escape times in the absence of killing. With these relations formula \eqref{eq:expected_1st} becomes
\beq\label{eq:esp_numeric}
\begin{split}
& \mathbb{E}\left(n\left(\tau^{\rm exit}\right)\right) = \\
& N_0\Biggl(1 - V \int\limits_0^{+\infty}\left(1 - e^{-Vt}F_e(t\mid\x_0) - V \int\limits_0^t e^{- V s}F_e(s\mid\x_0) ds\right)^{N_0-1}e^{-Vt} \left(1 - F_e(t\mid\x_0) \right) dt \Biggr)\,.
\end{split}
\eeq
Similarly substituting $k(\x) = V$ in Eq.~\eqref{eq:proba_escape_vs_killing} for the probability of observing an escape first gives
\beq\label{eq:proba_escape_before_killing_integral}
\Pr\left(\tau_{(N_0)}^e < \tau_{(N_0)}^k\right) = 1 - V N_0 \int\limits_0^{+\infty} e^{-N_0Vt}\left(1 - F_e(t\mid\x_0)\right)^{N_0} dt \,.
\eeq
The integrals in Eq.~\eqref{eq:esp_numeric} and \eqref{eq:proba_escape_before_killing_integral} are evaluated with the cumulative distribution function $F_e(t \mid \x_0)$ of escape times for a disk of radius $R$, starting point at the origin and narrow target of length $2\eps$: following Eq.~\eqref{eq:intJsmalltime} we use the piecewise smooth approximation
\beq\label{approx_F}
F_e(t\mid(0,0)) \approx \left\{
\begin{array}{ll}
Cte^{-\frac{a}{t}} \quad & t \leq t_0 \\
1  & t > t_0
\end{array}\right. , \quad \text{with} \quad  C = \frac{2D}{R^2\log\left(\frac{R}{\eps}\right)}, \quad a = \frac{R^2}{4D}, \quad t_0 = \frac{a}{W(Ca)}\,,
\eeq
where $W(z)$ is the Lambert function \cite{corless1996}. In particular we can obtain a formula for the probability of observing an escape as a first event when $N_0 \gg 1$: substituting Eq.~\eqref{approx_F} in Eq.~\eqref{eq:proba_escape_before_killing_integral} yields
\beq\label{eq:proba_escape_before_killing_approx}
\Pr\left(\tau_{(N_0)}^e < \tau_{(N_0)}^k\right) = 1 - V N_0\int\limits_0^{t_0} \exp\left(-N_0\left(V t - \log\left(1 - Cte^{-\frac{a}{t}}\right)\right)\right) dt \,,
\eeq
which can be approximated by
\beq
\Pr\left(\tau_{(N_0)}^e < \tau_{(N_0)}^k\right)\approx 1 - V N_0 \int\limits_0^{t_0} \exp\left(-N_0\left(V t + Cte^{-\frac{a}{t}}\right)\right) dt\,.
\eeq
Setting $u = N_0t$ and taking the large $N_0$ limit then give
\begin{align}
\Pr\left(\tau_{(N_0)}^e < \tau_{(N_0)}^k\right) &\approx 1 - V \int\limits_0^{N_0t_0} \exp\left(-\left(Vu + C u e^{-\frac{a}{u}N_0}\right)\right) du \\
&\approx 1 - V \int\limits_0^{+\infty} e^{- V u}\left( 1 - Cue^{-\frac{a}{u}N_0} \right) du\,,
\end{align}
which can be evaluated as
\beq
\Pr\left(\tau_{(N_0)}^e < \tau_{(N_0)}^k\right) \approx V C \int\limits_0^{+\infty} ue^{-\left(V u + \frac{a}{u}N_0\right)} du = 2 C a N_0 K_2\left(2\sqrt{aVN_0}\right)\,,
\eeq
where $K_2(x)$ is the second order modified Bessel function of the second kind. Using the large argument expansion $K_2(x) \sim \sqrt{\frac{\pi}{2x}} e^{-x}$ for $x \to \infty$ \cite{stegun1964} then yields the explicit formula below for the probability of observing an escape first rather than a killing,
\beq
\Pr\left(\tau_{(N_0)}^e < \tau_{(N_0)}^k\right) \approx \sqrt{\pi} C \left(\frac{\left(a N_0\right)^{3}}{V}\right)^{1/4} e^{-2\sqrt{a V N_0}}\,, \quad \text{for} \quad N_0 \gg 1\,,
\eeq
with the original parameters this is
\beq\label{eq:proba_escape_before_killing_equiv}
\Pr\left(\tau_{(N_0)}^e < \tau_{(N_0)}^k\right) \approx \frac{N_0}{\log\left(R/\eps\right)}\sqrt{\frac{\pi}{2R}}\left(\frac{D}{VN_0}\right)^{1/4}\exp\left( -R\sqrt{\frac{VN_0}{D}} \right)\,, \quad \text{for} \quad N_0 \gg 1\,.
\eeq
We then simulate the distribution of surviving particles at first exit for some killing rates $V$ chosen in the range $[0.5,\,10]$, with $N_0 = 1000$ initial particles released at the center of a disk of radius $R=2$ and narrow absorbing arc of length $\pi/9$ (Fig.~\ref{graph4}\textbf{(A)}). The empirical distributions shown in Fig.~\ref{graph4}\textbf{(A.1)} are average results over $5000$ repeated simulations: the strong peak in $n\left(\tau^{\rm exit}\right) = 0$ observed for large killing reflects the low survival probability with most simulations ending without successful exit. This problem is addressed in Fig.~\ref{graph4}\textbf{(A.2)}, where simulations ending in the killing of all initial particles are discarded. Hence by conditioning on at least one successful exit per run, smoother distributions are obtained, especially for large killing rate. \\
For small killing rate we report a linear relationship (Fig.~\ref{graph4}\textbf{(B)}) between the expected number of survival particles $\mathbb{E}\left(n\left(\tau^{\rm exit}\right)\right)$ and the initial population size $N_0$. Nonlinear effects only become significant at large killing rate (or very low escape probability). Note that these trends are both predicted by the expectation formula \eqref{eq:esp_numeric} and our stochastic simulation results. We also explore how the probability of observing an exit first rather than a killing decreases with respect to the initial population size (Fig.~\ref{graph4}\textbf{(C)}). This probability is very low: hence for computational convenience small killing rates $V=0.001$ and $0.01$ are employed (Fig.~\ref{graph4}\textbf{(C.1)}), with good agreements obtained between analytical (formula \eqref{eq:proba_escape_before_killing_integral}) and stochastic simulations. These results are further compared with formula \eqref{eq:proba_escape_before_killing_equiv} predicting the exponential decay of the probability $\Pr\left(\tau_{(N_0)}^e < \tau_{(N_0)}^k\right)$ (Fig.~\ref{graph4}\textbf{(C.2)}). We conclude that the present integral and asymptotic formulas can be used in a large parameter range to predict the mean number of particles at first exit and the probability of the first event being an escape. \\
\subsection{Escape probability for killing restricted to a subregion}\label{sec:restricted}
\begin{figure}[H]
\centering
\includegraphics[width=0.75\textwidth]{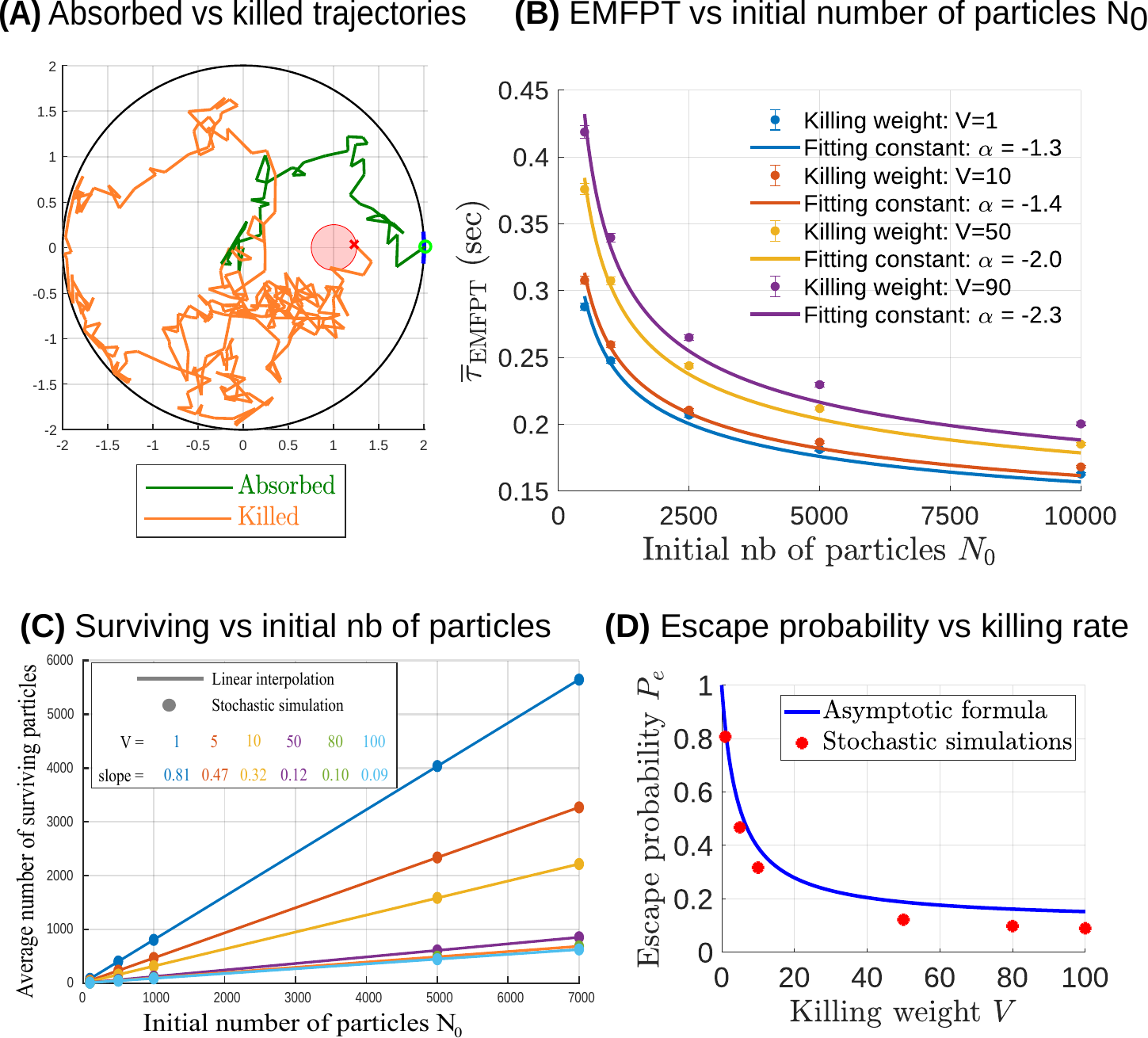}
\caption{\textbf{Narrow exit and killing restricted to a subregion.} \textbf{(A)}  Killing subregion $\Omega_k$ (pink shaded) of radius $r_k = 0.25$ centered in $\x_k=(1,0)$, with target $\p\Omega_a$ (blue segment). \textbf{(B)} EMFPT vs $N_0$ obtained from stochastic simulations (colored disks) and by fitting to the logarithmic law (Eq.~\eqref{eq:log_law}, continuous lines). \textbf{(C)} Survival vs initial number of particles for different killing rates $V$. \textbf{(D)} Escape probability vs killing weight $V$: simulation (red markers) and asymptotic (Eq.~\eqref{eq:split_prob_x0_0}, blue line). Stochastic simulations are averaged over $1000$ runs.}
\label{graph5}
\end{figure}

\begin{figure}[H]
\centering
\includegraphics[width=\textwidth]{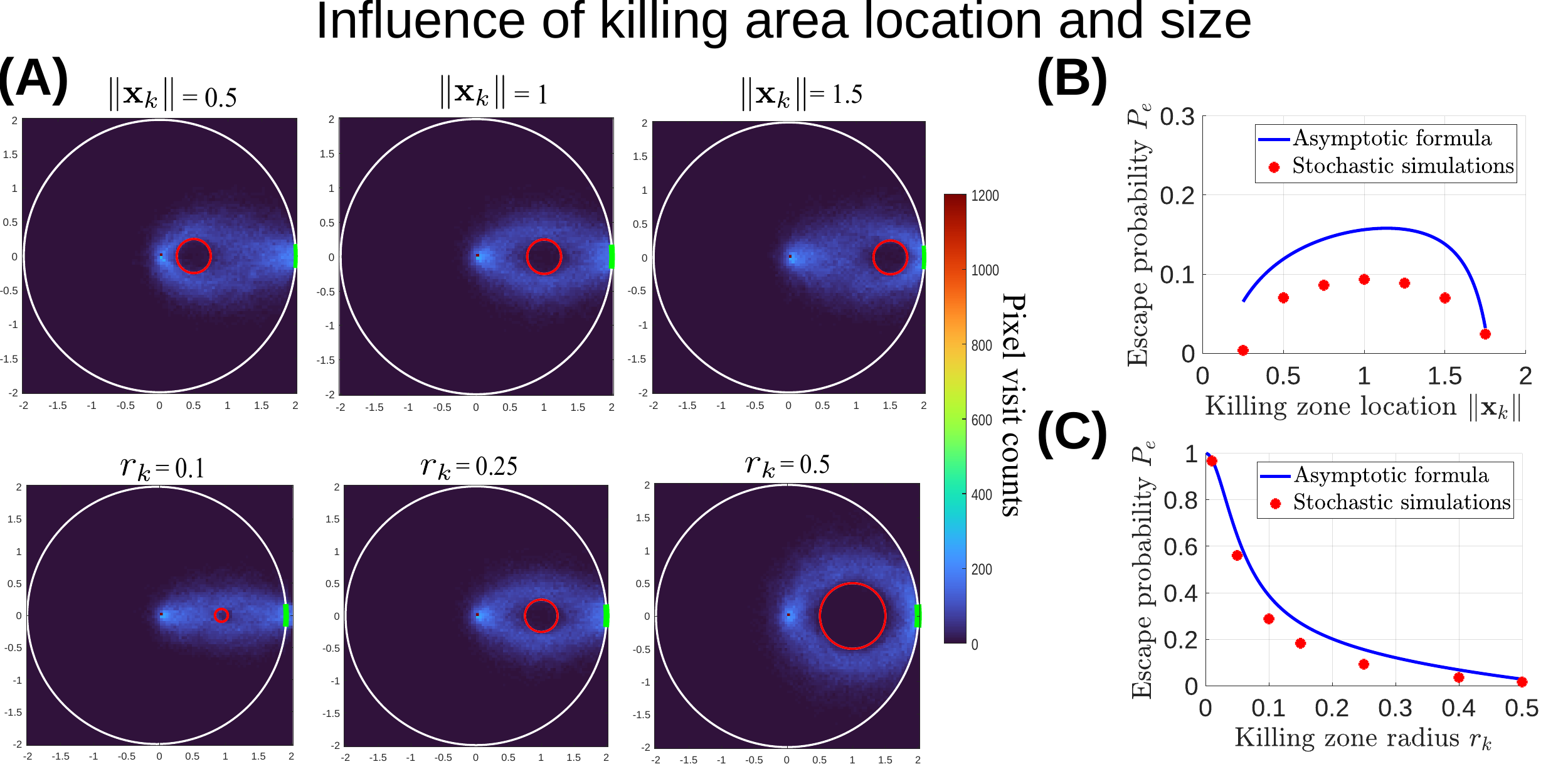}
\caption{\textbf{Effect of killing zone location and size on distribution of fastest exit paths and on the escape probability.} \textbf{(A)} Spatial distribution of fastest trajectories for different killing area geometrical configurations. \textbf{(B)} Escape probability vs distance from the source $\|\x_k\|$, showing asymptotic (blue curve, Eq.~\eqref{eq:split_prob_x0_0}) and simulation results (red dots), with killing zone radius $r_k = 0.25$. \textbf{(C)} Escape probability vs killing zone radius $r_k$ showing asymptotic (blue curve, Eq.~\eqref{eq:split_prob_x0_0}) and simulation results (red dots) with distance $\|\x_k\| = 1$. Other parameter values are $N_0 = 10^4$, $R=2$, $D=1$, $\eps = \pi/18$ and $V=90$.}
\label{graph6}
\end{figure}

We analyze the impact on the single-particle escape probability of a killing field restricted to a narrow subdisk $\Omega_k = B_{r_k}(\x_k)$ of radius $r_k\ll R$ centered in $\x_k$. We first set $r_k = 0.25$ and $\x_k = (1,0)$ (Fig.~\ref{graph5}\textbf{(A)}) and gradually increase the killing weight by a factor hundred (Fig.~\ref{graph5}\textbf{(B)}). We fit the average escape time of the fastest against the initial number of particles, using the following logarithmic formula
\beq\label{eq:log_law}
\bar\tau_{\rm EMFPT}(N_0) = \frac{\ell^2}{4D \left(\log\left( \frac{N_0}{2\log\left(\frac{R}{\eps}\right)(1 - (1-P_e(0,0))^{N_0})} \right) + \alpha \right)}\,,
\eeq
where $\alpha$ is a correcting factor, $P_e(0,0)$ is the escape probability starting from the origin (Eq.~\eqref{eq:split_prob_x0_0}) and $\ell$ measures the length of source-target connecting geodesics obliged to avoid the zone $\Omega_k$ at high killing rate (further details provided in \S \ref{sec:rate-function-killing}): we use the approximation $\ell \approx R$ due to the small killing zone size with $r_k \ll R$. As observed for the uniform killing case (Fig.~\ref{graph2}), increasing the killing rate reduces the amount of survival particles and therefore increases the extreme mean first passage time (Fig.~\ref{graph5}\textbf{(B)}). For $N_0$ fixed, we further calculate the average number $N_e$ of escape particles by averaging over $1000$ simulations (Fig.~\ref{graph5}\textbf{(C)}). The slopes of $N_e$ versus $N_0$ corresponds to the splitting probability, reported in Fig.~\ref{graph5}\textbf{(D)} and compared to formula \eqref{eq:split_prob} with $\x_0$ at the origin,
\beq\label{eq:split_prob_x0_0}
P_e(0,0) = \frac{\frac{D}{V r_k^2} + \frac{1}{2}\log\left(\frac{R}{r_k}\right) + \frac{1}{4} + \log\left( \frac{\|\x_a - \x_k\|}{R} \right) - \frac{1}{2}\log\left( \frac{R}{\|\x_k\|} - \frac{\|\x_k\|}{R} \right)}{\frac{D}{V r_k^2} + \frac{1}{2}\log\left(\frac{R}{r_k}\right) + \log\left(\frac{R}{\eps}\right) + \frac{5}{4} + 2\log\left( \frac{\|\x_a - \x_k\|}{R}  \right) - \frac{1}{2}\log\left(1 - \left(\frac{\|\x_k\|}{R}\right)^2 \right)}\,.
\eeq
To conclude, restricting the killing area to a small subsection of the domain significantly increases the survival probability (Fig.~\ref{graph2}\textbf{(B)}).
\noindent We then study how the position and the size of the killing area affect the escape probability and the spatial density of most likely paths used by the trajectory associated to the fastest (Fig.~\ref{graph6}). By shifting the killing area away the origin where particles are released, our analysis and simulations suggest the escape probability to be maximized half-way between the starting point and the target (Fig.~\ref{graph6}\textbf{(B)}). One limitation is that asymptotic formula \eqref{eq:split_prob_x0_0} does not hold when the killing area overlaps with the origin or the absorbing target as this configuration violates the well-spaced assumption. \\
We further analyze the spatial distribution of the fastest exit trajectories and find that the optimal paths avoid the killing area (Fig.~\ref{graph6}\textbf{(A)}), due to the high killing probability within the zone. The escape probability is inversely proportional to the size of the killing zone (as revealed by Eq.~\eqref{eq:split_prob_x0_0}). To illustrate this effect, we fix the center of the zone half-way between the origin and the target and gradually increase its radius (Fig.~\ref{graph6}\textbf{(C)}). At high mortality rate the killing area is equivalent to an obstacle between the origin and the target, causing geodesics to deviate from straight lines (Fig.~\ref{graph6}\textbf{(A)}).
\section{Path-integral representation, optimal control problem and paths for the fastest particles to exit}\label{sec:optimal_path}
Stochastic simulations from section \S \ref{sec:restricted} suggest that the trajectories used by the fastest among $N_0$ particles to escape alive concentrate in a tubular neighborhood of a trajectory. To study the optimal paths associated with the fastest escape particle, we shall use a formal argument based on the large deviation principle \cite{freidlin1998random,bouchet2016generalisation}, starting with the stochastic equation
\beq
\dot{\x}(t) =\sqrt{2D} \dot{\w}(t),
\eeq
with standard Brownian motion $\w(t)$ and diffusion coefficient $D$. In the small noise amplitude limit $D \ll 1$, the probability of observing large fluctuations of the diffusion process $\x(t)$ around a deterministic path ${\bf F}(t)$ is given by the action functional on the time interval $[0,T]$,
\beq\label{eq:large}
\Pr \left \lbrace \max_{t \in [0,T]}\|\x(t) - {\bf F}(t)\| > D \right\rbrace \asymp \exp \left \lbrace -\frac{\mathcal{S}_{\bf F}(T)}{D}\right\rbrace, \quad \text{where} \quad \mathcal{S}_{\bf F}(T) = \frac{1}{4} \int_{0}^{T} \|\dot {\bf F}(t)\|^2 dt \,.
\eeq
With the Feynman-Kac formula we can express the probability density function of a Brownian path subject to a random degradation field,
\beq\label{FKformula}
p\left(\x,T\mid\x_0\right) = \eE\left[ f(\x(T))\exp \left(-\int_{0}^{T} k(\x(t))dt\right) \right]\,,
\eeq
where $f(\x)$ is the initial Dirac distribution centered at $\x_0$. Integrating over $\Omega$ then yields the survival probability $S\left(T\mid\x_0\right)$
\beq\label{eq:survival}
S\left(T\mid\x_0\right) = \int_\Omega p\left(\x,T\mid\x_0\right) d\x = \eE \left[\left. \exp \left(-\int_{0}^{T} k(\x(t))dt\right)\right|\x(0)=\x_0 \right]\,,
\eeq
which follows from the Feynman-Kac formula \eqref{FKformula}. Then, using the large deviation formula \eqref{eq:large} we reformulate the expectation in Eq.~\eqref{eq:survival} as a path integral over the space of all trajectories starting in $\x_0$ and ending on the narrow target $\p\Omega_a$,
\beq
S\left(T\mid\x_0\right) &=&  \int_{\left\lbrace\underset{\x(0)=\x_0,\,\, \x(T) \in \p\Omega_a}{\x \,\, \hbox{ paths}}\right\rbrace}\exp \left\lbrace -\left(\frac{\mathcal{S}_{\x}(T)}{D}+\int_{0}^{T} k(\x(t))dt \right) \right\rbrace d\mathcal{D}(\x) \nonumber \\
&\asymp& \int_{\left\lbrace\underset{\x(0)=\x_0,\,\, \x(T)\in\p \Omega_a}{\x \,\, \hbox{ paths}}\right\rbrace} \exp \left\lbrace - \int_{0}^{T} \left( \frac{1}{4D} \|\dot \x(t)\|^2 + k(\x(t)) \right)dt \right\rbrace d\mathcal{D}(\x)\,. \label{eq:S_path_integral}
\eeq
We then rewrite the extreme mean first passage time $\bar\tau_{\rm EMFPT}(N_0)$, computed with Eq.~\eqref{eq:tau1}, to show that the most contributing path comes from the shortest path avoiding the killing area: the key term to analyze is the numerator of the integrand,
\beq
N(T) = \left(1 - F(T \mid \x_0)\right)^{N_0} - \left(1-P_e(\x_0)\right)^{N_0}\,,
\eeq
and using identity \eqref{eq:identity}, in the short time regime we get
\beq\label{eq:N_approx}
N(T) = \left(S(T\mid\x_0) + K(T\mid\x_0)\right)^{N_0} - \left(1 - P_e(\x_0)\right)^{N_0} \sim \left(S(T\mid\x_0)\right)^{N_0}, \quad N_0 \gg 1, \quad T \downarrow 0\,.
\eeq
Substituting Eq.~\eqref{eq:S_path_integral} within the approximation \eqref{eq:N_approx} and integrating with respect to time yields the EMFPT,
\begin{align}
&\bar\tau_{\rm EMFPT}(N_0) \sim \int_{0}^{\infty} \exp \left\lbrace N_0 \log \left( \int_{\left\lbrace\underset{\x(0)=\x_0,\,\, \x(T)\in\p \Omega_a}{\x \,\, \hbox{ paths}}\right\rbrace} \hspace{-10mm} \exp \left\lbrace - \int_{0}^{T}\left( \frac{1}{4D} \|\dot \x(t)\|^2 + k(\x(t))\right)dt \right\rbrace d\mathcal{D}(\x) \right)\right\rbrace dT\,,
\end{align}
and by Laplace's method the contribution of the integral for $N_0 \gg 1$ large \cite{Schuss2009Book,touchette2025} occurs at the minimum of the functional,
\beq\label{minimization}
\min_{\left\lbrace\underset{\x(0)=\x_0,\,\, \x(T)\in\p \Omega_a}{\x \,\, \hbox{ paths}}\right\rbrace} \left\lbrace \int_{0}^{T} \left( \frac{1}{4D} \|\dot \x(t)\|^2 + k(\x(t)) \right) dt\right\rbrace\,.
\eeq
\subsection{Minimum--action formulation for the fastest trajectory as an exit--time optimal control problem}\label{sec:oc_exit}
We first solve the minimization problem \eqref{minimization} using a controlled dynamics approach, with
\begin{equation}\label{eq:dyn}
\dot \x(t)=\bu(t), \qquad \x(0)=\x_0 \in \Omega\,,
\end{equation}
where $\bu(\cdot)$ is a given control. The exit time from the domain $\Omega$ is defined as follows:
\begin{equation}\label{eq:tauOmega}
\tau_\Omega(\x;\bu) = \inf\{t>0:\ \x(t) \in \p\Omega_a\}\,.
\end{equation}
Consequently we have $\x(\tau_\Omega(\x;\bu))\in \partial\Omega_a$ and the trajectory exits through the absorbing window. The cost functional associated with the control problem \eqref{eq:dyn} is defined as
\begin{equation}\label{eq:cost}
\mathcal{J}(\x;\bu) = \int\limits_0^{\tau_\Omega(\x;\bu)} \left(\frac{1}{4D}\|\bu(t)\|^2 + k(\x(t))\right)\,dt\,,
\end{equation}
while the value function $\mathcal{V}(\x)$ is
\begin{equation}\label{eq:value_window}
\mathcal{V}(\x) = \inf_{\bu(\cdot)} \left\{ \mathcal{J}(\x;\bu)\, \Big| \, \x(\tau_\Omega)\in\partial\Omega_a \right\}\,.
\end{equation}
Using dynamic programming \cite{schuss2011nonlinear} for the value function $\mathcal{V}(x)$ then leads to a stationary Hamilton--Jacobi--Bellman (HJB) equation,
\begin{equation}\label{eq:HJB_min}
0=\inf_{\bu\in\R^2}\left\{ \nabla \mathcal{V}(\x)\cdot \bu + \frac{1}{4D}\|\bu\|^2 + k(\x)\right\}, \qquad \x\in\Omega\,,
\end{equation}
with explicit minimizing control given by
\begin{equation}\label{eq:opt_control}
\bu^*(\x)=-2D\,\nabla \mathcal{V}(\x)\,,
\end{equation}
and by substituting Eq.~\eqref{eq:opt_control} into \eqref{eq:HJB_min} we get an eikonal--type HJB equation
\begin{equation}\label{eq:HJB_eikonal}
D\|\nabla \mathcal{V}(\x)\|^2 = k(\x), \qquad \x\in\Omega\,.
\end{equation}
Boundary conditions to this equation encode both narrow absorbing and reflecting boundaries: the exit cost vanishes on the target $\partial\Omega_a$,
\begin{equation}\label{eq:BC_absorb}
\mathcal{V}(\x)=0,\qquad \x \in \partial\Omega_a\,,
\end{equation}
while trajectories are reflected on $\partial\Omega_r$, thus leading to
\begin{equation}\label{eq:BC_reflect_surrogate}
\n(\x)\cdot \nabla \mathcal{V}(\x)=0, \qquad \x\in\partial\Omega_r\,.
\end{equation}
To find optimal trajectories we use a Hamiltonian formulation
\begin{equation}\label{eq:Hamiltonian}
H(\x,\bp) = \inf_{\bu\in\R^2}\left\{\bp\cdot \bu+\frac{1}{4D}\|\bu\|^2+k(\x)\right\} = -D\|\bp\|^2 + k(\x)\,,
\end{equation}
such that the HJB equation is $H(\x,\nabla\mathcal{V}(\x)) = 0$. Along an optimal trajectory $\x^*(t)$, the optimal control \eqref{eq:opt_control} yields the characteristic dynamics
\begin{equation}\label{eq:char}
\dot \x^*(t) = -2D\,\nabla \mathcal{V}(\x^*(t)),\qquad \x^*(0) = \x_0\,.
\end{equation}
Equivalently, by defining the costate (or momentum) variable as $\bp(t)=\nabla \mathcal{V}(\x^*(t))$ we obtain the system
\begin{equation}\label{eq:canonical}
\dot \x^*(t) = -2D\,\bp(t),\qquad \dot \bp(t)= -\nabla k(\x^*(t))\,,
\end{equation}
with boundary conditions
\beq
\x^*(0) = \x_0, \quad  \x^*(\tau_\Omega) \in \p\Omega_a\,,
\eeq
to which we add the following transversality condition since the terminal point is free along the target $\partial\Omega_a$,
\begin{equation}\label{eq:transversality_window}
\bp(\tau_\Omega)\ \parallel\ \n(\x(\tau_\Omega))\,.
\end{equation}
At the exit point the costate $\bp(t)$ is therefore normal to $\p\Omega_a$ and the optimal trajectory hits the absorbing window orthogonally. In the next section we show that optimal paths $\x^*(t)$ are characterized by a fixed Hamiltonian $H(\x^*(t),\bp(t)) = H_0$ level determined by initial costate and killing magnitudes.
\subsection{Point-target shooting formulation via parametrized initial direction}\label{sec:shooting_point_target}
An alternative approach would be to enforce the optimal trajectory to exit at a single prescribed point, for instance at the center $\x_a$ of the absorbing window $\p\Omega_a$. For this case, we can formulate the minimum--action trajectory search \eqref{minimization} as a shooting problem with unknown initial velocity orientation. We start from the following fixed point problem with final time $T$,
\beq\label{eq:point_target_action}
\min_{T>0}\ \min_{\x(\cdot)}\ \int_0^T \left(\frac{1}{4D}\|\dot \x(t)\|^2+k(\x(t))\right) \,dt, \qquad \x(0)=\x_0 \in \Omega, \quad \x(T)=\x_a \in \p\Omega_a\,,
\eeq
and by applying the Euler-Lagrange principle $\x(t)$ is an optimal path if and only if it is the solution of
\beq
\left[\frac{\p}{\p \x} - \frac{\p}{\p t} \frac{\p}{\p \dot \x}\right]\left(\frac{1}{4D}\|\dot \x(t)\|^2+k(\x(t))\right) = 0, \qquad \x(0) = \x_0,\ \ \x(T) = \x_a\,,
\eeq
which yields the second order boundary value problem,
\begin{equation}\label{eq:EL_point_target}
- \ddot \x(t) + 2D\,\nabla k(\x(t)) = 0, \qquad \x(0) = \x_0, \quad \x(T) = \x_a\,.
\end{equation}
With the momentum $\bp(t) = - \dot \x(t)/(2D)$ we get
\begin{equation}\label{eq:canonical_point}
\dot \x(t)=-2D\,\bp(t),\qquad \dot \bp(t)=-\nabla k(\x(t))\,,
\end{equation}
which is equivalent to system \eqref{eq:canonical} obtained via optimal control theory. Optimal solutions correspond to the level curves of the Hamiltonian $H(\x,\bp) = -D\|\bp\|^2 + k(\x)$, as seen by differentiating with respect to time,
\[
\frac{d}{dt}H(\x(t),\bp(t)) = \left. \left( \frac{\p H}{\p \x}, \frac{\p H}{\p \bp} \right)\right|_{\left(\x(t),\bp(t)\right)} \cdot \left( \dot \x(t) , \dot \bp(t)\right) = - \nabla k(\x(t)) \cdot 2D \bp(t) + 2D \bp(t) \cdot \nabla k(\x(t)) = 0\,.
\]
To define initial conditions \eqref{eq:canonical_point} we normalize the velocity $\|\dot \x(0)\| = 1$ and use its orientation $\theta$ to parametrize the optimal path $(\x_\theta(t),\bp_\theta(t))$, leading to
\begin{equation}\label{eq:IC_theta}
\x_\theta(0) = \x_0,\qquad \bp_\theta(0) = -\frac{1}{2D}\,(\cos\theta,\sin\theta), \quad \theta \in [0,2\pi)\,,
\end{equation}
with target point hitting time $T(\theta)$ given by
\begin{equation}\label{eq:T_theta}
T(\theta) = \inf\{t>0:\ \x_\theta(t) = \x_a\}\,.
\end{equation}
In the next section the optimal $\theta$ will be shown to depend on the killing rate and geometrical properties of killing and absorbing zones. Finally the initial and terminal conditions in Eq.~\eqref{eq:IC_theta} and \eqref{eq:T_theta} define the constant Hamiltonian level $H_0$,
\beq\label{eq:cte_hamiltonian}
H(\x_\theta(t),\bp_\theta(t)) = H_0, \quad \text{for} \quad 0 < t < T(\theta), \quad \text{with} \quad H_0 = - \frac{1}{4D} + k(\x_0)\,.
\eeq
If initially $k(\x_0)=0$ then $H_0 = -1/(4D)$ is the conserved Hamiltonian quantity along the optimal path in classical geometry.
\subsection{Numerical solution of shooting problem and comparison with stochastic simulations}\label{sec:num_shooting}
\begin{figure}[H]
\centering
\includegraphics[width=0.75\textwidth]{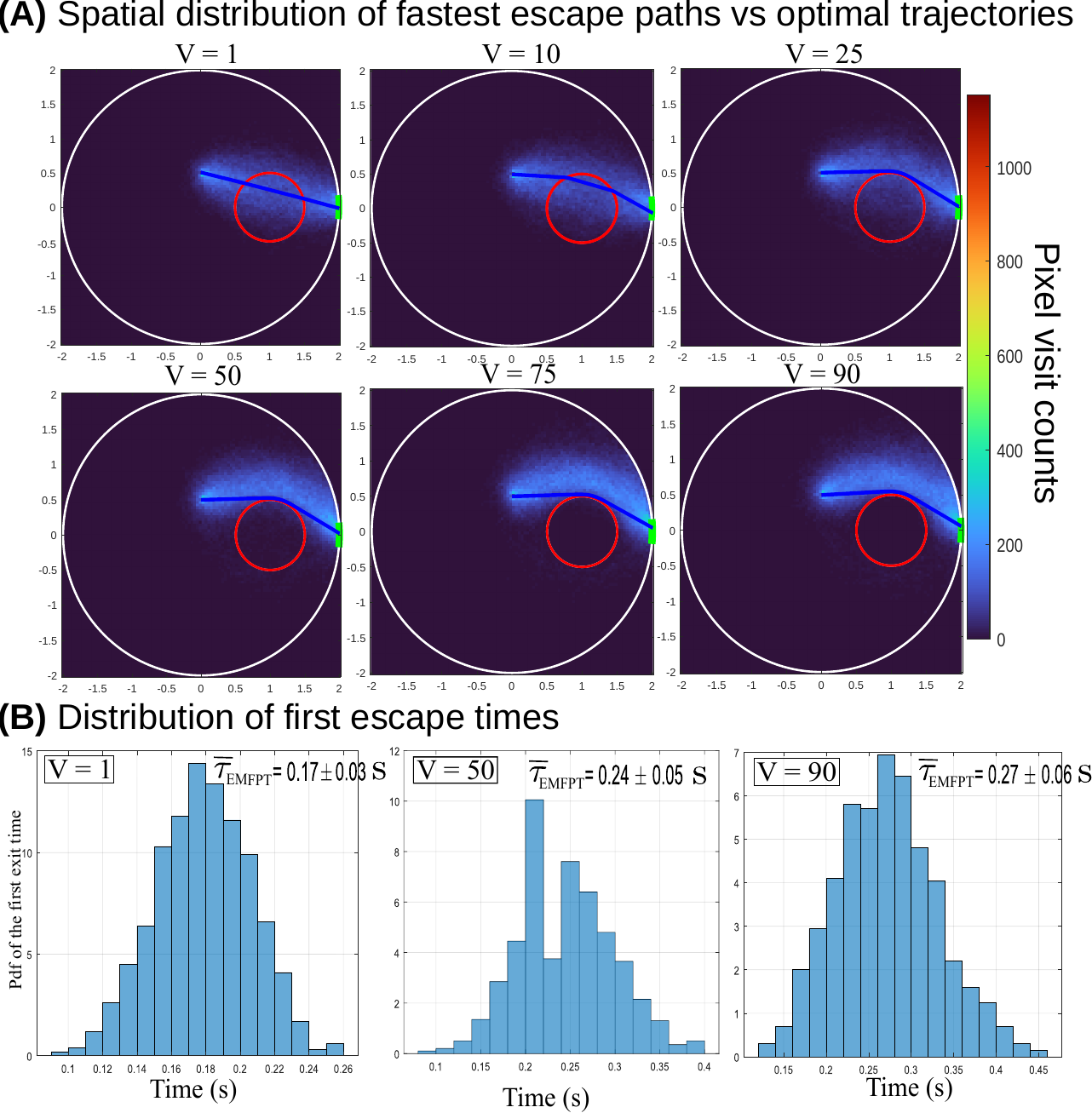}
\caption{\textbf{Geodesic trajectories due to a localized killing field.} \textbf{(A)} Comparison between optimal paths solving Eq.~\eqref{eq:Eulersys} (blue curves) and spatial distribution of optimal trajectories from stochastic simulations (blue shaded area). \textbf{(B)} Distribution of first exit times. Here $\Omega$ is a disk of radius $R=2$ and other parameters are $\x_0 = \left(0,\,0.5\right)$, $r_k = 0.5$, $\x_k = \left(1,0\right)$, and $\x_a = (2,0)$ and $\eps = \pi/18$. First, the shooting system \eqref{eq:Eulersys} is solved using $D=0.01$, $\Gamma = 0.01$, and initial velocity angle $\theta = \arctan\left(R/r_k\right) - \pi/2$ when $V=1$, $\theta = \arctan\left(R/r_k\right)/2 - \pi/4$ when $V=10$, and $\theta \approx 0.03$ for larger killing values. On the other hand, stochastic simulation results are obtained from $1000$ runs with at least one successful escape from $N_0 = 10000$ initial particles, with $D=1$ and $\Delta t = 0.01$ in the Euler's scheme \eqref{euler}.}
\label{graph7}
\end{figure}

We now numerically solve the minimization problem \eqref{minimization} using the point-target shooting approach developed in section \S \ref{sec:shooting_point_target}. To illustrate how optimal exit paths are affected by a localized killing zone we take a source $k(\x)$ with effect limited to the subdisk $B_{r_k}(\x_k)$ of radius $r_k$, centered in $\x_k$ (as in section \ref{sec:restricted}) and located between the starting point $\x_0$ and the absorbing target $\p\Omega_a$. Such a killing source $k(\x)$ can be modeled using a smooth phase field formulation,
\begin{equation}\label{eq:killing_source}
k(\x) = \frac{V}{2}\left(1 + \tanh\left(\frac{r_k - \|\x_k - \x\|}{\Gamma}\right)\right)\,,
\end{equation}
with $\Gamma \ll r_k$ controlling the thickness of the transition layer. Thus $V$ is the maximal killing rate inside the disk-like region and the sharp-indicator model used in the asymptotic analysis is recovered in the singular limit $\Gamma\to0$ (at fixed $V$ and $r_k$). This allows us to compute the gradient $\nabla k(\x(t))$ and reformulate the dynamical system \eqref{eq:canonical_point} as,
\beq\label{eq:Eulersys}
\dot{\x}(t) = -2D\bp(t), \quad \dot{\bp}(t) = -\left(\frac{V}{2\Gamma}\right) \operatorname{sech}^2\left( \frac{r_k - \|\x_k - \x(t)\|}{\Gamma}\right) \frac{\left( \x_k - \x(t) \right)}{\|\x_k - \x(t)\|}\,,
\eeq
with initial values
\beq
\x(0) = \x_0, \quad \bp(0) = -\frac{1}{2D}\left(\cos(\theta), \sin(\theta) \right)\,.
\eeq
and initial velocity orientation $\theta$. Because $\Gamma \ll r_k$ the ODE system \eqref{eq:Eulersys} is stiff near the edge of the killing region, where $\|\x - \x_k\| = r_k$, causing trajectories to be highly sensitive to initial velocity angle variations. To reduce stiffness we solve \eqref{eq:Eulersys} in the small diffusion coefficient limit $D \ll 1$ (also corresponding to the regime where the large deviation principle holds \cite{freidlin1998random,bouchet2016generalisation}), since this only affects optimal travel times and not the shape of the two-dimensional trajectories. \\
We compare solutions of the system \eqref{eq:Eulersys} with spatial distributions of optimal paths obtained from stochastic simulations (Fig.~\ref{graph7}), where we gradually increase the killing rate $V$. Note the asymmetric domain configuration, with a killing zone of radius $r_k = 0.5$ centered in $\x_k = (1,0)$ and starting point shifted away from the origin to $\x_0 = (0,r_k)$. \\
When the killing probability is small, we find that optimal paths concentrate along the straight line from the initial release site to the narrow exit window, as shown in Fig.~\ref{graph7}\textbf{(A)} for $V=1$. These trajectories gradually deviate from the straight line as the killing rate increases: both deterministic and stochastic results reveal that for larger $V$ the fastest trajectories avoid the killing zone because of high termination probabilities. In this regime optimal paths concentrate along the geodesic connecting the starting point to the absorbing target inside the perforated domain $\Omega \setminus B_{r_k}(\x_k)$. However, these also correspond to longer escape times as seen from the rightward shift of the distribution of first exit times between $V = 1$ and $V=90$ (Fig.~\ref{graph7}\textbf{(B)}). Note that the discrepancy between optimal paths solving the minimization problem \eqref{minimization} and their stochastic counterparts is likely due to the diffusion coefficient employed in the Euler's scheme, which is described in appendix \S \ref{sec:algo_simu} and is used for the results shown in Fig.~\ref{graph7}. For computational convenience, we set $D = 1$, but this does not correspond to the small noise amplitude regime from the large deviation principle.
\section{Distribution of first exit points along a narrow absorbing arc on a disk}\label{sec:explicit_exitpoint}
In this section we apply the general theory developed in \S \ref{sec:exitpoint_flux} to derive explicit solutions for the distribution of first exit points along a narrow absorbing boundary arc $\p\Omega_a$ of length $2\eps$, parametrized as
\begin{equation}\label{eq:arc_def}
\partial\Omega_a=\left\{\y(\theta) = R \left(\cos(\theta),\sin(\theta)\right) \left| \,\, |\theta| \leq \frac{\eps}{R} \right. \right\}, \qquad 0 < \eps \ll R\,,
\end{equation}
with $\Omega = B_R(0)$ a disk of radius $R$ and starting point $\x_0$ defined as
\begin{equation}\label{eq:x0_polar_disk}
\x_0 = r_0\left(\cos(\theta_0),\sin(\theta_0)\right), \qquad 0 \leq r_0 < R \,.
\end{equation}
The time-slice exit location density $\rho(\theta\mid t,\x_0)$ is obtained from Eq.~\eqref{eq:rho_formula} by normalizing the outward normal flux density $j(\theta,t\mid\x_0)$,
\begin{equation}\label{eq:rho_def_disk}
\rho(\theta\mid t,\x_0) = \frac{j(\theta,t\mid \x_0)}{\int_{-\eps/R}^{\eps/R} j(\phi,t \mid \x_0)\,R\,d\phi}\,,
\end{equation}
and the first exit-point density for $N_0$ large satisfies,
\beq\label{eq:pN_rho_tN_disk}
p_{N_0}(\theta\mid \x_0) \approx \rho(\theta\mid t_{N_0},\x_0), \quad N_0 \gg 1\,,
\eeq
with $t_{N_0}$ the characteristic of fastest time. We first characterize this distribution for the case of a weak uniform killing $k(\x) = V$ (\S \ref{sec:rho_uniform_killing}), and then for a nonuniform killing $k(\x)$ (\S \ref{sec:rate-function-killing}). In \S \ref{sec:tangent_two_optimal} we investigate how a localized killing zone tangent to the absorbing boundary can perturb $p_{N_0}(\theta\mid \x_0)$. Stochastic simulation results are presented in \S \ref{sec:simulated_exit_point}.
\subsection{Uniform killing field and boundary flux half-plane approximation}\label{sec:rho_uniform_killing}
For a uniform killing field $k(\x) = V$ we compute the exit flux by mapping the local neighborhood close to the narrow target $\p\Omega_a$ to a half-plane. This approximation holds since the trajectories that contribute to the flux must travel from the starting point $\x_0$ to a neighborhood of a point $\y(\theta) \in \p\Omega_a$ within a short time $t_{N_0}$. Hence the corresponding diffusive length $\sqrt{Dt_{N_0}}$ is small and we can neglect the boundary curvature. \\
We start by introducing the tangent $\boldsymbol\tau(\theta)$ and outward normal $\n(\theta)$ unit vectors at $\y(\theta) \in \p\Omega_a$, with which any point $\x$ close to $\y(\theta)$ can be written as
\begin{equation}\label{eq:local_coords_def}
\x = \y(\theta) + a\,\boldsymbol\tau(\theta) - b\,\n(\theta)\,,
\end{equation}
where $(a,b)$ is a set of local Cartesian coordinates: $a$ is the tangential offset and $b>0$ is the inward normal distance from the tangent line. This allows a decomposition of the total displacement as $\x_0 - \y(\theta)$ as
\begin{equation}\label{eq:local_decomp_repeat}
\x_0 - \y(\theta) = a(\theta)\,\boldsymbol\tau(\theta)-b(\theta)\,\n(\theta)\,,
\end{equation}
where
\beq
a(\theta) = (\x_0 - \y(\theta)) \cdot \boldsymbol\tau(\theta), \qquad b(\theta) = (\y(\theta) - \x_0) \cdot \n(\theta)\,,
\eeq
and distance
\begin{equation}\label{eq:a2b2_equals_d2}
d(\theta)^2 = \|\x_0 - \y(\theta)\|^2 = a(\theta)^2 + b(\theta)^2 \,.
\end{equation}
In half-space geometry $\{(a,b) \in \R^2 \mid b \geq 0 \}$ we solve the Fokker-Planck Eq.~\eqref{FPEp} with diffusion coefficient $D$, uniform killing $V$, starting point $(a_0,b_0)$ and absorbing boundary condition on the line $b=0$. Applying the method of images yields the pdf $p_{\mathrm{hp}}(a,b,t\mid a_0,b_0)$,
\begin{equation}\label{eq:hp_kernel_images}
p_{\mathrm{hp}}(a,b,t\mid a_0,b_0) = p_{\R^2}(a,b,t\mid a_0,b_0) - p_{\R^2}(a,b,t\mid a_0,-b_0)\,,
\end{equation}
where $p_{\R^2}(a,b,t\mid a_0,b_0)$ is the free-space heat kernel
\begin{equation}\label{eq:free_kernel_2d}
p_{\R^2}(a,b,t\mid a_0,b_0) = \frac{1}{4\pi Dt}\exp\!\left(- Vt - \frac{(a-a_0)^2+(b-b_0)^2}{4Dt}\right)\,.
\end{equation}
The outward flux density $j_{\mathrm{hp}}(a,t\mid a_0,b_0)$ in local cartesian coordinates is then computed as,
\begin{equation}\label{eq:hp_flux_def}
j_{\mathrm{hp}}(a,t\mid a_0,b_0) = \mb{J}(a,t\mid a_0,b_0)\cdot \n(a,b) = -D\,\partial_{\n} p_{\mathrm{hp}}\Big|_{b=0} = D\,\partial_b p_{\mathrm{hp}}\Big|_{b=0}\,,
\end{equation}
and differentiating Eq.~\eqref{eq:hp_kernel_images} with respect to $b$ gives the following results
\begin{align}
j_{\mathrm{hp}}(a,t\mid a_0,b_0) &= D\left( \partial_b p_{\R^2}(a,0,t\mid a_0,b_0) - \partial_b p_{\R^2}(a,0,t\mid a_0,-b_0) \right) \nonumber \\
&= D\left( \frac{b_0}{2Dt}\,p_{\R^2}(a,0,t\mid a_0,b_0) +\frac{b_0}{2Dt}\,p_{\R^2}(a,0,t\mid a_0,-b_0) \right) \nonumber \\
&= \frac{b_0}{4\pi D\,t^2}\exp\!\left(-Vt - \left(\frac{(a-a_0)^2+b_0^2}{4Dt}\right)\right), \qquad t>0, \quad b_0 > 0\,. \label{eq:halfplane_flux_final}
\end{align}
The terms $(a-a_0)^2 + b_0^2$ correspond to the square distance $d(\theta)^2$ from the boundary point $\y(\theta) \in \p\Omega_a$, with $a-a_0$ as the tangential boundary coordinate. Up to a smooth Jacobian between boundary arclength and cartesian coordinate $a$ elements, the outward flux density \eqref{eq:halfplane_flux_final} can be mapped to
\begin{equation}\label{eq:j_smalltime_Aexp_repeat}
j(\theta,t\mid \x_0) = A(\theta,t\mid \x_0) \, \exp\!\left(-Vt-\frac{d(\theta)^2}{4Dt}\right), \quad |\theta| \leq \frac{\eps}{R}, \quad t\downarrow 0\,,
\end{equation}
where the prefactor $A(\theta,t\mid \x_0)$ is smooth in $\theta$ and varies at most algebraically in $t$. In contrast, the exponential term varies on the scale $e^{-a/t}$ and will therefore dominate the angular dependence. With Eq.~\eqref{eq:arc_def}-\eqref{eq:x0_polar_disk} we write the distance $d(\theta)^2$ as,
\beq\label{eq:dtheta_explicit_disk}
d(\theta)^2 = \|\x_0-\y(\theta)\|^2 = r_0^2 + R^2 - 2 r_0 R \cos(\theta - \theta_0), \quad |\theta| \leq \frac{\eps}{R}\,,
\eeq
and upon normalizing the flux in Eq.~\eqref{eq:j_smalltime_Aexp_repeat}, the prefactor cancels and we obtain the distribution of first exit points along the narrow absorbing arc in the large $N_0$-limit,
\begin{equation}\label{eq:pN_exp_cos}
p_{N_0}(\theta \mid \x_0) \approx \rho(\theta\mid t_{N_0},\x_0) \approx \frac{\exp\!\left(\frac{r_0R\cos(\theta-\theta_0)}{2D\,t_{N_0}}\right)}{{\int_{-\eps/R}^{\eps/R}\exp\!\left(\frac{r_0R\cos(\phi-\theta_0)}{2D\,t_{N_0}}\right)R\,d\phi}}, \quad \text{with} \quad |\theta| \leq \frac{\eps}{R}, \quad N_0 \gg 1\,,
\end{equation}
with selected exit time $t_{N_0}$ computed in \S \ref{sec:FPE},
\beq
t_{N_0} \approx \frac{s^2}{4D\, \log\!\left(\dfrac{N_0}{2\log(R/\eps) \left(1 - (1-P_e(\x_0))^{N_0} \right)}\right)}, \quad \text{with} \quad s = d(\theta^*) = \min\left\{ d(\theta) \,\, \Big| \,\, |\theta| \leq \frac{\eps}{R} \right\}\,,
\eeq
thus $\theta^*$ is the optimal exit angle and $s$ is the shortest distance. However if $r_0 = 0$ then Eq.~\eqref{eq:pN_exp_cos} reduces to a uniform distribution,
\beq
p_{N_0}(\theta \mid \x_0) \approx \frac{1}{2\eps}, \quad |\theta| \leq \frac{\eps}{R}, \quad r_0 = 0, \quad N_0 \gg 1\,,
\eeq
since starting from the origin we get the same probability to reach any $\theta$ on the narrow absorbing arc. \\
When $r_0 \neq 0$ the shape of the distribution depends on the position of the optimal angle $\theta^*$ along the narrow arc. For the distance $d(\theta)$ to be minimized a an interior point we must have
\beq
|\theta_0| < \frac{\eps}{R}: \quad \theta^* = \theta_0, \quad s = d(\theta^*) = R - r_0\,,
\eeq
and from the expansion $\cos(\theta-\theta_0) \approx 1 - \frac{1}{2}(\theta-\theta_0)^2$ Eq.~\eqref{eq:pN_exp_cos} reduces to a truncated Gaussian density around the most likely exit angle,
\begin{equation}\label{eq:pN_trunc_gauss_final}
p_{N_0}(\theta \mid \x_0) \approx \frac{\exp\!\left(-\frac{\left(\theta-\theta_0\right)^2}{2\sigma_{N_0}^2}\right)}{{\int_{-\eps/R}^{\eps/R}\exp\!\left(-\frac{\left(\phi-\theta_0\right)^2}{2\sigma_{N_0}^2}\right)R\,d\phi}} \approx \frac{\exp\!\left(-\frac{(\theta-\theta_0)^2}{2\sigma_{N_0}^2}\right)}{\sqrt{2\pi}R\,\sigma_{N_0}}, \quad |\theta| \leq \frac{\eps}{R}, \quad N_0 \gg 1\,,
\end{equation}
where $\sigma_{N_0}^2$ is the variance proportional to the selected time $t_{N_0}$, defined as
\begin{equation}\label{eq:tN_explicit_disk}
\sigma_{N_0}^2 = \frac{2D\,t_{N_0}}{r_0R}, \quad t_{N_0} \approx \frac{(R-r_0)^2}{4D\, \log\!\left(\dfrac{N_0}{2\log(R/\eps) \left(1 - (1-P_e(\x_0))^{N_0} \right)}\right)}, \quad N_0 \gg 1\,.
\end{equation}
The following remark summarizes how the variance modifies the shape of the exit-point distribution from a uniform to a Gaussian distribution as $N_0\to\infty$.
\begin{remark}[\textbf{Uniform versus Gaussian exit-point distribution}] For $r_0 \neq 0$, $|\theta_0| < \frac{\eps}{R}$ and $N_0 \gg 1$, the minimizing angle $\theta^* = \theta_0$ is an interior point and the exit-point density $p_{N_0}(\theta\mid \x_0)$ on the absorbing arc is approximately Gaussian with variance
\begin{equation}\label{eq:sigma_thetaN_explicit_remark}
\sigma_{N_0}^2 \approx \frac{(R - r_0)^2}{2 r_0R\,\log\!\left(\dfrac{N_0}{2\log(R/\eps)\left(1 - (1-P_e(\x_0))^{N_0} \right)}\right)}\,.
\end{equation}
The shape observed on the finite arc $\left[-\eps,\eps\right]$ is controlled by the ratio between the intrinsic angular width $R\sigma_{N_0}$ and the arc half-width $\eps$:

\smallskip
\noindent\textbf{(i) Almost uniform regime.} If $\sigma_{N_0} \gg \eps/R$, then the exponential factor varies weakly across the arc and
\begin{equation}\label{eq:uniform_regime}
p_{N_0}(\theta\mid \x_0)\approx \frac{1}{2\eps},\qquad |\theta| \leq \frac{\eps}{R}\,.
\end{equation}
Using \eqref{eq:sigma_thetaN_explicit_remark}, this corresponds to
\begin{equation}\label{eq:uniform_condition}
\log\!\left(\frac{N_0}{2\log(R/\eps)\left(1 - (1-P_e(\x_0))^{N_0} \right)}\right) \ll \frac{R(R-r_0)^2}{2r_0\eps^2}.
\end{equation}

\smallskip
\noindent\textbf{(ii) Gaussian regime.}
If $\sigma_{N_0} \ll \eps/R$, the truncation is negligible and
\begin{equation}\label{eq:peaked_regime}
p_{N_0}(\theta\mid \x_0) \approx
\frac{1}{\sqrt{2\pi}R\,\sigma_{N_0}}\exp\!\left(-\frac{(\theta-\theta_0)^2}{2\sigma_{N_0}^2}\right), \quad |\theta| \leq \frac{\eps}{R}\,,
\end{equation}
and the distribution concentrates near the most likely exit point $\theta_0$. Equivalently,
\begin{equation}\label{eq:peaked_condition}
\log\!\left(\frac{N_0}{2\log(R/\eps)\left(1 - (1-P_e(\x_0))^{N_0} \right)}\right) \gg \frac{R(R-r_0)^2}{2r_0\eps^2}\,.
\end{equation}

\smallskip
\noindent In particular, at fixed $\eps$ the distribution becomes more concentrated as $N_0$ increases, whereas decreasing $r_0$ (starting closer to the origin) broadens the distribution through the scaling $\sigma_{N_0}\propto \sqrt{R/r_0}$.
\end{remark}

\noindent Lastly we consider cases where $d(\theta)$ is minimized at the corners, which occur when the initial angle satisfies
\beq\label{eq:theta0out}
|\theta_0| \geq \frac{\eps}{R}: \quad \theta^*= \frac{\eps\theta_0}{R|\theta_0|}, \quad \text{and} \quad s = d(\theta^*) = r_0^2 + R^2 - 2r_0R \cos\left(\frac{\eps\theta_0}{R|\theta_0|} - \theta_0\right)\,.
\eeq
By splitting and expanding $\cos(\theta - \theta_0)$ near the minimizing angle $\theta^*$ we obtain
\begin{align}
\cos(\theta - \theta_0) & = \cos(\theta - \theta^*)\cos(\theta^* - \theta_0) - \sin(\theta - \theta^*)\sin(\theta^* - \theta_0) \nonumber \\
&\approx \cos(\theta^* - \theta_0) - (\theta - \theta^*)\sin(\theta^* - \theta_0) - \frac{1}{2}(\theta - \theta^*)^2\cos(\theta^* - \theta_0) + O((\theta - \theta^*)^3)\,. \label{eq:cos_expan}
\end{align}
For $|\theta_0| = \eps/R$ (or very close to the corners with $\left|\theta_0\right| - \frac{\eps}{R} \ll 1$), we have $\sin(\theta^* - \theta_0) \approx 0$ in Eq.~\eqref{eq:cos_expan} and expect the distribution to converge to a half-Gaussian density,
\begin{equation}\label{eq:half_peaked_regime}
p_{N_0}(\theta\mid \x_0) \approx
\frac{\sqrt{2}}{\sqrt{\pi}R\,\sigma_{N_0}}\exp\!\left(-\frac{\left(\theta-\frac{\eps\theta_0}{R|\theta_0|}\right)^2}{2\sigma_{N_0}^2}\right), \quad |\theta| \leq \frac{\eps}{R}, \quad N_0 \gg 1\,.
\end{equation}
Alternatively for $|\theta_0| \gg \frac{\eps}{R}$ further away from the narrow corners, the linear expansion term in Eq.~\eqref{eq:cos_expan} dominates and the first-exit point distribution reduces to a one-sided exponential,
\begin{equation}\label{eq:pN_one_sided_exp}
p_{N_0}(\theta \mid \x_0) \approx \frac{\exp\!\left(-\lambda_{N_0}\theta\right)}{{\int_{-\eps/R}^{\eps/R}\exp\!\left(-\lambda_{N_0}\phi\right)R\,d\phi}} = \frac{\lambda_{N_0}\exp\!\left(-\lambda_{N_0}\theta\right)}{2R\sinh\left(\lambda_{N_0}\frac{\eps}{R}\right)}, \quad \text{with} \quad \theta_0 \notin \left[-\frac{\eps}{R},\frac{\eps}{R}\right], \quad N_0 \gg 1\,,
\end{equation}
and $|\theta| \leq \frac{\eps}{R}$. The parameter $\lambda_{N_0}$ controlling the slope is defined as
\beq
\lambda_{N_0} = \frac{r_0R}{2D\,t_{N_0}}\sin\left(\frac{\eps\theta_0}{R|\theta_0|} - \theta_0\right), \quad \text{with} \quad t_{N_0} \approx \frac{\left(r_0^2 + R^2 - 2r_0R \cos\left(\frac{\eps\theta_0}{R|\theta_0|} - \theta_0\right)\right)^2}{4D\, \log\!\left(\dfrac{N_0}{2\log(R/\eps) \left(1 - (1-P_e(\x_0))^{N_0} \right)}\right)}\,.
\eeq
\subsection{Non-uniform spatial killing and WKB asymptotic approximation of the boundary flux}\label{sec:rate-function-killing}
We shall now study the distribution of exit points for the fastest particle with a non-uniform killing field $k(\x)$. Our approach uses a WKB method to approximate the leading order term of the distribution, showing how killing influences through the geodesic associated with the control problem. This is in contrast with the previous case (section \S \ref{sec:rho_uniform_killing}) where geodesics are given by the Euclidean distance due to uniform killing. Lastly we explore various regimes where the distribution of first exit points can be uniform or peaked at the most probable exit point. \\
In the short-time asymptotic relevant to compute statistics of the fastest, we use the following WKB ansatz for the pdf $p(\x,t\mid\x_0)$ solution of the Fokker-Planck equation \eqref{FPEp} with no drift,
\begin{equation}\label{eq:WKB_ansatz}
p(\x,t\mid \x_0)\;\sim\;A(\x,t\mid \x_0)\,\exp\!\big(-\Phi(\x,t\mid \x_0)\big), \qquad t \downarrow 0\,,
\end{equation}
where $\Phi(\x,t\mid\x_0)$ is the reduced action or rate function, and $A(\x,t \mid\x_0)$ is a subexponential prefactor term. To derive equations satisfied by $\Phi(\x,t\mid\x_0)$ and $A(\x,t\mid\x_0)$, we insert Eq.~\eqref{eq:WKB_ansatz} into \eqref{FPEp} to get
\beq
\frac{\p}{\p t}(Ae^{-\Phi}) = D\Delta(Ae^{-\Phi})-k(\x)Ae^{-\Phi}\,,
\eeq
which becomes
\beq\label{eq:WKB_substitution}
\frac{\p A}{\p t} - A \frac{\p \Phi}{\p t} = D\left(\Delta A - 2\nabla A\cdot\nabla\Phi + A\big(\|\nabla\Phi\|^2-\Delta\Phi\big)\right) - k(\x)A\,.
\eeq
The WKB separation proceeds by matching the dominant exponential orders \cite{schuss2011nonlinear,Schuss2009Book}, which yields an eikonal equation for the reduced action $\Phi(\x,t\mid\x_0)$
\begin{equation}\label{eq:HJ}
\frac{\p\Phi}{\p t} + D\|\nabla\Phi\|^2 = 0 \,,
\end{equation}
together with a transport equation for the prefactor $A(\x,t\mid\x_0)$,
\begin{equation}\label{eq:transport_leading}
\frac{\p A}{\p t} + 2D\,\nabla\Phi\cdot\nabla A + \big(D\Delta\Phi + k(\x)\big)A \approx 0 \,.
\end{equation}
As shown in section \S \ref{sec:oc_exit} the solution of the eikonal equation is given by
\begin{equation}\label{eq:Phi_action}
\Phi(\x,t\mid \x_0) = \inf_{\substack{\bgamma(0)=\x_0\\ \bgamma(t)=\x}} \int_0^t\left(\frac{\|\dot\bgamma(s)\|^2}{4D}+ k(\bgamma(s)) \right)\,ds,
\end{equation}
with minimizer $\bgamma^*$ satisfying
\begin{equation}\label{eq:EL_gamma_star}
\ddot\bgamma^*(s)=2D\,\nabla k(\bgamma^*(s)), \qquad \bgamma^*(0) = \x_0, \quad \bgamma^*(t) = \x\,,
\end{equation}
while the solution of the transport equation is obtained with the characteristic path $s\mapsto \bgamma^*(s)$. By defining the momentum as
\begin{equation}\label{eq:momentum_def}
\bp^*(s) = \frac{\dot\bgamma^*(s)}{2D}\,,
\end{equation}
then the endpoint gradient relation for the action gives
\begin{equation}\label{eq:gradPhi_momentum}
\nabla \Phi(\x,t\mid \x_0)\Big|_{\x = \bgamma^*(t)} = \bp^*(t)\,,
\end{equation}
and the characteristic velocity in \eqref{eq:transport_leading} is $2D\nabla\Phi = \dot\bgamma^*$. Along $\bgamma^*$ the transport equation reduces to an ODE:
\begin{equation}\label{eq:transport_along_gamma}
\frac{d}{ds}A(\bgamma^*(s),s\mid \x_0) + \big(D\Delta\Phi(\bgamma^*(s),s\mid \x_0)+k(\bgamma^*(s))\big)\,
A(\bgamma^*(s),s\mid \x_0) \approx 0\,.
\end{equation}
A more explicit representation is obtained by expressing $A$ in terms of a Jacobi determinant. We use the one-parameter family of optimal trajectories $\bgamma_\theta(s)$ generated
by changing the initial shooting angle $\theta$ (as in the numerical scheme described in \S \ref{sec:num_shooting}), and define
\begin{equation}\label{eq:m_def}
\bbm(s) = \frac{\p}{\p\theta}\bgamma_\theta(s)\Big|_{\theta=\theta^*},
\qquad \y(\theta) = \bgamma_\theta(t)\in\partial\Omega_a\,.
\end{equation}
Linearizing \eqref{eq:EL_gamma_star} around $\bgamma^*$ yields the Jacobi variational equation
\begin{equation}
\ddot \bbm(s) = 2D\,\mathrm{Hess}\,k(\bgamma^*(s))\,\bbm(s),
\label{eq:Jacobi_m}
\end{equation}
with initial conditions inherited from the shooting parametrization. For example, if the initial velocity is
$\dot\bgamma_\theta(0)=(\cos\theta,\sin\theta)$, then
\begin{equation}\label{eq:Jacobi_IC}
\bbm(0)=0, \qquad \dot \bbm(0)=(-\sin\theta^*,\cos\theta^*)\,.
\end{equation}
In two dimensions, the leading-order WKB transport solution can be written (up to a constant factor) as
\begin{equation}\label{eq:A_geo}
A(\bgamma^*(t),t \mid \x_0)\;\propto\;\frac{1}{\sqrt{\big\|\p_\theta \y(\theta^*)\big\|}} \;= \;\frac{1}{\sqrt{\|\bbm(t)\|}},
\end{equation}
known as the ``inverse square-root of ray spreading'' rule and corresponds to the prefactor structure as discussed in \cite{schuss2011nonlinear}.
\subsubsection{WKB approximation of the boundary flux.}
Using \eqref{eq:WKB_ansatz}--\eqref{eq:gradPhi_momentum}, the boundary flux at $\y\in\partial\Omega_a$ is
\begin{align}
j(\y,t\mid \x_0) &=-\n(\y)\cdot\big( D\nabla\big(Ae^{-\Phi}\big) \big) \\
&\approx D\,A(\y,t\mid \x_0)\,\big(\n(\y) \cdot \nabla \Phi(\y,t\mid \x_0)\big)\,\exp\left(-\Phi(\y,t\mid \x_0)\right) \nonumber \\
&= \frac{1}{2}\,\big(\dot\bgamma^*(t)\cdot n(\y)\big)\, A(\y,t\mid \x_0)\,\exp\left(-\Phi(\y,t\mid \x_0)\right)\,. \label{eq:flux_WKB_final}
\end{align}
Thus the short-time flux is controlled by the optimal action $\Phi(\y,t\mid \x_0)$ and the transport prefactor $A(\y,t\mid \x_0)$, where the latter can be computed by integrating the linearized Jacobi system \eqref{eq:Jacobi_m}--\eqref{eq:Jacobi_IC} in parallel with the optimal ODE \eqref{eq:EL_gamma_star}. \\
For a disk of radius $R$ with narrow absorbing arc parametrized as $\y(\theta) = R\left(\cos(\theta),\sin(\theta) \right)$ with $|\theta| \leq \frac{\eps}{R}$, we define the boundary flux, restricted action and prefactor as
\begin{equation}\label{eq:Phi_A_theta}
j(\theta,t\mid\x_0) = j(\y(\theta),t\mid \x_0), \qquad \Phi(\theta,t\mid\x_0) = \Phi(\y(\theta),t\mid \x_0), \qquad A(\theta,t\mid\x_0) = A(\y(\theta),t\mid \x_0)\,.
\end{equation}
Along the trajectory $\theta^*(t)$ that minimizes $\Phi(\theta,t\mid\x_0)$ on $\left[-\frac{\eps}{R}, \frac{\eps}{R}\right]$, we can expand
\begin{equation}\label{eq:Phi_expand}
\Phi(\theta,t\mid\x_0)=\Phi(\theta^*(t),t\mid\x_0)+\frac12\,\Phi_{\theta\theta}(\theta^*(t),t\mid\x_0)\,(\theta-\theta^*(t))^2+o\big((\theta-\theta^*)^2\big)\,,
\end{equation}
and Laplace's method yields the approximation below for the total boundary flux,
\begin{equation}\label{eq:Jk_Laplace_theta}
J(t\mid \x_0) = \int_{-\frac{\eps}{R}}^{\frac{\eps}{R}} j(\theta,t\mid \x_0)\,R\,d\theta \approx R\, j(\theta^*(t),t\mid \x_0)\, \sqrt{\frac{2\pi}{\Phi_{\theta\theta}(\theta^*(t),t\mid\x_0)}}\,.
\end{equation}
Using \eqref{eq:flux_WKB_final} and \eqref{eq:A_geo} at $\y^* = \y(\theta^*(t))$ gives
\begin{equation*}
J(t\mid \x_0) \approx \mathcal{C}(t\mid \x_0) \exp \!\big(-\Phi(\theta^*(t),t\mid\x_0)\big), \quad \text{with} \quad \mathcal{C}(t\mid \x_0) = \frac{R\left(\dot\bgamma^*(t)\cdot n(\y^*)\right)}{2\sqrt{|\partial_\theta \y(\theta^*)|}}\, \sqrt{\frac{2\pi}{\Phi_{\theta\theta}(\theta^*(t),t\mid\x_0)}}\,,
\end{equation*}
a subexponential factor. Matching with the boundary layer near a narrow absorbing arc introduces, at leading order, the standard 2-D narrow-escape factor $1/\big(2\log(R/\eps) \big)$,
\begin{equation}\label{eq:Jk_narrow_escape}
J(t\mid \x_0) \asymp \frac{\mathcal{C}(t\mid \x_0)}{2\log(R/\eps)} \exp\!\big(-\Phi(\theta^*(t),t\mid\x_0)\big), \qquad t\downarrow 0\,.
\end{equation}
To conclude, formula \ref{eq:Jk_narrow_escape} generalizes the short-time flux to a non-uniform killing field, where the Euclidean cost is replaced by the minimizing action and with the prefactor given by the WKB transport \cite{schuss2011nonlinear}.
\subsubsection{Extreme mean first passage time with a killing field}
We show that for $N_0$ independent particles, the extreme mean first passage time with killing satisfies
\begin{equation}\label{CDFexteme}
\bar\tau_{\rm EMFPT}(N_0) = \int\limits_0^{+\infty}\left(\frac{\left(1 - F(t\mid\x_0)\right)^{N_0} -\left(1-P_e(\x_0)\right)^{N_0}}{1-\left(1-P_e(\x_0)\right)^{N_0}}\right)\, dt \sim t_{N_0}, \quad N_0 \gg 1\,,
\end{equation}
with $t_{N_0}$ the characteristic of fastest time. Starting from the approximation,
\begin{equation}\label{eq:mean_laplace_form_detail}
\bar\tau_{\rm EMFPT}(N_0) \approx \int_0^\infty \exp\left(- \frac{N_0}{Q_{N_0}}F(t \mid \x_0)\right)\,dt, \quad \text{with} \quad Q_{N_0} = 1-\left(1-P_e\left(\x_0\right)\right)^{N_0}, \quad N_0 \gg 1\,,
\end{equation}
we perform the change of variables $u=N_0F(t\mid\x_0)/Q_{N_0}$ to obtain
\begin{equation}\label{eq:mean_u_integral_detail}
\bar\tau_{\rm EMFPT}(N_0) \approx \int_0^{N_0P_e\left(\x_0\right)/Q_{N_0}} \frac{Q_{N_0}e^{-u}}{N_0\,J(t(u)\mid \x_0)}\,du \approx \frac{Q_{N_0}}{N_0 \,J(t_{N_0}\mid \x_0) }, \quad N_0 \gg 1\,,
\end{equation}
and since dominant contributions come from short times for which $u = O(1)$, the selected time $t_{N_0}$ satisfies $N_0F(t_{N_0}\mid\x_0)/Q_{N_0} \asymp 1$. Expanding near $t_{N_0}$ yields
\begin{equation}\label{eq:linearize_Fk}
F(t\mid\x_0)\approx \frac{Q_{N_0}}{N_0}+J(t_{N_0}\mid\x_0)(t-t_{N_0})
\quad \text{near} \quad t_{N_0}, \quad N_0 \gg 1\,,
\end{equation}
and evaluating in $t=0$ gives
\begin{equation}\label{eq:J_tN0}
J(t_{N_0}\mid\x_0) = \frac{Q_{N_0}}{N_0 t_{N_0}}, \quad N_0 \gg 1\,,
\end{equation}
and by substituting within Eq.~\eqref{eq:mean_u_integral_detail} we recover the standard extreme-statistics estimate
\begin{equation}\label{eq:mean_asymp_tNk_detail}
\bar\tau_{\rm EMFPT}(N_0) \sim t_{N_0},
\qquad N_0\to\infty\,.
\end{equation}
The asymptotic estimate \eqref{eq:J_tN0} can be used to estimate the short-time structure of the optimal action, as seen by setting $t = t_{N_0}$ in Eq.~\eqref{eq:Jk_narrow_escape},
\begin{equation}\label{eq:Phi_balance}
\Phi(\theta^*(t_{N_0}),t_{N_0}\mid\x_0) = \log\left( \frac{N_0}{2\log(R/\eps)Q_{N_0}} \right) + \log \left(\mathcal{C}(t_{N_0}\mid \x_0 \right)) + o(1)\,.
\end{equation}
\subsubsection{Short-time structure of the optimal action}
To select the optimal trajectory, we use the optimal action
\begin{equation}\label{eq:Phi_final_def_clean}
\Phi(t\mid \x_0) = \min_{\y \in \p\Omega_a}\,
\inf_{\substack{\bgamma(0) = \x_0\\ \bgamma(t) = \y}}
\int_0^t\left(\frac{\|\dot\bgamma(s)\|^2}{4D}+k(\bgamma(s))\right)\,ds\,,
\end{equation}
and assume that there is a unique minimizing trajectory $\bgamma^*(s)$ joining $\x_0$ to $\y^*(t)\in\p\Omega_a$. Along $\bgamma^*$ we use the time-averaged energy and killing
\begin{equation}\label{eq:ell_kbar_def_clean}
\int_0^t \|\dot\bgamma^*\|^2 ds = \ell_*^2/t, \qquad \bar{k}_*(t) = \frac{1}{t}\int_0^t k(\bgamma^*(s))\,ds\,,
\end{equation}
such that in the short-time regime, we have
\begin{equation}\label{eq:a_over_t_plus_bt_general_clean}
\Phi(t\mid \x_0) \approx \int_0^t\left(\frac{\|\dot\bgamma^*(s)\|^2}{4D}+k(\bgamma^*(s))\right)\,ds \approx \frac{\ell_*(t)^2}{4D\,t}+t\,\bar{k}_*(t),
\qquad t\downarrow 0\,.
\end{equation}
Upon evaluating Eq.~\eqref{eq:a_over_t_plus_bt_general_clean} at the selected time $t_{N_0}$ and using the estimate for the short-time action in Eq.~\eqref{eq:Phi_balance}, we obtain a quadratic relation,
\begin{equation}\label{eq:balance_ab_pi_clean}
\frac{a}{t_{N_0}}+b\,t_{N_0}
\approx L_{N_0}, \qquad a=\frac{\ell_*^2}{4D}, \qquad b = \bar{k}_*, \qquad L_{N_0} = \log\!\left( \frac{N_0}{2\log(R/\eps)Q_{N_0}} \right) + O(1)\,,
\end{equation}
with $\ell_*=\ell_*(t_{N_0})$, $\bar{k}_*=\bar{k}_*(t_{N_0})$ and solution $t_{N_0}$ given by
\begin{equation}\label{eq:tNk_explicit_quadratic_clean}
t_{N_0}\approx \begin{cases}
\frac{L_{N_0}-\sqrt{L_{N_0}^2-4ab}}{2b} & \text{if} \quad b > 0 \quad \text{and} \quad L_{N_0}^2 > 4ab \\
\frac{a}{L_{N_0}} & \text{if} \quad b = 0
\end{cases}, \quad N_0 \gg 1\,.
\end{equation}
When $b\neq0$, we can further expand in the limit $L_{N_0}\to\infty$ to obtain
\begin{equation}\label{eq:tNk_expansion_clean}
t_{N_0} = \frac{a}{L_{N_0}} + \frac{a^2 b}{L_{N_0}^3} + O(L_{N_0}^{-5}), \qquad L_{N_0} = \log\!\left( \frac{N_0}{2\log(R/\eps)Q_{N_0}} \right) \to\infty\,,
\end{equation}
which reduces to the classical no-killing scaling when $b\to 0$.
\subsubsection{First exit-point distribution in the large initial population regime}\label{subsec:exitpoint_regimes_killing}
The exit-point density on the arc for the first successful absorber among $N_0$ is given by
\begin{equation}\label{eq:pNk_rhok}
p_{N_0}(\theta\mid \x_0)\approx \rho(\theta\mid t_{N_0},\x_0), \qquad \rho(\theta\mid t,\x_0)=\frac{j(\theta,t\mid \x_0)}{\int_{-\eps/R}^{\eps/R} j(\phi,t\mid \x_0)R\,d\phi}\,,
\end{equation}
where $t_{N_0}$ is given by Eq.~\eqref{eq:tNk_expansion_clean}. Using the WKB solution for $\rho(\theta\mid t_{N_0},\x_0)$ given in Eq.~\eqref{eq:flux_WKB_final} yields
\begin{equation}\label{eq:pNk_rate_theta}
p_{N_0}(\theta\mid \x_0) \approx \frac{\exp\!\big(-\Phi(\theta,t_{N_0}\mid \x_0)\big)}{\int_{-\eps/R}^{\eps/R}\exp\!\big(-\Phi(\phi,t_{N_0}\mid \x_0)\big)R\,d\phi}, \qquad |\theta| \leq \frac{\eps}{R}\,,
\end{equation}
and therefore the large-$N_0$ exit-point distribution is governed by the minimizers of $\Phi(\theta,t_{N_0}\mid \x_0)$. When $\theta^* = \theta^*(t_{N_0})$ is an interior, nondegenerate minimizer of $\Phi(\theta,t_{N_0} \mid \x_0)$, then
\begin{equation*}
\Phi(\theta,t_{N_0}\mid \x_0) = \Phi(\theta^*,t_{N_0}\mid \x_0) + \frac{1}{2} \Phi_{\theta\theta}(\theta^*,t_{N_0}\mid \x_0)\,(\theta-\theta^*)^2 + o\!\big((\theta-\theta^*)^2\big)\,,
\end{equation*}
and Eq.~\eqref{eq:pNk_rate_theta} yields a Gaussian distribution. Thus we have locally near the optimal exit point:
\begin{equation}\label{eq:Gaussian_general}
p_{N_0}(\theta\mid \x_0) \approx \frac{1}{\sqrt{2\pi}R\,\sigma_{N_0}} \exp\!\left(-\frac{(\theta-\theta^*)^2}{2\sigma_{N_0}^2}\right), \qquad \sigma_{N_0}^2=\frac{1}{\Phi_{\theta\theta}(\theta^*,t_{N_0}\mid \x_0)}\,.
\end{equation}
We identify two main regimes with respect to $N_0$:
\begin{enumerate}
\item \textbf{Window-dominated regime (moderate $N_0$):} if $\sigma_{N_0}\gtrsim \eps/R$, then the exponential weight varies weakly on $\left[- \frac{\eps}{R}, \frac{\eps}{R}\right]$ and the density is approximately uniform:
\begin{equation*}
p_{N_0}(\theta\mid \x_0)\approx \frac{1}{2\eps}\,.
\end{equation*}
\item \textbf{Localization regime (very large $N_0$):} if $\sigma_{N_0}\ll \eps/R$, the distribution is sharply peaked near $\theta^*$ and the Gaussian approximation \eqref{eq:Gaussian_general} applies. The crossover is determined by $\sigma_{N_0} \sim \eps/R$.
\end{enumerate}
To conclude, the minimum-action $\Phi(\theta,t\mid \x_0)$ interpolates between qualitatively different behaviors:
\begin{enumerate}
\item \textbf{Weak killing (perturbative).} If the typical fastest time satisfies
\[
\int_0^{t_{N_0}} k(\bgamma_{0}(s))\,ds \ll 1
\]
along the no-killing minimizer $\bgamma_0$, then the killing only produces pertubation corrections. In this regime, $\theta^*$ is close to the no-killing minimizer (closest point on the arc) and $t_{N_0}$ differs only through a second order correction.
\item \textbf{Intermediate killing (competition).} When
\[
\int_0^{t_{N_0}} k(\bgamma(s))\,ds=O(1)\,,
\]
then the minimizer of $\Phi(\theta,t\mid\x_0)$ can shift, producing an observable bias of the exit-point distribution away from high-killing regions. This corresponds to a deformation of the optimal paths computed from the Euler--Lagrange equation.
\item \textbf{Strong localized killing (obstacle-like).} When killing is large on a localized set intersecting the straight path to the window, the minimizer avoids that set and the dominant kinetic cost is controlled by the shortest geodesic in the perforated domain. In this regime, the Euclidean distance gets replaced by the geodesic length $\ell_*$.
\end{enumerate}
\subsection{Symmetric exit-point distribution resulting from an inner disk killing zone tangent to the narrow absorbing arc}\label{sec:tangent_two_optimal}
We analyze the impact of a tangent killing zone to the narrow target $\p\Omega_a$ on the exit-distribution of the first absorber among $N_0$ particles. The domain outer to the killing zone is a disk of radius $R$ centered at the origin, to which an inner disk $B_{r_k}(\x_k)$ is removed:
\begin{equation*}
\Omega \setminus \overline{B_{r_k}(\x_k)} \; = \; \bigl\{\x\in\R^2 \mid \ \|\x\|<R \quad \text{and} \quad \|\x - \x_k\| > r_k \bigr\}, \qquad \x_k = (x_k,0), \quad x_k + r_k = R\,,
\end{equation*}
such that the inner disk is tangent to the absorbing boundary $\p\Omega_a$ at $\x_a = (R,0)$, defined as
\begin{equation*}
\p\Omega_a \;=\;\left\{\,\y(\theta)=R(\cos\theta,\sin\theta)\ \left| \ \theta\in\left[-\frac{\eps}{R},\frac{\eps}{R}\right]\, \right. \right\}\,.
\end{equation*}
Here the starting point is $\x_0 = O = (0,0)$. We recall the successful exit-point density on the absorbing arc for the first absorber among $N_0$,
\begin{equation}\label{eq:pNk_rate_again}
p_{N_0}(\theta\mid O) \approx \frac{\exp\!\big(-\Phi(\theta,t_{N_0})\big)}{\int_{-\frac{\eps}{R}}^{\frac{\eps}{R}}\exp\!\big(-\Phi(\phi,t_{N_0})\big)R\,d\phi}\,,
\end{equation}
where $\Phi(\theta,t_{N_0}) \equiv \Phi(\theta,t_{N_0}\mid O)$ is the short-time action defined as
\begin{equation}\label{eq:Phi_l_kbar}
\Phi(\theta,t_{N_0}) = \frac{\ell(\theta)^2}{4Dt_{N_0}}+t_{N_0}\,\bar k(\theta,t_{N_0}), \qquad \ell(\theta) = \text{shortest path length in } \Omega \setminus \overline{B_{r_k}(\x_k)} \text{ from } O \text{ to }\y(\theta)\,.
\end{equation}
In the strong obstacle regime, with killing zone tangent to the target, the angular dependence is dominated by $\ell(\theta)$; the transport (or prefactor) only yields subexponential corrections and does not change the leading order localization shape in $\theta$. Note that the straight radial segment $O \longrightarrow \y(\theta)$ will intersect the inner disk if and only if the distance from $\x_k$ to the line $\{s(\cos(\theta),\sin(\theta)) : s\ge 0\}$ is smaller than the killing zone radius $r_k$. The condition for such a blocked regime is
\begin{equation*}
\|\x_k\|\,|\sin\theta| = x_k\,|\sin\theta| < r_k\,,
\end{equation*}
which yields the following critical angle
\begin{equation}\label{eq:theta_c}
\theta_c=\arcsin\!\Big(\frac{r_k}{x_k}\Big)=\arcsin\!\Big(\frac{r_k}{R-r_k}\Big)\,.
\end{equation}
Taking the target half-width $\eps$ to satisfy
\begin{equation}\label{eq:eps_blocked}
0<\eps<\theta_c\,,
\end{equation}
we obtain that each window point is ``shadowed'' by the obstacle: the shortest path must therefore go around the inner disk. In this case there are two symmetric families of minimizers (upper/lower). For $\theta\in\left[0,\frac{\eps}{R}\right]$ the minimizer goes above the disk; for $\theta\in\left[-\frac{\eps}{R}, 0\right]$ it goes below. Consequently, $\ell(\theta)=\ell(|\theta|)$ and $\Phi(\theta,t)=\Phi(|\theta|,t)$ and the distribution is symmetric.
\subsubsection{Computation of the shortest length}
We fix $\theta\in\left[0,\frac{\eps}{R}\right]$ such that the shortest path in $\Omega$ is the concatenation of two line segments tangent to the killing zone $B_{r_k}(\x_k)$ plus one circular arc segment above it. The length of each segment can be computed as follows:
\begin{enumerate}
\item \textbf{Tangency from $O$ to $B_{r_k}(\x_k)$.} With $A \in \p B_{r_k}(\x_k)$ denoting the tangency point connecting the shortest path to the origin, we compute the length $L_O$ of the tangent segment $O \longrightarrow A$ as
\begin{equation}\label{eq:LO}
L_O = \|O - A\|=\sqrt{x_k^2-r_k^2}\,,
\end{equation}
with
\begin{equation}\label{eq:beta_phiA}
\varphi_A = \pi - \arccos\!\Big(\frac{r_k}{x_k}\Big) \in (\pi/2,\pi)\,,
\end{equation}
the polar angle, around the center $\x_k$, of the upper tangency point $A$.
\item \textbf{Tangency from $\y(\theta)$ to $B_{r_k}(\x_k)$.} Here $C(\theta) \in \p B_{r_k}(\x_k)$ denotes the tangency point connecting $\y(\theta)$ on the target to the killing zone. To compute the length of the tangent segment $C(\theta)\longrightarrow \y(\theta)$, we use the distance from $\y(\theta)$ to the center $\x_k$ of the killing zone,
\begin{equation}\label{eq:s_theta}
s(\theta)=\|\y(\theta)-\x_k\| = \sqrt{R^2+x_k^2-2Rx_k\cos\theta}\,,
\end{equation}
and thus the tangent length from $\y(\theta)$ is
\begin{equation}\label{eq:LY}
L_Y(\theta)=\|C(\theta) - \y(\theta)\|=\sqrt{s(\theta)^2-r_k^2}\,.
\end{equation}
We define $\psi(\theta)$ and $\gamma(\theta)$ as the following angles
\begin{equation}\label{eq:psi_gamma}
\psi(\theta)=\arg\big(\y(\theta) - \x_k\big) = \arctan\!\Big(\frac{R\sin\theta}{R\cos\theta-x_k}\Big), \qquad \gamma(\theta)=\arccos\!\Big(\frac{r_k}{s(\theta)}\Big)\,,
\end{equation}
such that the polar angle around $\x_k$ of the upper tangency point $C(\theta)$ becomes
\begin{equation}\label{eq:phiC}
\varphi_C(\theta)=\psi(\theta)+\gamma(\theta)\,.
\end{equation}
\item \textbf{Circular arc $A \longrightarrow C(\theta)$ above the killing zone.} This is the clockwise obstacle arc from $A$ down to $C(\theta)$ on the upper side, with angle span given by
\begin{equation}\label{eq:Dphi}
\Delta\varphi(\theta)=\varphi_A-\varphi_C(\theta)\in(0,2\pi),
\end{equation}
and arc length equals $r_k\,\Delta\varphi(\theta)$. \\
\end{enumerate}
The total shortest length $\ell(\theta)$ is obtained by summing the length of these three pieces:
\begin{equation}\label{eq:l_theta_explicit}
\ell(\theta) = L_O + L_Y(\theta) + r_k\bigl(\varphi_A-\varphi_C(\theta)\bigr) = L_O + L_Y(\theta) + r_k\bigl(\varphi_A-\psi(\theta)-\gamma(\theta)\bigr), \quad \theta \in \left[0, \frac{\eps}{R} \right]\,,
\end{equation}
and substituting Eq.~\eqref{eq:LO} -- Eq.~\eqref{eq:phiC} yields
\begin{equation}\label{eq:l_theta}
\ell(\theta) = \sqrt{x_k^2-r_k^2} + \sqrt{s(\theta)^2 - r_k^2} + r_k\left(\pi - \arccos\!\Big(\frac{r_k}{x_k}\Big) - \arctan\!\Big(\frac{R\sin\theta}{R\cos\theta-x_k}\Big) - \arccos\!\Big(\frac{r_k}{s(\theta)}\Big)\right)\,,
\end{equation}
where $s(\theta) = \|\y(\theta) - \x_k\|$. This formula extends to $\theta\in\left[- \frac{\eps}{R}, \frac{\eps}{R}\right]$ by symmetry $\ell(\theta)=\ell(|\theta|)$. A direct Taylor expansion in the narrow window limit as $\theta\downarrow 0$ then yields,
\begin{equation}\label{eq:smalltheta_components}
s(\theta)=\sqrt{r_k^2+Rx_k\,\theta^2}+O(\theta^4), \qquad L_Y(\theta)=\sqrt{Rx_k}\,\theta+O(\theta^3), \qquad \varphi_C(\theta)=\frac{R+\sqrt{Rx_k}}{r_k}\,\theta+O(\theta^3)\,,
\end{equation}
which implies the cusp-like two-branch structure for the total shortest length:
\begin{equation}\label{eq:l_smalltheta}
\ell(\theta)=\ell_0-R|\theta|+O(|\theta|^3), \quad \text{with} \quad \ell_0=\sqrt{x_k^2-r_k^2}+r_k\left(\pi-\arccos\left(\frac{r_k}{x_k}\right)\right)\,.
\end{equation}
If $k(\x)$ does not create an $O(1)$ angular variation on the window at time $t_{N_0}$, or for a killing zone consisting of a pure geometrical obstacle, then we get the following approximation for the short-time action,
\begin{equation}\label{eq:Phi_smalltheta}
\Phi(\theta,t)\approx \frac{\ell(\theta)^2}{4Dt} \approx \frac{\ell_0^2}{4Dt} - \frac{\ell_0R}{2Dt}\,|\theta| + \frac{R^2}{4Dt}\,\theta^2 + O(|\theta|^3), \qquad |\theta|\le \frac{\eps}{R} \ll 1\,.
\end{equation}
To evaluate $l'(\theta)$ we differentiate \eqref{eq:s_theta}--\eqref{eq:psi_gamma} as follows,
\begin{equation}
L_Y'(\theta)
=\frac{Rx_k\sin\theta}{\sqrt{s(\theta)^2-r_k^2}}, \qquad
\psi'(\theta)
=\frac{R\bigl(R-x_k\cos\theta\bigr)}{s(\theta)^2}, \qquad
\gamma'(\theta)
=\frac{r_k\,Rx_k\sin\theta}{s(\theta)^2\sqrt{s(\theta)^2-r_k^2}}\,,
\end{equation}
and for $\theta\in(0,\frac{\eps}{R}]$ this gives,
\begin{equation}\label{eq:lprime_explicit}
\ell'(\theta)
=
L_Y'(\theta)-r_k\bigl(\psi'(\theta)+\gamma'(\theta)\bigr)
=
\frac{Rx_k\sin\theta}{\sqrt{s(\theta)^2-r_k^2}}
-
r_k\left[
\frac{R(R-x_k\cos\theta)}{s(\theta)^2}
+
\frac{r_k\,Rx_k\sin\theta}{s(\theta)^2\sqrt{s(\theta)^2-r_k^2}}
\right]\,.
\end{equation}
\subsubsection{Explicit geometrical formula for the first exit-point distribution}\label{sec:explicit_dis}
In the blocked regime \eqref{eq:eps_blocked}, the large-$N_0$ density \eqref{eq:pNk_rate_again} concentrates near $\theta=\pm\eps/R$ because the minimizers are attained at the narrow arc corners. By linearizing $\Phi(\theta,t)$ near $\theta=\eps/R$ we get,
\begin{equation*}
\Phi(\theta,t) \approx \Phi\left(\frac{\eps}{R},t\right) +
\Phi_{\theta}\left(\frac{\eps}{R},t\right)\,\left(\theta-\frac{\eps}{R}\right), \qquad \theta\uparrow\frac{\eps}{R},
\end{equation*}
with $\Phi_\theta\left(\frac{\eps}{R},t\right)<0$ a boundary minimum. This gives the formula below for the exit-point distribution,
\begin{equation}\label{eq:final_density_endpoint}
p_{N_0}(\theta \mid O) \approx \frac{\exp\!\big(-\lambda\,\left(\frac{\eps}{R}-|\theta|\right)\big)}{\displaystyle \int_{-\frac{\eps}{R}}^{\frac{\eps}{R}}\exp\left(-\lambda \,\left(\frac{\eps}{R}-|\phi|\right)\right)R\,d\phi} = \frac{\lambda\exp\left(-\lambda\left(\frac{\eps}{R}-|\theta|\right)\right)}{2R\bigl(1-e^{-\lambda\eps/R}\bigr)}, \qquad \lambda = -\Phi_\theta\left(\frac{\eps}{R},t_{N_0}\right)\,,
\end{equation}
with shape parameter $\lambda$. This distribution corresponds to a ``two-endpoint'' symmetric density that is exponentially concentrated within boundary layers of width $\sim R/\lambda$ around $\pm\eps/R$. To determine $\lambda$ we must evaluate
\begin{equation}\label{eq:kappa_general}
\lambda  = \left. - \frac{\p}{\p\theta} \left( \frac{\ell(\theta)^2}{4Dt_{N_0}} + t_{N_0} \bar k(\theta,t_{N_0}) \right)\right|_{\theta = \frac{\eps}{R}} = -\frac{\ell\left(\frac{\eps}{R}\right)\,\ell'\left(\frac{\eps}{R}\right)}{2Dt_{N_0}}
\;-\;t_{N_0}\,\partial_\theta \bar k\left(\frac{\eps}{R},t_{N_0 }\right)\,,
\end{equation}
with $\ell\left(\frac{\eps}{R}\right)$ and $\ell'\left(\frac{\eps}{R}\right)$ obtained from Eq.~\eqref{eq:l_theta},
\begin{equation}\label{eq:l_eps}
\begin{split}
\ell\left(\frac{\eps}{R}\right) &= \sqrt{x_k^2-r_k^2} + \sqrt{s\left(\frac{\eps}{R}\right)^2 - r_k^2} \\
& + r_k\left(\pi - \arccos\!\Big(\frac{r_k}{x_k}\Big) - \arctan\!\left(\frac{R\sin\left(\frac{\eps}{R}\right)}{R\cos\left(\frac{\eps}{R}\right) - x_k}\right) - \arccos\!\left(\frac{r_k}{s\left(\frac{\eps}{R}\right)}\right)\right)\,,
\end{split}
\end{equation}
and Eq.~\eqref{eq:lprime_explicit},
\begin{equation}\label{eq:lprime_eps}
\ell'\left(\frac{\eps}{R}\right) = \frac{Rx_k\sin\left(\frac{\eps}{R}\right)}{\sqrt{s\left(\frac{\eps}{R}\right)^2-r_k^2}} - r_k\left[ \frac{R\left(R-x_k\cos\left(\frac{\eps}{R}\right)\right)}{s\left(\frac{\eps}{R}\right)^2} + \frac{r_k\,Rx_k\sin\left(\frac{\eps}{R}\right)}{s\left(\frac{\eps}{R}\right)^2\sqrt{s\left(\frac{\eps}{R}\right)^2-r_k^2}}
\right]\,,
\end{equation}
with $s(\eps/R) = \left\|\y\left( \frac{\eps}{R} \right) - \x_k \right\| = \sqrt{R^2+x_k^2 - 2Rx_k \cos\left( \frac{\eps}{R} \right)}$. We use the characteristic of fastest time $t_{N_0}$ found in Eq.~\eqref{eq:tNk_explicit_quadratic_clean},
\begin{equation}\label{eq:tNk_again}
t_{N_0} \approx \frac{L_{N_0}-\sqrt{L_{N_0}^2-4ab}}{2b}, \quad b > 0, \qquad t_{N_0} \approx \frac{a}{L_{N_0}}, \quad b = 0
\end{equation}
with
\begin{equation*}
a=\frac{\ell\left(\frac{\eps}{R}\right)^2}{4D},
\quad
b=\bar k\left(\frac{\eps}{R},t_{N_0 }\right)\,,
\quad
L_{N_0} \approx
\log\!\left( \frac{N_0}{2\log(R/\eps)Q_{N_0}} \right),
\quad Q_{N_0} = 1 - (1-P_e(\x_0))^{N_0}\,.
\end{equation*}
Note that if $\p_\theta\bar k$ is negligible on the window, then $\lambda$ as defined in Eq.~\eqref{eq:kappa_general} reduces to
\begin{equation}\label{eq:kappa_clean}
\lambda \;\approx\; -\frac{\ell\left(\frac{\eps}{R}\right)\,\ell'\left(\frac{\eps}{R}\right)}{2D\,t_{N_0 }}\,.
\end{equation}
To get a simplified expression for $\lambda$ in the narrow absorbing window limit with $\eps\ll R$, we use the cusp law for the total shortest length $\ell(\theta)$ (Eq.~\eqref{eq:l_smalltheta} with $\theta > 0$)
\begin{equation*}
\ell(\theta) = \ell_0 - R\theta + O(\theta^3), \qquad \ell_0 = \sqrt{x_k^2-r_k^2} + r_k\left(\pi - \arccos\left(\frac{r_k}{x_k}\right)\right)\,,
\end{equation*}
and its derivative $\ell'(\theta)=-R + O(\theta^2)$. By substituting $\theta = \eps/R$ we get
\begin{equation*}
\ell\left(\frac{\eps}{R}\right)=\ell_0 - R\left(\frac{\eps}{R}\right)+O\left(\left(\frac{\eps}{R}\right)^3\right),
\qquad \ell'\left(\frac{\eps}{R}\right) = -R + O\left(\left(\frac{\eps}{R}\right)^2\right)\,.
\end{equation*}
If additionally $b=0$, or if $b$ only affects $t_{N_0}$ but not the $\theta$-shape of the exit-point distribution, then we have
\begin{equation}
t_{N_0} \approx \frac{a}{L_{N_0}} \approx \frac{\ell\left(\frac{\eps}{R}\right)^2}{4D\log\!\left(\frac{N_0}{2\log(R/\eps)Q_{N_0}}\right)}
\end{equation}
and then upon substituting expressions for $t_{N_0}$, $\ell(\eps/R)$ and $\ell'(\eps/R)$ within Eq.~\eqref{eq:kappa_clean} we obtain
\begin{equation}\label{eq:kappa_small_eps_final}
\lambda \;\approx\;
\frac{\ell\left(\frac{\eps}{R}\right)\,R}{2D} \left( \frac{4D}{\ell\left(\frac{\eps}{R}\right)^2} \log\!\left(\frac{N_0}{2\log(R/\eps)Q_{N_0}}\right) \right) = \frac{2R}{\ell\left(\frac{\eps}{R}\right)} \log\!\left(\frac{N_0}{2\log(R/\eps)Q_{N_0}}\right), \qquad \eps \ll R\,,
\end{equation}
and using the approximation $\ell\left(\frac{\eps}{R}\right) \approx \ell_0 = \sqrt{x_k^2-r_k^2} + r_k\left(\pi - \arccos\left(\frac{r_k}{x_k}\right)\right)$ we obtain an asymptotic formula revealing how the dimension of a tangent killing zone to the target affects the slope $\lambda$ of the first-exit point distribution,
\begin{equation}\label{eq:lambda}
\lambda \;\approx\; \frac{2R\log\!\left(\frac{N_0}{2\log(R/\eps)(1-(1-P_e(\x_0))^{N_0})}\right)}{\sqrt{x_k^2-r_k^2} + r_k\left(\pi - \arccos\left(\frac{r_k}{x_k}\right)\right)}, \quad \text{with} \quad r_k = R - x_k, \quad \eps \ll R\,.
\end{equation}
\subsection{Simulated first exit-point distribution and link with asymptotic theory}\label{sec:simulated_exit_point}
\begin{figure}[H]
\centering
\includegraphics[width=0.75\textwidth]{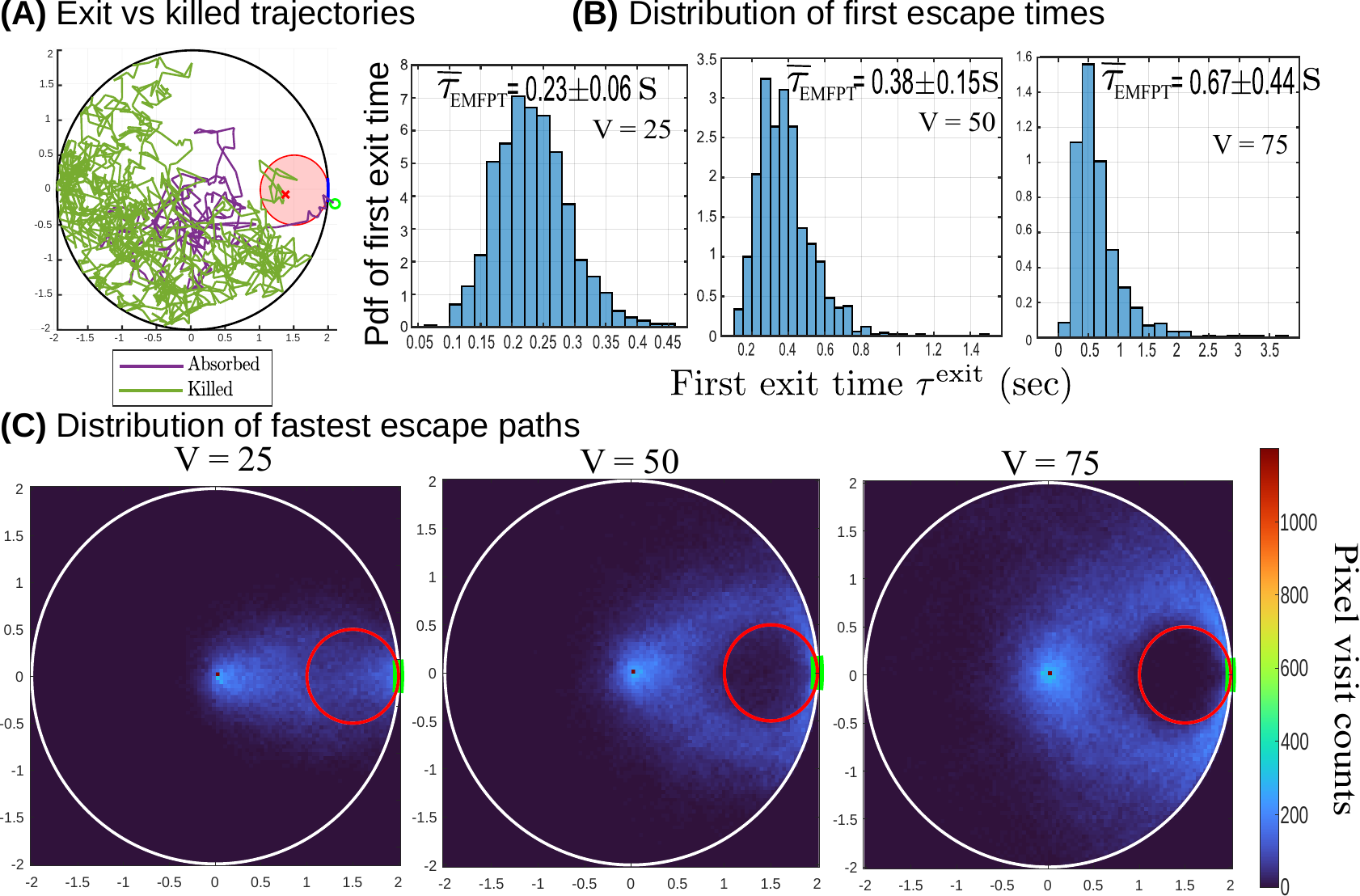}
\caption{\textbf{Killing zone tangent to narrow target.} \textbf{(A)} Example of killed (green path) and absorbed (purple path) trajectories in a domain with killing zone tangent to the target. \textbf{(B)} Distribution of exit times for the fastest from $N_0 = 10^4$ initial particles. \textbf{(C)} Spatial distribution of fastest escape trajectories. Other parameter values are $R=2$, $D=1$, $\eps = \pi/18$, $\x_0 = \left(0,0\right)$, $r_k = 0.5$, $\x_k = (1.5,0)$.}
\label{graph8}
\end{figure}

\begin{figure}[H]
\centering
\includegraphics[width=\textwidth]{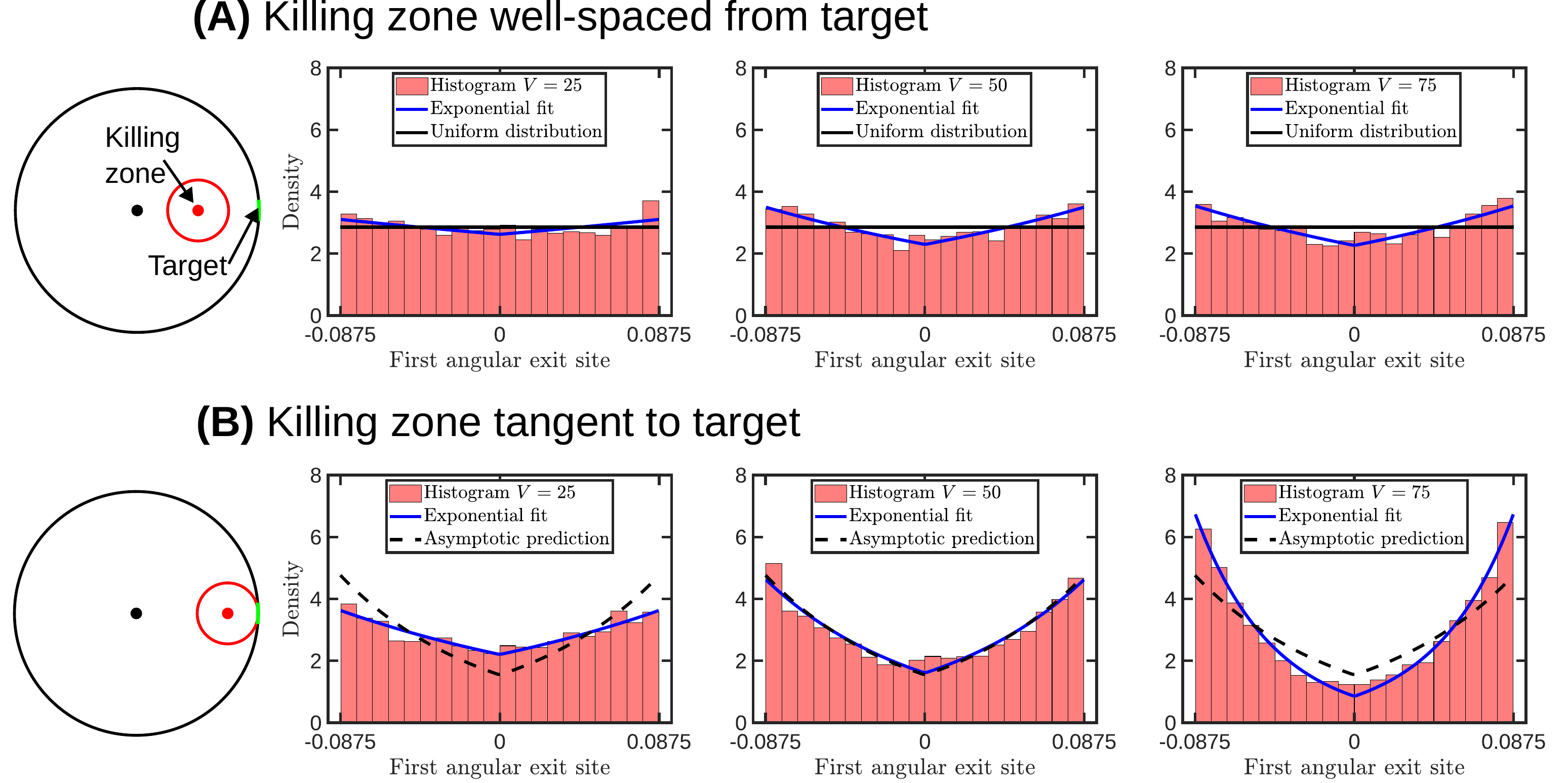}
\caption{\textbf{Distribution of first exit sites along the target.} \textbf{(A)} Well-spaced killing zone from the target, with $\x_k = (1,0)$ and $r_k = 0.5$. Distributions are almost uniform (black full curves). \textbf{(B)} Tangent killing zone to the absorbing target, with $\x_k = (1.5,0)$ and $r_k = 0.5$. Blue curves indicate exponential fit (Eq.~\eqref{eq:sim_exp} with $\lambda$ given in Table \ref{table:fit}) of simulated distributions $p^{\rm sim}(\theta)$ (red histogram densities) obtained from $5000$ repeated releases of $10^4$ particles at the origin. Black dashed curves in \textbf{(B)} indicate asymptotic prediction with $\lambda$ computed using the leading-order formula \eqref{eq:lambda}. Other parameters are $R=2$, $D=1$, $\eps = \pi/18$.}
\label{graph9}
\end{figure}

We now simulate a geometrical configuration in which the killing zone is tangent to the narrow absorbing arc (Fig.~\ref{graph8}\textbf{(A)}): on a disk of radius $R=2$ with release site at the origin this can be achieved by choosing the center and radius of the zone to be $\x_k=(1.5,0)$ and $r_k=0.5$. This geometry violates the ``well-separated'' assumption underlying the standard minimization picture \eqref{minimization}: when the killing is strong (large $V$), the surviving fastest trajectories are constrained to a thin ``safe corridor'' between the killing region and the boundary, which causes the empirical density of fastest paths to concentrate along the boundary (Fig.~\ref{graph8}\textbf{(C)}). Consistently, increasing $V$ shifts the distribution of fastest exit times to the right and flattens its peak (Fig.~\ref{graph8}\textbf{(B)}), reflecting an increase of both the mean fastest time and its variability among successful escapes. For the largest $V$, the spatial distribution also spreads over a wider portion of the domain (Fig.~\ref{graph8}\textbf{(C)}, last panel), a direct consequence of the reduced safe region: successful trajectories diversify by skirting the killing zone at different offsets. \\
\noindent The location of the killing zone relative to the absorbing arc has a decisive impact on the exit-point law. When the killing zone is well-spaced from the target (Fig.~\ref{graph9}\textbf{(A)} with $\x_k=(1,0)$), the boundary flux along the small arc is only weakly perturbed, and the simulated empirical distribution $p^{\rm sim}(\theta)$ of exit points remains close to the uniform density $p^{\rm unif}(\theta) \approx 1/(2\eps)$. In contrast, when the killing zone is tangent to the target (Fig.~\ref{graph9}\textbf{(B)} with $\x_k=(1.5,0)$), a large killing weight $V$ induces a pronounced distortion: exits near the center of the arc become comparatively unlikely and the distribution develops a characteristic U-shaped profile well captured by the symmetric normalized exponential density of shape parameter $\lambda$ derived in \S \ref{sec:explicit_dis} (Eq.~\eqref{eq:final_density_endpoint}), here denoted as
\beq\label{eq:sim_exp}
p_{\lambda}(\theta) = \frac{\lambda}{2R}\frac{e^{\lambda\left(|\theta| - \frac{\eps}{R}\right)}}{1 - e^{-\frac{\lambda\eps}{R}}}, \quad \text{with} \quad |\theta| \leq \frac{\eps}{R}, \quad \int_{-\eps/R}^{\eps/R} p_{\lambda}(\theta) Rd\theta = 1\,.
\eeq
Accordingly, Fig.~\ref{graph9} reports fits of the density $p_{\lambda}(\theta)$ against simulated distributions $p^{\rm sim}(\theta)$ as compact summaries of the well-spaced versus tangent regimes for different killing weight values, where the shape parameter $\lambda^*$ yielding the best fit is recovered via a minimization problem
\beq\label{eq:lambda_opt}
\lambda^* = \text{argmin}_{\lambda \in \R} \left(\int_{-\frac{\eps}{R}}^{\frac{\eps}{R}} \left( p_\lambda(\theta) - p^{\rm sim}(\theta)\right)^2 Rd\theta\right)^{1/2}, \qquad \mathcal{R}^2 = 1 - \frac{\int_{-\frac{\eps}{R}}^{\frac{\eps}{R}} \left( p_{\lambda^*}(\theta) - p^{\rm sim}(\theta)\right)^2 Rd\theta}{\int_{-\frac{\eps}{R}}^{\frac{\eps}{R}} \left(p^{\rm unif}(\theta) - p^{\rm sim}(\theta)\right)^2 Rd\theta}\,,
\eeq
with $\mathcal{R}^2$ the coefficient of determination measuring the goodness-of-fit and how much empirical distributions deviate from the uniform density $p^{\rm unif}(\theta) = \lim_{\lambda \to 0} p_{\lambda}(\theta) = 1/(2\eps)$. Finally Table \ref{table:fit} reports values for $\lambda^*$ and $\mathcal{R}^2$ corresponding to the blue curves fitted densities in Fig.~\ref{graph9}. For a tangent killing zone to the target we also compare with the leading-order asymptotic prediction in the blocked regime with $\lambda$ computed with Eq.~\eqref{eq:lambda} (dashed black curves in Fig.~\ref{graph9}\textbf{(B)}). We find a very good agreement when $V = 50$, deviations observed outside of an intermediate killing range are likely to be caused by higher order killing dependent terms neglected from Eq.~\eqref{eq:lambda}.
\begin{table}[H]
\centering
\begin{tabular}{|c|c|c|}
\hline
\textbf{Rate} $V$ & \textbf{Well-spaced killing zone} (Fig.~\ref{graph9}\textbf{(A)}) & \textbf{Tangent killing zone} (Fig.~\ref{graph9}\textbf{(B)}) \\
\hline
$25$ & $\lambda^* \approx 1.9$ and $\mathcal{R}^2 \approx 0.32$ & $\lambda^* \approx 5.7$ and $\mathcal{R}^2 \approx 0.82$ \\
\hline
$50$ & $\lambda^* \approx 4.8$ and $\mathcal{R}^2 \approx 0.77$ & $\lambda^* \approx 12.1$ and $\mathcal{R}^2 \approx 0.92$ \\
\hline
$75$ & $\lambda^* \approx 5.1$ and $\mathcal{R}^2 \approx 0.75$ & $\lambda^* \approx 23.7$ and $\mathcal{R}^2 \approx 0.98$ \\
\hline
\end{tabular}
\caption{\textbf{Exponential fit of empirical distributions from Fig.~\ref{graph9}.} Comparison between $\lambda^*$ (obtained via the minimization problem \eqref{eq:lambda_opt}) and coefficient $\mathcal{R}^2$ for two regimes (well-spaced vs tangent killing zones to target). In the tangent regime we compute $\lambda \approx 12.88$ using the leading-order asymptotic result (formula \eqref{eq:lambda}). Parameter values the same as in Fig.~\ref{graph9}.}
\label{table:fit}
\end{table}
\section{Discussion and conclusion}\label{sec:discuss}
In this manuscript we analyzed how extreme statistics of stochastic Brownian particles searching for hidden targets in confined domains are affected by degradation sources. In particular we modeled killing with a uniform Poissonian rate on the full domain or over a subregion of it. Such loss phenomena are involved in many physiological processes, from buffers removing calcium ions during signaling \cite{DUPONT201715} to the rapid degradation of transcription factors in the nucleus to regulate responses to external stimuli \cite{cooper2007cell} and to sperm degradation during egg search \cite{sakkas2015}. \\
We introduced the stochastic model in section \S \ref{sec:def_stoch_model}, where we showed how solutions of the Fokker-Planck equation are used to compute escape probabilities, the distribution of escape times and the distribution of first-exit points along a narrow absorbing arc (\S \ref{sec:EMFPT}-\ref{sec:exitpoint_flux}). In section \S \ref{sec:FPE} we used a short-time boundary flux approximation to derive an asymptotic expression for the extreme mean first passage time (EMFPT) to escape alive from a large initial population of size $N_0$, revealing how the $1/\log(N_0)$ scaling law is affected by single particle escape probabilities, with asymptotic formulas for these probabilities derived in appendix \S \ref{sec:escapeProba}. Using stochastic simulations to explore various killing regimes--section \S \ref{sec:simulations}--, we found (\S \ref{sec:uniform}) that killing trajectories do slow down the EMFPT, but one key point is that limit inversion between increasingly large initial populations and variations of other model parameters (killing rate for instance) is not possible. Thus to explore how the EMFPT is affected by variations of the degradation rate, we integrated numerically the full integral formula (Eq.~\eqref{eq:tau_quad}) for two different scenarios, keeping fixed either the number $N_0$ of particles released initially or the number $N_e$ of successful arrivals. Interestingly these two scenarios lead to opposite trends as the killing probability increases: fixing $N_0$ increases the EMFPT, whereas fixing $N_e$ decreases it. In \S \ref{sec:restricted} we report how killing restricted to a small subdisk hiding the target can significantly perturb the distribution of fastest trajectories at large killing intensity: these deviate from straight paths connecting the source and target to the shortest geodesics circumventing the killing zone, as if it were a pure geometrical obstacle \cite{basnayake2018PLA}. Interestingly in section \S \ref{sec:optimal_path} we show that these optimal trajectories are solution of a variational problem \cite{freidlin1998random}. We then investigated in section \S \ref{sec:explicit_exitpoint} the distribution of first-exit points along a narrow absorbing arc, considering both uniform (\S \ref{sec:rho_uniform_killing}) and non-uniform (\S \ref{sec:rate-function-killing}) killing cases. For degradation fields restricted to narrow subregions, their location with respect to the target has a decisive impact on the shape of the distribution (\S \ref{sec:simulated_exit_point}): almost uniform at low killing rates or for well-spaced configurations, while for tangent configurations the first exit point density is exponentially distributed with shape parameter depending on symmetric geodesic paths avoiding the killing zone. \\
To conclude, in addition to extend to 3-D domains and degradation areas, it would be interesting to study the competition between killing and absorption for multiple small interior absorbing traps replacing a single boundary target. Another relevant problem would be to analyze how various spatially heterogeneous killing fields, modeled with exponential \cite{lagache2009} or quadratic \cite{mazzolo2022} rate functions, affect the distribution of fastest escape trajectories. \\
A possible application of our study concerns the molecular dynamics inside the endoplasmic reticulum (ER) lumen, where the competition between transport and degradation regulate time scales of protein exits. Newly synthesized proteins diffuse through the ER network toward narrow exit sites, where export takes place, while misfolded proteins are retained and eliminated by ER-associated degradation before they can reach these sites \cite{tsai2010unfolded,zhang2020endoplasmic}. Similarly, calcium ions travelling toward release sites are continuously captured by buffers such as calreticulin, which acts as an effective killing field \cite{basnayake2019fastplos}. In this context, the opposite trends reported in Fig.~\ref{graph3} acquire a direct physiological interpretation: when degradation increases at fixed production (fixed $N_0$), the arrival of the first molecule at an exit site is delayed, whereas when the cell compensates by upregulating synthesis to maintain the number $N_e$ of successful arrivals, as during the unfolded protein response, the first arrival is accelerated. Increasing the number of copies would thus not only restore the export flux but could also shorten the activation time of downstream processes, a generalization of the redundancy principle operating under stress \cite{schuss2019redundancy} that we call the stress redundancy principle. When degradation machinery is spatially restricted, for instance concentrated in specialized ER subdomains, our analysis predicts that the fastest molecules follow geodesics avoiding these regions (\S\ref{sec:optimal_path}), and that the location of the degradation zone relative to an exit site shapes the distribution of first arrival points (\S\ref{sec:simulated_exit_point}). Testing these predictions would require extending the present analysis from two-dimensional domains to the tubular and sheet geometry of the ER network, where luminal transport occurs by diffusion, as was started to be addressed in \cite{dora2023}.
\section*{Acknowledgements}
F.~P.-L.~gratefully acknowledges the support from the Natural Sciences and Engineering Research Council of Canada (NSERC) (Award No 578183-2023) and from the ARC Foundation (Award No ARCPDF120200-20001505) through postdoctoral fellowships. D.~H.~was supported by the European Research Council (ERC) under the European Union’s Horizon 2020 Research and Innovation Program (grant agreement No 882673), and ANR AstroXcite.
\begin{appendix}
\section{Computation of escape probabilities from steady mixed boundary value problems}\label{sec:escapeProba}
In this appendix we derive expressions for the single particle escape probability $P_e(\x_0)$ to reach a narrow absorbing target for a uniform killing field over the full domain (\S \ref{sec:escapeProbaUnif}) or a localized domain subregion (\S \ref{sec:escapeProbaSub}). These derivations rely on solving the boundary value problem for the time-invariant pdf $\pt(\x\mid\x_0) = \int_0^\infty p(\x,t\mid\x_0) dt$ (Eq.~\eqref{eq:BVP} without drift),
\begin{subequations}\label{eq:BVPnodrift}
\begin{align}
D\Delta\pt(\x\mid\x_0) - k(\x)\pt(\x\mid\x_0) = -\delta(\x-\x_0)\,, \quad \x, \in \Omega, \quad \x_0 \in \Omega\,,
\end{align}
with diffusion coefficient $D$ and mixed boundary conditions
\begin{align}
\pt(\x\mid\x_0) = 0, \quad \x \in \p\Omega_a, \quad \jt(\x\mid\x_0) = 0, \quad \x \in \p\Omega_r\,.
\end{align}
\end{subequations}
Note that when the target is a narrow subsection of the full boundary, our computations rely on asymptotic and Green's function methods. We use a Neumann Green's function $G(\x;\y)$ satisfying
\beq\label{gfe_pde}
D\Delta G(\x;\y) = \frac{1}{|\Omega|} - \delta(\x-\y), \quad \x, \y \in \Omega, \quad \n \cdot \left( D \nabla G(\x;\y)\right) = 0, \quad \x \in \p\Omega, \y \in \Omega\,,
\eeq
with the constraint $\int_\Omega G(\x;\y)d\x = 0$, allowing us to express $\pt(\y\mid\x_0)$ for $\y \in \p\Omega$ and $\x_0$ outside of a boundary layer near the target as
\beq\label{eq:green_id}
\pt(\y\mid\x_0) = \overline{\pt}(\x_0) - \int_{\p\Omega_a} G(\x;\y) \jt(\x\mid\x_0) d\x - \int_{\Omega} G(\x;\y) k(\x) \pt(\x\mid\x_0) d\x + G(\x_0;\y) \,,
\eeq
where $\overline{\pt}(\x_0) = \frac{1}{|\Omega|}\int_\Omega \pt(\x\mid\x_0) d\x$.
\subsection{Case 1: Uniform killing field}\label{sec:escapeProbaUnif}
We consider a domain $\Omega$ with a small absorbing target $\p\Omega_a$, with $|\p\Omega_a|/|\p\Omega| \ll 1$, that is a narrow arc of radius $\eps$ centered at $\x_a \in \p\Omega$. The killing field is uniform with $k(\x) \equiv V$ and the initial stochastic particle position release site is outside of a small boundary layer near the exit, thus we have $\|\x_0 - \x_a\| > \eps$. Substituting $\y = \x_a$ and $k(\x) = V$ into Eq.~\eqref{eq:green_id} yields
\beq\label{eq:unifKilling}
\overline{\pt}(\x_0) = \int_{\p\Omega_a} G(\x;\x_a) \jt(\x\mid\x_0) dS_{\x} + V \int_{\Omega} G(\x;\x_a) \pt(\x\mid\x_0) d\x - G(\x_0;\x_a)\,.
\eeq
To evaluate the first integral above we follow \cite{Holcman2015} and approximate the flux by a constant,
\beq\label{eq:fluxcte}
\jt(\x\mid\x_0) \approx \jt(\x_a\mid\x_0), \quad \x \in \p\Omega_a \,,
\eeq
the compatibility condition of system \eqref{eq:BVPnodrift} then gives the constraint
\beq\label{eq:constraint}
\jt(\x_a\mid\x_0) = \frac{1}{2\eps}\left(1 - \int_\Omega k(\x)\pt(\x\mid\x_0) d\x \right) = \frac{1}{2\eps}\left( 1 - V\pi R^2 \overline{\pt}(\x_0) \right) \,.
\eeq
Near the singularity $\x=\x_a$ we split the Green's function into singular and regular parts \cite{ward2005},
\beq\label{eq:expan_bdy_sec}
G(\x;\x_a) \approx - \frac{1}{\pi D}\log\left( \frac{1}{R}\|\x-\x_a\| \right) + U(\x_a;\x_a), \quad \|\x - \x_a\| \ll R, \quad \x \neq \x_a, \quad \|\x_a\| = R\,,
\eeq
and using Eq.~\eqref{eq:fluxcte}-\eqref{eq:expan_bdy_sec} yields
\begin{align}
\int_{\p\Omega_a} G(\x;\x_a) \jt(\x\mid\x_0) d\x &\approx \frac{1}{\eps}\left( 1 - V\pi R^2 \overline{\pt}(\x_0) \right) \int_0^\eps \left(-\frac{1}{\pi D}\log\left(\frac{\rho}{R}\right) + U(\x_a;\x_a)\right) d\rho \,,
\end{align}
with a direct evaluation this is
\beq\label{eq:int1}
\int_{\p\Omega_a} G(\x;\x_a) \jt(\x\mid\x_0) d\x \approx \left(\frac{1}{\pi D} - \frac{VR^2}{D} \overline{\pt}(\x_0)\right) \left(\log\left(\frac{R}{\eps}\right) + 1 + \pi D U(\x_a;\x_a) \right)\,.
\eeq
To approximate the second integral in Eq.~\eqref{eq:unifKilling}, we decompose the domain into an inner region consisting of a half-disk $\Omega \cap B_{\eps}(\x_a)$ of radius $\eps$ around the target and an outer region $\Omega\backslash B_{\eps}(\x_a)$,
\begin{align}
\int_{\Omega} G(\x;\x_a) \pt(\x\mid\x_0) d\x &= \int_{\Omega \backslash B_{\eps}(\x_a)} G(\x;\x_a) \pt(\x\mid\x_0) d\x + \int_{\Omega \cap B_{\eps}(\x_a)} G(\x;\x_a) \pt(\x\mid\x_0) d\x \,, \\
&\approx \overline{\pt}(\x_0) \int_{\Omega \backslash B_{\eps}(\x_a)} G(\x;\x_a) d\x + \pt(\x_a\mid\x_0)\int_{\Omega \cap B_{\eps}(\x_a)} G(\x;\x_a) d\x \,,
\end{align}
but since $\pt(\x_a\mid\x_0) = 0$,
\beq\label{eq:int2}
\int_{\Omega} G(\x;\x_a) \pt(\x\mid\x_0) d\x \approx \overline{\pt} \pi\left(R^2 - \frac{\eps^2}{2}\right) G(\x_0;\x_a) \approx \overline{\pt} \pi R^2 G(\x_0;\x_a) \,.
\eeq
By substituting Eq.~\eqref{eq:int1} and \eqref{eq:int2} within Eq.~\eqref{eq:unifKilling} and solving for $\overline{\pt}$, we obtain
\beq
\overline{\pt}(\x_0) = \frac{ \log\left( \frac{R}{\eps}\right) + 1 + \pi D \left(U(\x_a;\x_a) - G(\x_0;\x_a)\right) }{\pi D + V \pi R^2 \left( \log\left( \frac{R}{\eps} \right) + 1 + \pi D\left(  U(\x_a;\x_a) - G(\x_0;\x_a) \right) \right)} \,,
\eeq
and then for the splitting probability we get
\beq\label{eq:escapeProbaNeuTemp}
P_e(\x_0) = 1 - V\pi R^2 \overline{\pt}(\x_0) = \frac{\frac{D}{VR^2}}{\frac{D}{VR^2} + \log\left( \frac{R}{\eps} \right) + 1 + \pi D\left(  U(\x_a;\x_a) - G(\x_0;\x_a) \right)}\,.
\eeq
To get an explicit expression for this probability, we use the exact solution for the Green's function $G(\x;\y)$ on a disk of radius $R$ \cite{ward2005},
\beq\label{eq:gfun}
G(\x;\y) = \frac{1}{2\pi D}\left(-\log\left(\frac{1}{R}\|\x-\y\|\right) - \log\left(\left\|\frac{1}{R^2}\x\|\y\| - \frac{\y}{\|\y\|}\right\|\right) + \frac{1}{2R^2}\left(\|\x\|^2 + \|\y\|^2\right) - \frac{3}{4}\right)\,,
\eeq
and with regular part $U(\y;\y)$,
\beq\label{eq:regularpart}
U(\y;\y) =
\begin{cases}
-\frac{1}{2\pi D}\left(\log\left(1-\frac{\|\y\|^2}{R^2}\right) - \frac{1}{R^2}\|\y\|^2 + \frac{3}{4}\right) &  \text{ if } \quad \|\y\| < R  \\
\frac{1}{8\pi D} & \text{ if } \quad \|\y\| = R
\end{cases}\,.
\eeq
Since $\|\x_a\| = R$ we calculate
\beq
\pi D \left(U(\x_a;\x_a) - G(\x_0;\x_a)\right) = \frac{1}{4} + \log\left(\frac{1}{R} \|\x_0 - \x_a\| \right) - \frac{\|\x_0\|^2}{4R^2}\,,
\eeq
and then substituting into Eq.~\eqref{eq:escapeProbaNeuTemp} yields the specific expression for $P_e(\x_0)$:
\beq\label{eq:escapeProbaNeu}
P_e(\x_0) = \frac{\frac{D}{VR^2}}{\frac{D}{VR^2} + \log\left( \frac{\|\x_0 - \x_a\|}{\eps} \right) + \frac{5}{4} - \frac{\|\x_0\|^2}{4R^2}} \,.
\eeq
\subsection{Case 2: Localized killing restricted to a small subregion}\label{sec:escapeProbaSub}
We then compute the escape probability when the killing area consists of a subdisk $\Omega_k = B_{r_k}(\x_k) \subset \Omega$ of small radius $r_k \ll R$ and centered around $\x_k$: inside $\Omega_k$,  Brownian particles are terminated at a uniform rate $V$ while outside the rate is zero. The killing area could be located between the initial release site $\x_0$ and the target, thus acting as an obstacle. The killing is $k(\x) = V\mathbbm{1}_{\Omega_k}(\x)$, with $\mathbbm{1}_{\Omega_k}(\x)$ an indicator function. \\
To obtain an asymptotic expression, we substitute $\y = \x_a$ and $\y = \x_k$ within Eq.~\eqref{eq:green_id} and  get a system of equations:
\begin{subequations}\label{eq:sys_loc}
\begin{align}
0 &= \overline{\pt}(\x_0) - \int_{\p\Omega_a} G(\x;\x_a) \jt(\x\mid\x_0) d\x - \int_{\Omega} k(\x) G(\x;\x_a) \pt(\x\mid\x_0) d\x + G(\x_0;\x_a)\,, \\
\pt(\x_k\mid\x_0) &= \overline{\pt}(\x_0) - \int_{\p\Omega_a} G(\x;\x_k) \jt(\x\mid \x_0) d\x - \int_{\Omega} k(\x) G(\x;\x_k) \pt(\x\mid\x_0) d\x + G(\x_0;\x_k)\,,
\end{align}
\end{subequations}
with the constraint (Eq.~\eqref{eq:constraint}) for $\jt(\x_a\mid\x_0)$ that becomes,
\beq\label{eq:constraint_loc}
\jt(\x_a\mid\x_0) = \frac{1}{2\eps}\left(1 - V\int_{\Omega_k} \pt(\x\mid\x_0) d\x \right) \approx \frac{1}{2\eps}\left(1 - V \pi r_k^2 \pt(\x_k\mid\x_0) \right)\,.
\eeq
We first evaluate the integrals involving the killing field, starting with
\beq\label{eq:temp0}
\int_{\Omega} k(\x) G(\x;\x_k) \pt(\x\mid\x_0) d\x = V \int_{\Omega_k} G(\x;\x_k) \pt(\x\mid\x_0) d\x \approx V \pt(\x_k\mid\x_0) \int_{\Omega_k} G(\x;\x_k) d\x \,,
\eeq
and since $\x_k \notin \p\Omega$ we use a different expansion near the singularity than in Eq.~\eqref{eq:expan_bdy_sec},
\beq
G(\x;\x_k) \approx -\frac{1}{2\pi D}\log\left( \frac{1}{R}\|\x-\x_k\| \right) + U(\x_k;\x_k), \quad \|\x-\x_k\| \ll R, \quad \x \neq \x_k, \quad \|\x_k\| < R\,,
\eeq
with which we can approximate the integral \eqref{eq:temp0} as follows
\begin{align}
\int_{\Omega} k(\x) G(\x;\x_k) \pt(\x\mid\x_0) d\x &\approx V \pt(\x_k\mid\x_0) 2\pi \int_0^{r_k} \left( -\frac{\log\left( \frac{\rho}{R} \right)}{2\pi D} + U(\x_k;\x_k) \right) \rho d\rho\,, \nonumber \\
&\approx V \pt(\x_k\mid\x_0)\left(\frac{r_k^2}{D} \left( \frac{1}{2}\log\left( \frac{R}{r_k} \right) + \frac{1}{4} + \pi D U(\x_k;\x_k) \right)\right)\,. \label{eq:temp1}
\end{align}
The initial release and exit sites are well-spaced, thus we have
\beq\label{eq:temp2}
\int_{\Omega} k(\x) G(\x;\x_a) \pt(\x\mid\x_0) d\x \approx V \pi r_k^2 G(\x_k;\x_a) \pt(\x_k\mid\x_0)\,,
\eeq
while for the integrals along the absorbing boundary segment,  we have
\beq\label{eq:temp3}
\int_{\p\Omega_a} G(\x;\x_a) \jt(\x\mid\x_0) d\x = \left( \frac{1}{\pi D} - \frac{V r_k^2}{D}\pt(\x_k\mid\x_0)  \right) \left(\log\left(\frac{R}{\eps} \right) + 1 + \pi D U(\x_a;\x_a) \right)\,,
\eeq
and
\beq\label{eq:temp4}
\int_{\p\Omega_a} G(\x;\x_k) \jt(\x\mid\x_0) d\x = \left(1 - V r_k^2 \pt(\x_k\mid\x_0) \right) G(\x_a;\x_k)\,.
\eeq
By then substituting Eq.~\eqref{eq:temp1}-\eqref{eq:temp4} within system \eqref{eq:sys_loc} we get
\begin{subequations}
\begin{align}
0 =& \,\, \overline{\pt}(\x_0) - \left( \frac{1}{\pi D} - \frac{V r_k^2}{D}\pt(\x_k\mid\x_0)  \right) \left(\log\left(\frac{R}{\eps} \right) + 1 + \pi D U(\x_a;\x_a) \right) \nonumber \\
& - V \pi r_k^2 G(\x_k;\x_a) \pt(\x_k\mid\x_0) + G(\x_0;\x_a)\,, \\
\pt(\x_k\mid\x_0) =& \,\, \overline{\pt}(\x_0) - \left(1 - V r_k^2 \pt(\x_k\mid \x_0) \right) G(\x_a;\x_k) \nonumber \\
& - \frac{Vr_k^2}{D} \pt(\x_k\mid\x_0) \left( \frac{1}{2}\log\left( \frac{R}{r_k} \right) + \frac{1}{4} + \pi D U(\x_k;\x_k) \right) + G(\x_0;\x_k)\,,
\end{align}
\end{subequations}
and upon solving for $\pt(\x_k\mid\x_0)$ we obtain
\begin{equation}
\begin{split}
&\pt(\x_k\mid\x_0) = \\
&\frac{\log\left(\frac{R}{\eps}\right) + 1 + \pi D \left(U(\x_a;\x_a) + G(\x_0;\x_k)  - G(\x_0;\x_a) - G(\x_a;\x_k)\right)}{\pi D + V\pi r_k^2 \left( \log\left(\frac{R}{\eps}\right) + \frac{1}{2}\log\left(\frac{R}{r_k}\right) + \frac{5}{4} + \pi D\left(U(\x_a;\x_a) + U\left(\x_k;\x_k\right) - G\left(\x_a;\x_k\right) - G\left(\x_k;\x_a\right)\right) \right)}\,. 
\end{split}
\end{equation}
To recover the escape probability we calculate
\beq
P_e(\x_0) = 1 - \int_\Omega k(\x)\pt(\x\mid\x_0) d\x \approx 1 - V \pi r_k^2 \pt(\x_k \mid \x_0)\,,
\eeq
which yields
\begin{equation}
\begin{split}
& P_e(\x_0) = \\
& \frac{\frac{D}{V r_k^2} + \frac{1}{2}\log\left(\frac{R}{r_k}\right) + \frac{1}{4} + \pi D\left(U(\x_k;\x_k) + G(\x_0;\x_a) - G(\x_0;\x_k) - G(\x_k;\x_a)\right)}{\frac{D}{V r_k^2} + \frac{1}{2}\log\left(\frac{R}{r_k}\right)+ \log\left(\frac{R}{\eps}\right) + \frac{5}{4} + \pi D\left(U(\x_a;\x_a) + U(\x_k;\x_k) - G(\x_a;\x_k) - G(\x_k;\x_a)\right)}\,. 
\end{split}
\end{equation}
Using exact solutions of the Green's function (Eq.~\eqref{eq:gfun}) and its regular part (Eq.~\eqref{eq:regularpart}) on a disk of radius $R$ leads to the formula:
\begin{equation}\label{eq:split_prob}
\begin{split}
&P_e(\x_0) = \\
&\frac{\frac{D}{V r_k^2} + \frac{1}{2}\log\left(\frac{R}{r_k}\right) + \frac{1}{4} + \log\left( \frac{\|\x_a - \x_k\|}{\|\x_a - \x_0\|} \right) - \frac{1}{2}\log\left( \frac{R^2 - \|\x_k\|^2}{\|\x_0 - \x_k\| R} \right) + \frac{1}{2}\log\left( \left\| \frac{\x_0\|\x_k\|}{R^2} - \frac{\x_k}{\|\x_k\|} \right\| \right)}{\frac{D}{V r_k^2} + \frac{1}{2}\log\left(\frac{R}{r_k}\right) + \log\left(\frac{R}{\eps}\right) + \frac{5}{4} + 2\log\left( \frac{\|\x_a - \x_k\|}{R}  \right) - \frac{1}{2}\log\left(1 - \left(\frac{\|\x_k\|}{R}\right)^2 \right)}\,,
\end{split}
\end{equation}
which shows the influence on the escape probability of the killing rate and of the local configuration and sizes of the release site, absorbing target and killing zone.
\section{Description of stochastic simulation algorithm}\label{sec:algo_simu}
Here we describe the simulation scheme. The simulation domain is a disk of radius $R = 2$ with a narrow absorbing boundary corresponding to a $10^\circ$ arc centered around $\x_a = (R,0)$, thus the arc radius is $\eps = \pi/18$. Starting from a point $\x_0$, stochastic trajectories are discretized with an Euler's scheme,
\beq\label{euler}
\x(t+ \Delta t) =\left\lbrace
\begin{matrix}
\x(t) + \sqrt{2D} \Delta \w(t) & \text{w. p.} \quad 1-k(\x)\mathbbm{1}_{\Omega_k}(\x) \Delta t \\
& \\
\text{Terminated } & \text{w. p.} \quad k(\x)\mathbbm{1}_{\Omega_k}(\x) \Delta t
\end{matrix}\right., \quad \x(0) = \x_0\,,
\eeq
with diffusion coefficient $D=1$, discrete time-step $\Delta t = 0.01$ and indicator function $\mathbbm{1}_{\Omega_k}(\x)$ of killing area, with $\Omega_k$ that could coincide with the full domain. For a constant killing measure $k(\x) = V$, live particles are destroyed at a Poissonian rate $V$ with probability $V \Delta t$ at each time step inside the killing area \cite{schuss2011nonlinear}. \\
\end{appendix}
\bibliographystyle{ieeetr}
\bibliography{bibliography}
\end{document}